\documentclass[11pt,twoside]{article}
\usepackage[pdftex]{graphicx}
\usepackage{amssymb}
\usepackage{amsthm}
\usepackage{times,cite,bm}
\usepackage{graphics,color}
\usepackage{float}
\newcommand{\be}{\begin{eqnarray}}
\newcommand{\ee}{\end{eqnarray}}
\newcommand{\bez}{\begin{eqnarray*}}
\newcommand{\eez}{\end{eqnarray*}}
\newcommand{\bbN}{\mathbb{N}}
\newcommand{\bbT}{\mathbb{T}}
\newcommand{\pa}{\partial}

\theoremstyle{plain}
\newtheorem{theorem}{Theorem}[section]

\newtheorem{proposition}[theorem]{Proposition}
\newtheorem{corollary}[theorem]{Corollary}

\theoremstyle{definition}

\newtheorem{remark}[theorem]{Remark}
\newtheorem{example}[theorem]{Example}

\title{\bf KP line solitons and Tamari lattices\thanks{\copyright 2010 by A. Dimakis and F. M\"uller-Hoissen}}

\author{ {\scshape Aristophanes Dimakis} \\
 Department of Financial and Management Engineering, \\
 University of the Aegean, 41, Kountourioti Str., GR-82100 Chios, Greece \\
 E-mail: \emph{dimakis@aegean.gr}
             \and
 {\scshape Folkert M\"uller-Hoissen} \\
 Max-Planck-Institute for Dynamics and Self-Organization \\
 Bunsenstrasse 10, D-37073 G\"ottingen, Germany \\
 E-mail: \emph{folkert.mueller-hoissen@ds.mpg.de} 
       }

\date{ }

\begin{document}

\renewcommand{\theequation} {\arabic{section}.\arabic{equation}}

\maketitle

\begin{abstract}
The KP-II equation possesses a class of line soliton solutions which can be qualitatively 
described via a tropical approximation as a chain of rooted binary trees, except at ``critical'' events 
where a transition to a different rooted binary tree takes place. 
We prove that these correspond to maximal chains in Tamari lattices (which are poset structures 
on associahedra). We further derive results that allow to compute details of the evolution, 
including the critical events. 
Moreover, we present some insights into the structure of the more general line soliton solutions.
All this yields a characterization of possible evolutions of line soliton 
patterns on a shallow fluid surface (provided that the KP-II approximation applies). 
\end{abstract}

\section{Introduction}
The Kadomtsev-Petviashvili (KP) II equation possesses exact solutions consisting of 
an arbitrary number of line solitons \cite{Zakh+Shab74,Satsuma76,Anke+Free78JFM,Freeman79,Okhu+Wada83}. 
More comprehensive studies of the structure of the rather complex networks emerging in this way have been undertaken quite recently \cite{Bion+Koda03,Koda04,Bion+Chak06,Bion+Chak07,Bion07,Chak+Koda08JPA,Chak+Koda08CM,Chak+Koda09,Chak+Koda10,CLM10,KOT09,Kao+Kodama10,YLK10} 
(see also the review \cite{Koda10} and the references cited therein).
Whereas in these works a classification in terms of the asymptotic behavior at large negative and 
positive times, and large (positive or negative) values of the coordinate transverse to the main 
propagation direction, has been addressed, in the present work we proceed toward an 
understanding of the full evolution. 

It is rather difficult to generate specific line soliton patterns in a laboratory (but see 
\cite{YLK10,Koda10} for recent progress). In order to test the validity of the KP 
approximation, there is at least the possibility to generate such networks by chance and then to observe 
their evolution qualitatively, i.e. as a (time-ordered) sequence of certain patterns. 
For a subclass of KP line soliton solutions we demonstrate in this work that the allowed evolutions 
are in correspondence with maximal\footnote{A chain in a partially ordered set (poset) is called 
\emph{maximal} if it is not a proper subchain of another chain.} 
chains in a Tamari lattice \cite{Tama62} (see also
 \cite{Fried+Tamari67,Huang+Tamari72,Pallo86,Pallo87,Pallo03,Pallo09,Bennett+Birkhoff94,Geyer94,Bjoerner+Wachs97,Loday+Ronco98,Loday+Ronco02,Loday02,Aguiar+Sottile06,Early04,Sunic07,Deho10} 
for some work related to Tamari lattices). 

\begin{figure}[H] 
\begin{center} 
\resizebox{6.cm}{!}{
\includegraphics{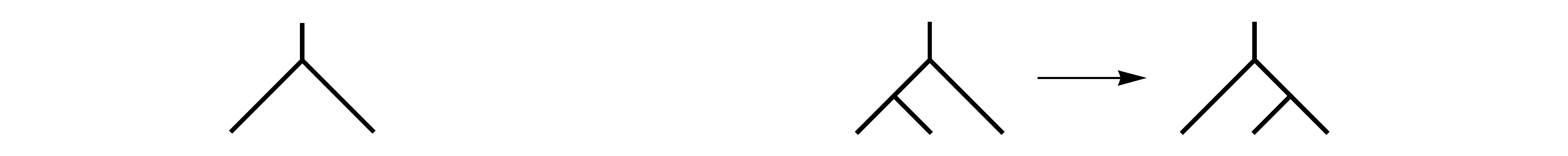}
}
\parbox{15cm}{
\caption{The tree on the left side represents the Tamari lattice $\bbT_1$ which consists of 
a single node. On the right side, a corresponding representation of $\bbT_2$ is shown.
\label{fig:btrees}  }
}
\end{center} 
\end{figure}

The Tamari lattice $\bbT_n$ can be defined as a partially ordered set (poset) in which the elements consist of different ways of grouping a sequence of $n+1$ objects into pairs using parentheses 
(binary bracketing).\footnote{A \emph{lattice} is a poset in which any two elements 
have a unique least upper (with respect to the ordering) 
and a unique greatest lower element. A finite lattice possesses a maximal and a minimal element. 
In case of the Tamari lattices, related to line soliton evolutions in this work, these elements 
correspond to the asymptotic line soliton patterns for large negative, respectively large positive time. }
The partial order is imposed by allowing only a \emph{rightward} application of the associativity law: $(ab)c \to a(bc)$. 
$\bbT_1$ has a single element, $(ab)$, which can also be represented as the rooted binary tree\footnote{In 
this work, a rooted binary tree will always assumed to be planar and proper, i.e. each 
node has exactly two leaves. In counting nodes we only consider ``internal nodes''. 
We draw trees upside down. 
} 
on the left side in Fig.~\ref{fig:btrees}.
$\bbT_2$ is given by $(ab)c \to a(bc)$, which corresponds to the two rooted binary trees on the right 
of Fig.~\ref{fig:btrees}. 
For a sequence of four objects $abcd$, the five possible groupings are $((ab)c)d$, $(a(bc))d$, $a((bc)d)$, $a(b(cd))$ and $(ab)(cd)$. The Tamari lattice $\bbT_3$ then consists of the two chains 
$((ab)c)d \to (a(bc))d \to a((bc)d) \to a (b (cd))$ 
and $((ab)c)d \to (ab)(cd) \to a (b (cd))$, and thus forms a pentagon. 

In section~\ref{sec:simplest_class} we specify the class of KP line soliton solutions, which is 
the central object of this work, and demonstrate their rooted tree structure. 
In section~\ref{sec:evolution} we make further steps toward a classification of such  
solutions as evolutions of rooted trees. This somewhat pedagogical approach is supplemented 
by general results derived in Appendix~\ref{AppA}. Section~\ref{sec:general} presents some insights 
concerning the understanding of general line soliton solutions. 
Section~\ref{sec:conclusions} draws some conclusions and 
briefly summarizes further results, elaborated in additional appendices.

\section{Rooted tree structure of the simplest class of KP line soliton solutions}
\label{sec:simplest_class}
\begin{figure}[H] 
\begin{center} 
\begin{minipage}{5cm}
\resizebox{!}{3.5cm}{
\includegraphics{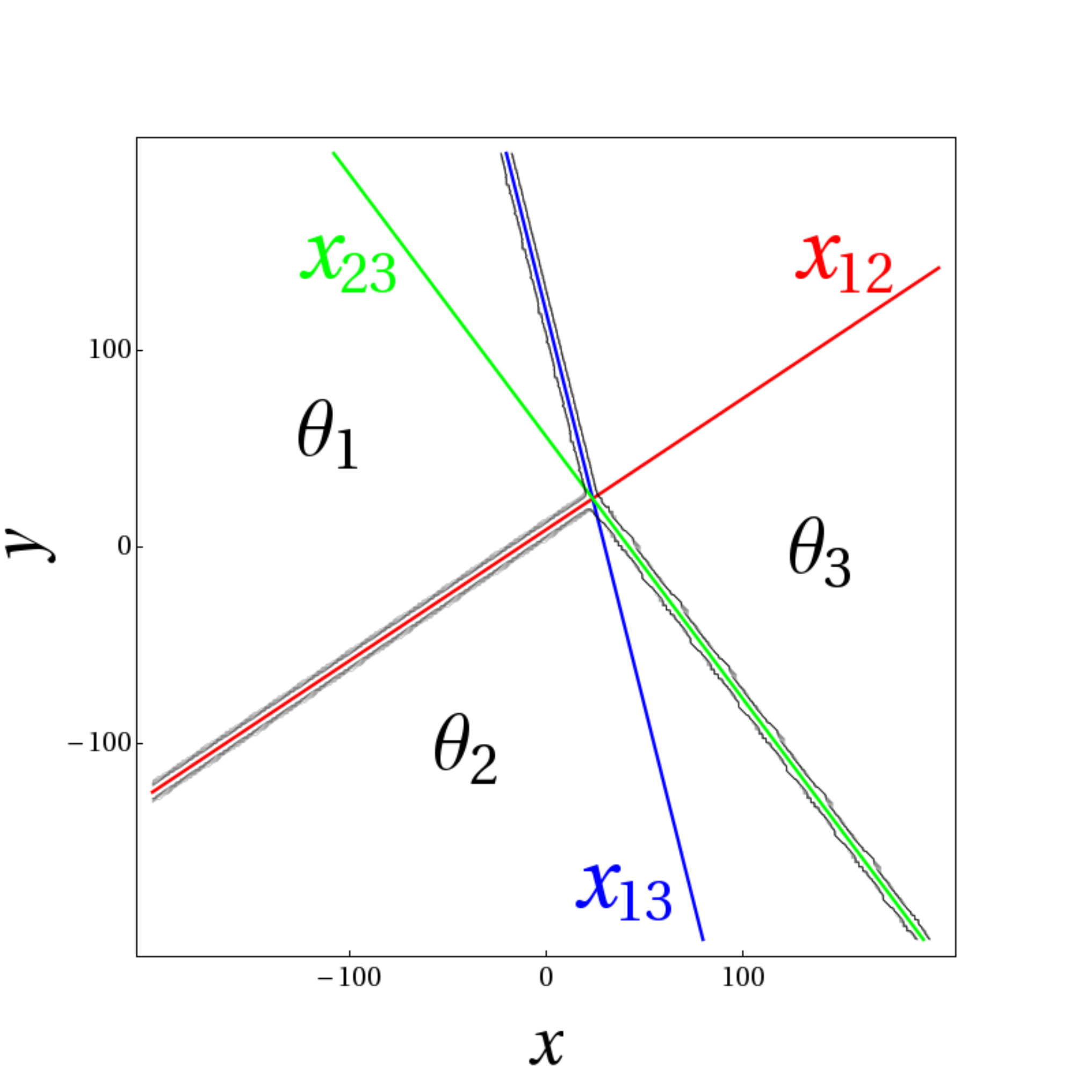}
}
\end{minipage}
\hspace{.5cm}
\begin{minipage}{5cm}
\vspace{-.3cm}
\resizebox{!}{2.5cm}{
\includegraphics{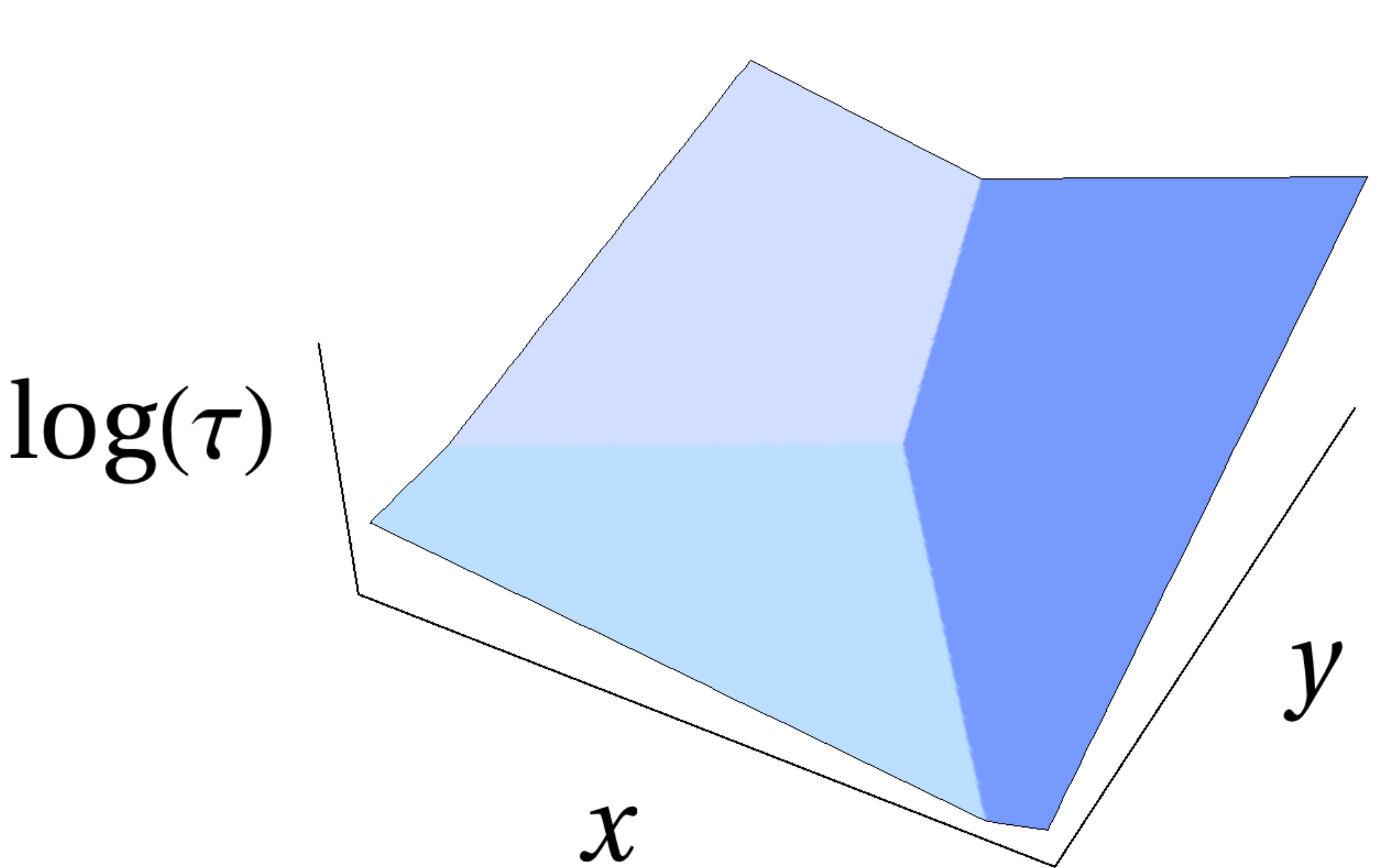}
}
\end{minipage}
\parbox{15cm}{
\caption{A contour plot of a line soliton solution with $M=2$ at a fixed time. This is an example 
of a ``Miles resonance'' \cite{Miles77}.  
The thin lines are boundary lines between two phase regions that are dominated by another phase, 
so that they are not visible in a line soliton plot. 
This solution keeps its form for all times (while moving from the right to the left) and can be 
represented by the trivial Tamari lattice $\bbT_1$. 
The right figure shows a plot of $\log(\tau)$ at the same time. It confirms the tropical description 
of the line soliton solution as the boundary between three planes determined by the tropical function 
$\max\{\theta_1,\theta_2,\theta_3\}$. 
\label{fig:Miles}  }
}
\end{center}
\end{figure}

\setcounter{equation}{0}
Writing the variable $u$ of the KP equation as 
\bez
      u = 2 \, \log(\tau)_{xx} \, , 
\eez 
with a function $\tau$, the KP equation 
\bez
      (-4 \, u_t + u_{xxx} + 6 \, u u_x )_x + 3 \, u_{yy} = 0
\eez
(where e.g. $u_x = \pa u/\pa x$) is transformed into the Hirota bilinear form 
\bez
    4 \, ( \tau \, \tau_{xt} - \tau_x \, \tau_t ) - 3 \, ( \tau \, \tau_{yy} - \tau_y{}^2 )
    - \tau \, \tau_{xxxx} + 4 \, \tau_x \, \tau_{xxx} - 3 \, \tau_{xx}{}^2 = 0 \; .
\eez
The simplest class of line soliton solutions is then given by\footnote{We note that the KP equation 
is invariant under $y \mapsto -y$. Hence for any solution there is another solution obtained by 
reflection with respect to the $x$-axis. This symmetry leaves the class of $\tau$-functions 
specified here, but acts within the more general class considered in section~\ref{sec:general},  
see Example~\ref{ex:y_to_-y}. } 
\be
    \tau = \sum_{k=1}^{M+1} e^{\theta_k} \, , \qquad
    \theta_k = p_k \, x + p_k^2 \, y + c_k \, ,     \label{tau_tree-class}
\ee
provided we make the replacement
\be
      c_k \mapsto p_k^3 \, t + c_k  \; .    \label{ck_t-shift}
\ee
$p_k,c_k$ are real constants and $M \in \bbN$. The absorption of the time variable $t$ 
into the parameter $c_k$ (via the inverse of the above redefinition) is very helpful 
at the moment. Without restriction of generality we can assume that $p_1 < \cdots < p_{M+1}$. 
The $xy$-plane is divided into regions dominated by one of the phases (see also \cite{Bion+Chak06}). 
Let us call the region where $\theta_i > \theta_k$, for all $k \neq i$, the \emph{$\theta_i$-region}. There we have 
\bez
    \log(\tau) = \theta_i + \log\Big( 1 + \sum_{j=1 \atop j \neq i}^{M+1} e^{-(\theta_i-\theta_j)} \Big) 
            \simeq \theta_i \, , 
\eez
where the approximation is valid sufficiently far away from the boundary. Hence
\bez
    \log(\tau) \simeq \max\{\theta_1, \ldots, \theta_{M+1} \} \, ,
\eez
where the right hand side can be regarded as a \emph{tropical} version of $\log(\tau)$ (see 
also Appendix~\ref{AppD}). 
Away from the boundary of a dominating phase region, $\max\{\theta_1, \ldots, \theta_{M+1} \}$ 
is linear in $x$, hence $u$ vanishes. A line soliton branch thus corresponds to a boundary line 
between two dominating phase regions. 
This is the picture that underlies our approach toward a classification of KP line soliton solutions.
For $M=1$ we have a single line soliton. Fig.~\ref{fig:Miles} shows the case $M=2$. 
For $i \neq j$, we have
\bez
    \theta_i - \theta_j = (p_i - p_j) [ x - x_{ij}(y) ] \, ,
\eez
where 
\bez
    x_{ij}(y) = -(p_i+p_j) \, y - c_{ij} = x_{ji}(y) \, , \qquad
    c_{ij} = \frac{c_i - c_j}{p_i - p_j} = c_{ji} \; .
\eez
Hence $\theta_i = \theta_j$ determines the boundary line $x= x_{ij}(y)$ 
between the region where $\theta_i$ dominates $\theta_j$ and the region where $\theta_j$ dominates $\theta_i$.  
Such a line cannot be parallel to the $x$-axis. 
Consequently it divides the plane into a left and a right part. 

\begin{proposition}
For $p_i < p_j$ we have
\bez
      \theta_i  \, \gtrless \, \theta_j \qquad 
      \mbox{for} \qquad x \lessgtr x_{ij}(y,t) \, ,
\eez
i.e. $\theta_i$ dominates $\theta_j$ on the left side of the line $x = x_{ij}(y,t)$, and 
vice versa on the right side. 
\hfill $\square$
\end{proposition}

A particular consequence is that, for all $i=1,\ldots,M+1$, the $\theta_i$-region is convex, 
and thus in particular \emph{connected}.
For $M>1$ we have the identity
\be
     x_{ij}(y) - x_{ik}(y) = (p_k -p_j)(y - y_{ijk}) \qquad \mbox{with} \quad
     y_{ijk} = - c_{ijk} \, ,    \label{x_ij-x_ik}
\ee
where
\bez
   c_{ijk} = \frac{c_{ij} - c_{ik}}{p_j-p_k} 
    &=& \frac{c_i}{ (p_i-p_j)(p_i-p_k)} + \frac{c_j}{ (p_j-p_k)(p_j-p_i)}
      + \frac{c_k}{ (p_k-p_i)(p_k-p_j)}  \\
    &=& \frac{c_i}{ (p_i-p_j)(p_i-p_j)} + \mbox{cyclic permutations}
\eez
is totally symmetric (i.e. invariant under arbitrary permutations of $i,j,k$). 
It follows that the boundary lines $(x_{ij}(y),y)$ and $(x_{ik}(y),y)$ meet in the point 
\bez
       P_{ijk} = (x_{ijk},y_{ijk}) \, ,
\eez
where 
\bez
  x_{ijk} = x_{ij}(y_{ijk}) 
          = - \frac{c_i (p_j^2-p_k^2) + c_j (p_k^2-p_i^2) 
            + c_k (p_i^2-p_j^2)}{(p_i-p_j)(p_j-p_k)(p_k-p_i)}  \; . 
\eez
It further follows that also the line $(x_{jk}(y),y)$ passes through $P_{ijk}$. 
At the ``critical point'' $P_{ijk}$ we have $\theta_i = \theta_j = \theta_k$ 
(see also Fig.~\ref{fig:Miles}).

\begin{proposition}
\label{prop:x-order}
Let $p_i < p_j < p_k$. Then
\bez
   \begin{array}{l}  x_{ij}(y) < x_{ik}(y) < x_{jk}(y)  \\
     x_{ij}(y) > x_{ik}(y) > x_{jk}(y)   \end{array} 
                    \quad \mbox{for} \quad  
   \begin{array}{l}   y < y_{ijk} \\ y > y_{ijk} \end{array} \; .
\eez
\end{proposition}
\noindent
\textit{Proof:} This is an immediate consequence of (\ref{x_ij-x_ik}).  \hfill $\square$
\vskip.2cm

A (part of a) boundary line $x=x_{ij}(y)$ is called \emph{non-visible} if it lies in a region where 
$\theta_i$ (or $\theta_j$) is dominated by another phase. Otherwise it is called 
\emph{visible}. Correspondingly, the critical points can be classified as 
visible or non-visible. Visibility of $P_{ijk}$ means that at this point the 
$\theta_i$-, $\theta_j$-, and $\theta_k$-region meet.

\begin{proposition}
\label{prop:non-vis_halflines}
Let $p_i < p_j < p_k$. Then the half-lines $\{ x=x_{ij}(y) \,|\, y>y_{ijk} \}$,  
$\{ x = x_{jk}(y) \,|\, y>y_{ijk}\}$ and $\{ x = x_{ik}(y) \,|\, y<y_{ijk}\}$ are non-visible.
\end{proposition}
\noindent
\textit{Proof:} The following identity is easily verified, 
\bez
    \theta_k - \theta_i = (p_k - p_i) [ x - x_{ij}(y) + (p_k - p_j)(y - y_{ijk}) ] \; .
\eez
Hence, along $x=x_{ij}(y)$, $y>y_{ijk}$, $\theta_k$ dominates $\theta_i$, so that 
this half-line is non-visible. The same identity written in the form 
\bez
    \theta_i - \theta_k = -(p_k - p_i) [ x - x_{jk}(y) - (p_j - p_i)(y - y_{ijk}) ] \, ,
\eez
respectively
\bez
    \theta_j - \theta_i = (p_j - p_i) [ x - x_{ik}(y) - (p_k - p_j)(y - y_{ijk}) ] \, ,
\eez
implies the non-visibility of the other two half-lines.
\hfill $\square$
\vskip.2cm

\begin{figure}[H] 
\begin{center} 
\resizebox{!}{5.cm}{
\includegraphics{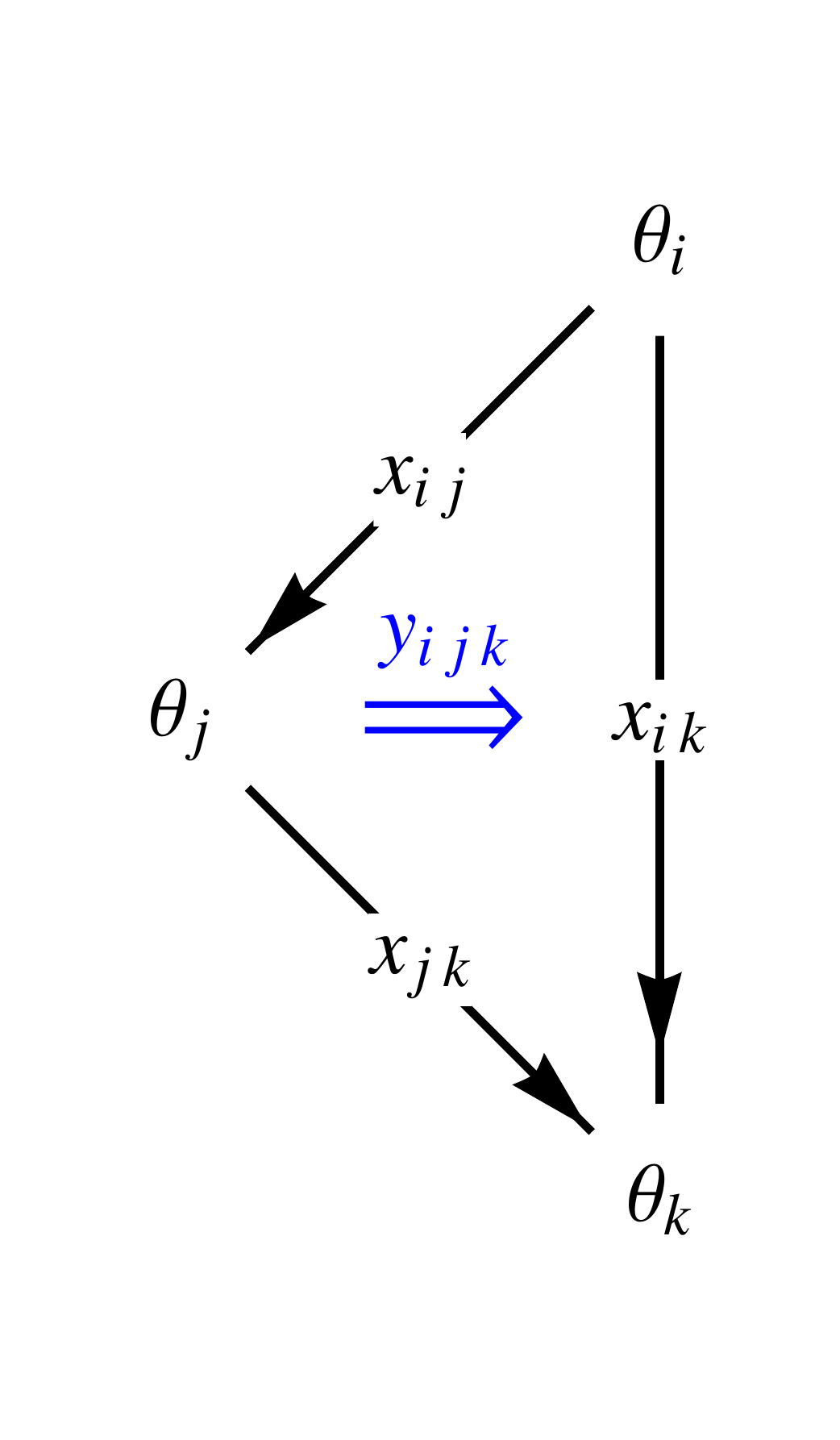}
}
\hspace{1cm}
\resizebox{!}{5.cm}{
\includegraphics{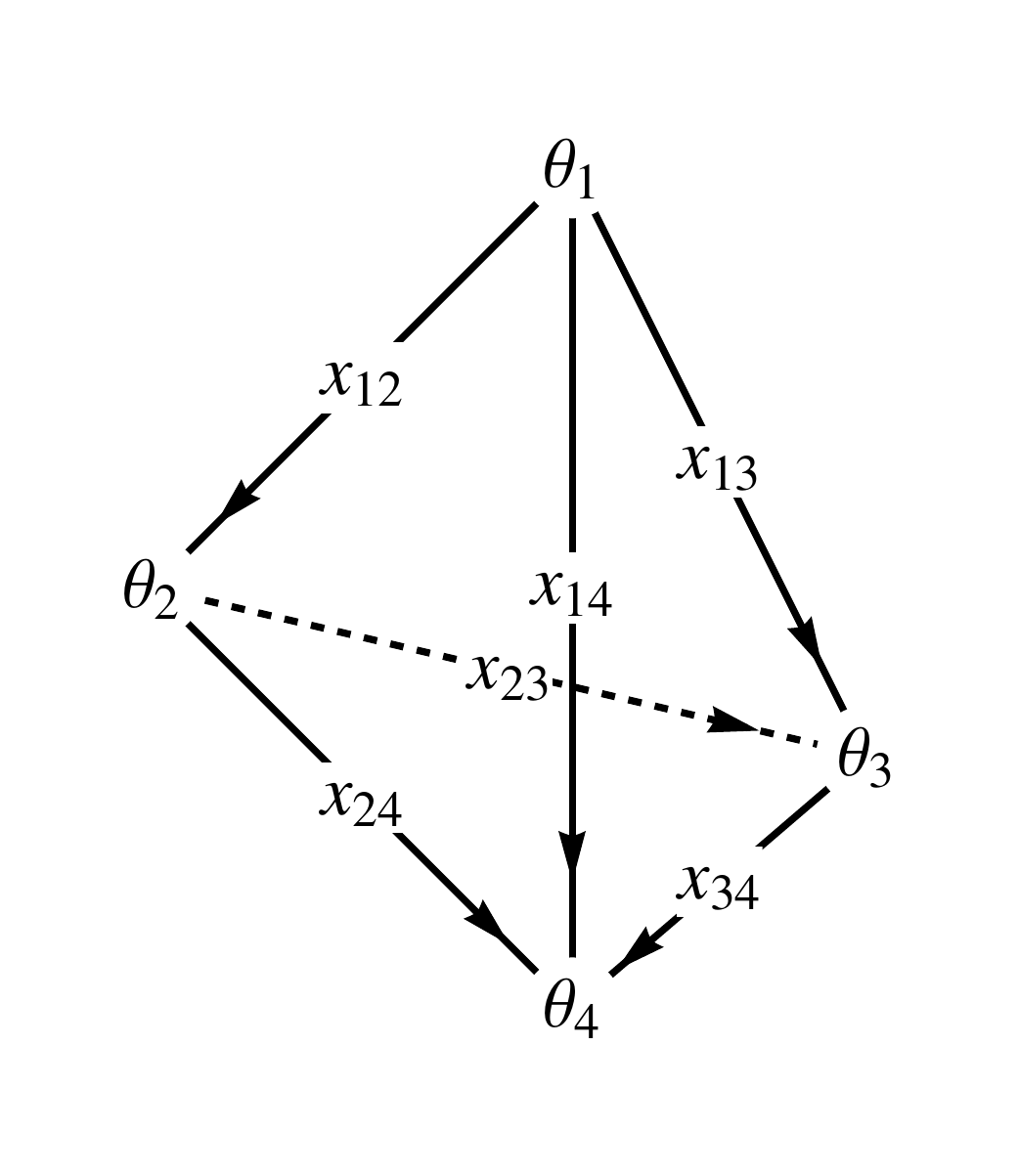}
}
\parbox{15cm}{
\caption{The left figure expresses what happens when $y$ passes a critical value $y_{ijk}$ 
corresponding to a \emph{visible} critical point $P_{ijk}$ with $p_i<p_j<p_k$. 
The nodes coincide if $y=y_{ijk}$. For $y < y_{ijk}$ (left side of the left figure), 
$x_{ij}$ and $x_{ik}$ are visible (but not $x_{ik}$) and their 
order $x_{ij} < x_{ik}$ is expressed by the direction of the edges. For $y > y_{ijk}$ (vertical 
chain in the left figure), $x_{ik}$ is visible, but not the other two. 
The right figure shows the tetrahedron poset obtained in this way for $M=3$ with $p_1<p_2<p_3<p_4$. 
With each of its faces a critical value of $y$ is associated (and a corresponding ``higher order'' arrow), 
as in the left figure. 
Hence, e.g. for $y < \min\{y_{ijk}\}$ we have the chain $x_{12} < x_{23} < x_{34}$. \label{fig:trigon} }
}
\end{center}
\end{figure}

As a consequence of the last proposition, if $P_{ijk}$ is visible, then 
only the half-lines $\{ x = x_{ij}(y) \,|\,  y<y_{ijk}\}$, 
$\{ x = x_{jk}(y) \,|\, y<y_{ijk}\}$ and $\{ x = x_{ik}(y) \,|\, y>y_{ijk}\}$ are visible 
in a neighborhood of $P_{ijk}$ (Fig.~\ref{fig:Miles} shows the case $M=2$). 
Fig.~\ref{fig:trigon} summarizes the \emph{process} connected with the passage of $y$ through a 
critical value, corresponding to a \emph{visible} critical point.\footnote{The 
alert reader will notice that Fig.~\ref{fig:trigon} uses a notation of higher category theory. Indeed, 
the structures appearing in this work provide corresponding examples. } 
This gives a rule to construct a poset for each $M>1$. The nodes are the phases and an edge 
is directed from $\theta_i$ to $\theta_j$ if $i<j$, assuming that $p_1 < \cdots < p_{M+1}$. 
We assign $x_{ij}$ to the corresponding edge. 
For $M=2$ this yields a poset structure on a triangle, for $M=3$ on a tetrahedron 
(see Fig.~\ref{fig:trigon}), 
and more generally on the complete graph on $M+1$ nodes, which can be viewed as an $M$-simplex. 

\begin{proposition}
Let $p_1 < p_2 < \cdots < p_{M+1}$. \\
(1) For $y > \max\{y_{ijk}\}$, only the half-line $x=x_{1,M+1}(y)$ is visible. \\
(2) For $y < \min\{y_{ijk}\}$, all the half-lines $x=x_{m,m+1}(y)$, $m=1,\ldots,M$, are visible, and no other. 
\end{proposition}
\noindent
\textit{Proof:} The following is a special case of the identity already used in the proof of Proposition~\ref{prop:non-vis_halflines}, 
\bez
    \theta_1 - \theta_n = (p_1 - p_n) [ x - x_{1,M+1}(y) - (p_{M+1} - p_n)(y - y_{1,n,M+1}) ] \; .
\eez
This implies $\theta_1 > \theta_n$ along $x = x_{1,M+1}(y)$, $y > \max\{y_{ijk}\}$, for 
$n =2,\ldots,M$. Hence $x=x_{1,M+1}(y)$ is visible for large enough $y$. According to
Proposition~\ref{prop:non-vis_halflines}, all other lines are non-visible for large enough $y$. 
This proves (1). We also have
\bez
    \theta_m - \theta_n = (p_m - p_n) [ x - x_{m,m+1}(y) - (p_{m+1} - p_n)(y - y_{m,m+1,n}) ] \; .
\eez
Along $x = x_{m,m+1}(y)$, $y < \min\{y_{ijk}\}$, it implies $\theta_m > \theta_n$ for all $n \neq m,m+1$. 
As a consequence, this line is visible for large enough negative $y$, and this holds for $m=1,\ldots,M$. 
Again, Proposition~\ref{prop:non-vis_halflines} forbids other lines to be visible for large enough negative $y$, 
and this proves (2). 
\hfill $\square$
\vskip.2cm

\begin{figure}[H] 
\begin{center} 
\resizebox{!}{3.5cm}{
\includegraphics{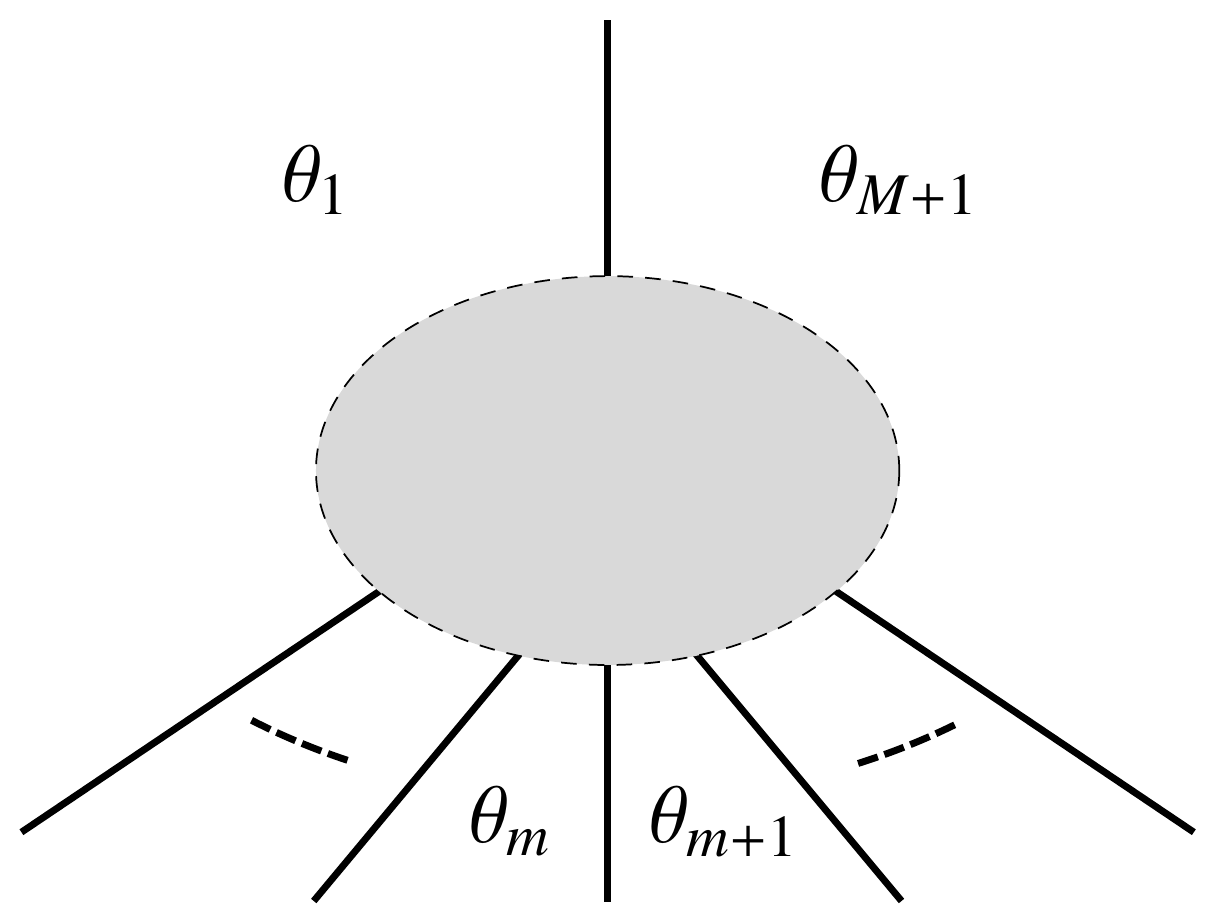}
}
\parbox{15cm}{
\caption{The asymptotic structure of a line soliton graph with $p_1 < p_2 < \cdots < p_{M+1}$.
The order of the dominating phase regions is a consequence of $x_{m,m+1} < x_{m+1,m+2}$, 
$m=1,\ldots,M-1$, for $y$ smaller than all of its critical values, according to 
Proposition~\ref{prop:x-order}. 
\label{fig:asymptotics}  }
}
\end{center}
\end{figure}

The last result (see also \cite{Chak+Koda09,Koda10}) implies the following asymptotic structure of a line 
soliton graph (from the restricted class considered in this section), see Fig.~\ref{fig:asymptotics}. 
For large enough $y$ there is only a single half-line. For large enough negative $y$ one observes $M$ lines. 
In particular, all $M+1$ regions of dominating phase extend to infinity in negative 
$y$-direction. This in turn implies that \emph{no bounded} dominating phase regions exist (since 
the dominating phase regions are connected).\footnote{This is not true for more general line soliton 
solutions (see also section~\ref{sec:general}), outside the class considered here.} 
Hence the graph has the structure of a rooted tree. 

\section{Time evolution of line soliton patterns}
\label{sec:evolution}
\setcounter{equation}{0}

\subsection{The first step}
\label{subsec:first_step}
\begin{figure}[H] 
\begin{center} 
\resizebox{12.cm}{!}{
\includegraphics{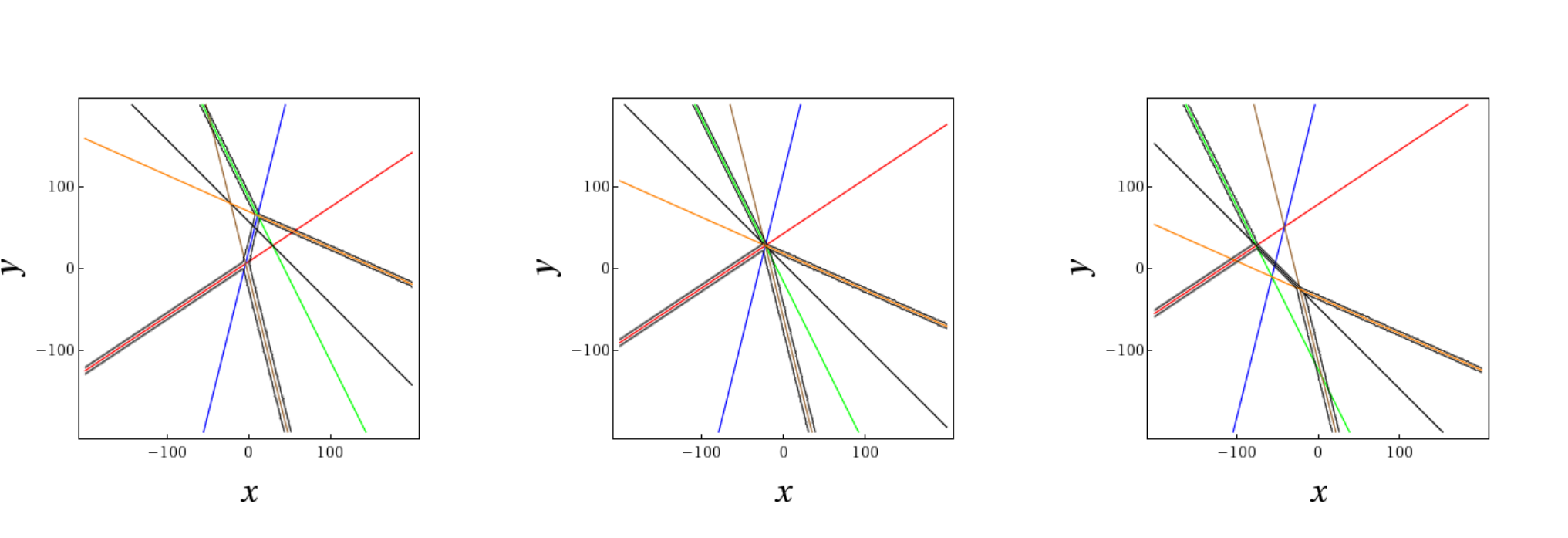}
}
\parbox{15cm}{
\caption{Evolution of a line soliton structure with $M=3$. These are snapshots at times 
$t < t_{1234}$ (left), $t = t_{1234}$ (middle) and $t>t_{1234}$ (right). Again, thin lines are 
boundary lines between two phase regions that are dominated by another phase and hence 
not visible in a line soliton plot.
Disregarding the degenerate configuration at $t = t_{1234}$, this evolution can obviously 
be represented by the single chain of which the Tamari lattice $\bbT_2$ consists 
(see Fig.~\ref{fig:btrees}).
\label{fig:M3}  }
}
\end{center} 
\end{figure}

Let us reintroduce the time variable $t$ (which we hid away in the preceding section) 
via the replacement (\ref{ck_t-shift}). Then we have
\bez
    x_{ij}(y,t) &=& -(p_i+p_j) \, y - (p_i^2 + p_i p_j + p_j^2) \, t - c_{ij} \, , \\
    y_{ijk}(t) &=& - (p_i + p_j + p_k) \, t - c_{ijk} \, , \qquad
    x_{ijk}(t) = x_{ij}(y_{ijk}(t),t) \, ,  
\eez
where $c_{ij}$ and $c_{ijk}$ are given by the previous formulae. The critical points 
$P_{ijk}$ now depend on $t$, hence they constitute ``critical lines'' in $\mathbb{R}^3$ 
(with coordinates $x,y,t$). For $M >2$ we have the identity
\be
    y_{ijk}(t) - y_{ijl}(t) = (p_l -p_k)(t - t_{ijkl}) \qquad \mbox{with} \quad
    t_{ijkl} = - c_{ijkl} \, ,    \label{t_ijkl=-c_ijkl}
\ee
where
\bez
     c_{ijkl}
   = \frac{c_{ijk} - c_{ijl}}{p_k-p_l} 
   &=& \frac{c_i}{(p_i-p_j)(p_i-p_k)(p_i-p_l)} + \frac{c_j}{(p_j-p_i)(p_j-p_k)(p_j-p_l)} \nonumber \\
   &&  + \frac{c_k}{(p_k-p_i)(p_k-p_j)(p_k-p_l)} + \frac{c_l}{(p_l-p_i)(p_l-p_j)(p_l-p_k)} \nonumber \\
   &=& \frac{c_i}{(p_i-p_j)(p_i-p_k)(p_i-p_l)} + \mbox{cyclic permutations}
\eez
is totally symmetric in the indices $i,j,k,l$. 
It follows that two critical points $P_{ijk}(t)$ and $P_{ijl}(t)$ 
coincide (only) at the time given by $t_{ijkl}$. 
Furthermore, at this value of time it turns out that also $P_{ikl}$ and
$P_{jkl}$ coincide with this point. 
Hence we actually have a coincidence of (at least) \emph{four} critical points. 
At the ``critical event''
\bez
      (P_{ijkl}, t_{ijkl}) \in \mathbb{R}^3  \qquad \mbox{with} \qquad   
       P_{ijkl} := P_{ijk}(t_{ijkl}) \, ,
\eez
where the four critical lines intersect, we have $\theta_i = \theta_j = \theta_k = \theta_l$. 
For $M=3$, there is only a single critical time, namely 
$t_{1234}$, and $P_{1234}$ is \emph{visible} at $t=t_{1234}$, i.e. a meeting 
point of line soliton branches (see Fig.~\ref{fig:M3}). 
For $M>3$, there are ${M+1 \choose 4}$ critical times, and the situation is more involved. 
\vskip.1cm

\begin{proposition}
\label{prop:y-order}
Let $p_i < p_j < p_k < p_l$. Then
\bez
   \begin{array}{l}  y_{ijk}(t) < y_{ijl}(t) < y_{ikl}(t) < y_{jkl}(t)  \\
     y_{ijk}(t) > y_{ijl}(t) > y_{ikl}(t) > y_{jkl}(t) 
   \end{array} 
                    \quad \mbox{for} \quad  
   \begin{array}{l}   t < t_{ijkl} \\ t > t_{ijkl} \end{array} \; .
\eez
\end{proposition}
\noindent
\textit{Proof:} This is an immediate consequence of the identity (\ref{t_ijkl=-c_ijkl}). \hfill $\square$

\begin{proposition}
\label{prop:P3_non-visible}
Let $p_i < p_j < p_k < p_l$. Then \\
(1) $P_{ijl}(t)$ and $P_{jkl}(t)$ are non-visible for $t < t_{ijkl}$. \\
(2) $P_{ijk}(t)$ and $P_{ikl}(t)$ are non-visible for $t > t_{ijkl}$.
\end{proposition}
\noindent
\textit{Proof:} An identity used in the proofs of some propositions in section~\ref{sec:simplest_class} 
generalizes via (\ref{ck_t-shift}) to
\bez
   \theta_i - \theta_k = (p_i-p_k)[ x-x_{ij}(y,t) + (p_k-p_j)(y-y_{ijl}(t)) 
                          + (p_k-p_j)(p_k-p_l)(t-t_{ijkl})] \; .
\eez
Evaluating this at $P_{ijl}(t)$, we obtain
\bez
    \theta_i - \theta_k = (p_k-p_i)(p_k-p_j)(p_l-p_k)(t-t_{ijkl}) \, ,
\eez
which is negative if $t < t_{ijkl}$, hence $P_{ijl}(t)$ is then non-visible. 
A similar argument applies in the other cases.
\hfill $\square$

\begin{proposition}
\label{prop:P3_asympt}
(1) For $t < \min\{t_{ijkl}\}$ only the critical points $P_{1,m,m+1}(t)$, $m=2,\ldots,M$, are visible. \\
(2) For $t > \max\{t_{ijkl}\}$ only the critical points $P_{m-1,m,M+1}(t)$, $m=2,\ldots,M$, are visible.
\end{proposition}
\noindent
\textit{Proof:} At $P_{1,m,m+1}(t)$ we have
\bez
     \theta_1 - \theta_n = -(p_n-p_1)(p_n-p_m)(p_n-p_{m+1})(t-t_{1,m,m+1,n}) \, ,
\eez
which, for $t$ smaller than all of its critical values, is positive for all $n$ different from $1,m,m+1$. 
Proposition~\ref{prop:P3_non-visible}, part 1, tells us that all other critical points are non-visible.

At $P_{m-1,m,M+1}(t)$ we have
\bez
     \theta_{M+1} - \theta_n = (p_{M+1}-p_n)(p_n-p_m)(p_n-p_{m-1})(t-t_{n,m-1,m,M+1}) \, ,
\eez
which, for $t$ greater than all of its critical values, is positive for all $n$ different from $m-1,m,M+1$. 
Proposition~\ref{prop:P3_non-visible}, part 2, shows that all other critical points are non-visible. 
\hfill $\square$
\vskip.2cm

\begin{figure}[H] 
\begin{center} 
\resizebox{!}{1.5cm}{
\includegraphics{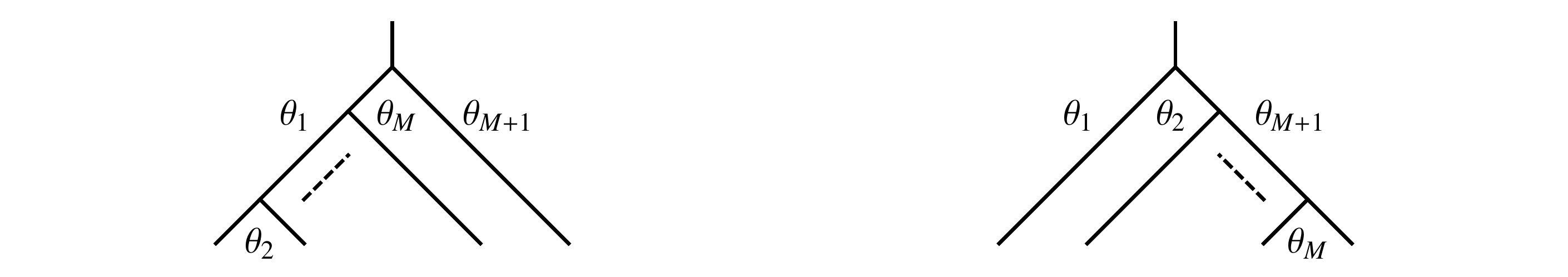}
}
\parbox{15cm}{
\caption{The structure of the solution for $t < \min\{t_{ijkl}\}$ (left tree) and $t > \max\{t_{ijkl}\}$ 
(right tree). For fixed $M$ these two trees form the maximal and the minimal element, respectively, 
of a Tamari lattice. 
\label{fig:graph_asympt}  }
}
\end{center} 
\end{figure}

Collecting our results, for $t$ smaller than all of its critical values the line soliton pattern  
can be represented by the left graph in Fig.~\ref{fig:graph_asympt} (note that 
$y_{1,m,m+1} < y_{1,m+1,m+2}$, $m=2,\ldots,M-1$, according to Proposition~\ref{prop:y-order}), 
and for $t$ greater than all of its critical values by the right graph (since then 
$y_{m,m+1,M+1} > y_{m+1,m+2,M+1}$, $m=1,\ldots,M-2$). 

\begin{figure}[H] 
\begin{center} 
\resizebox{!}{2.cm}{
\includegraphics{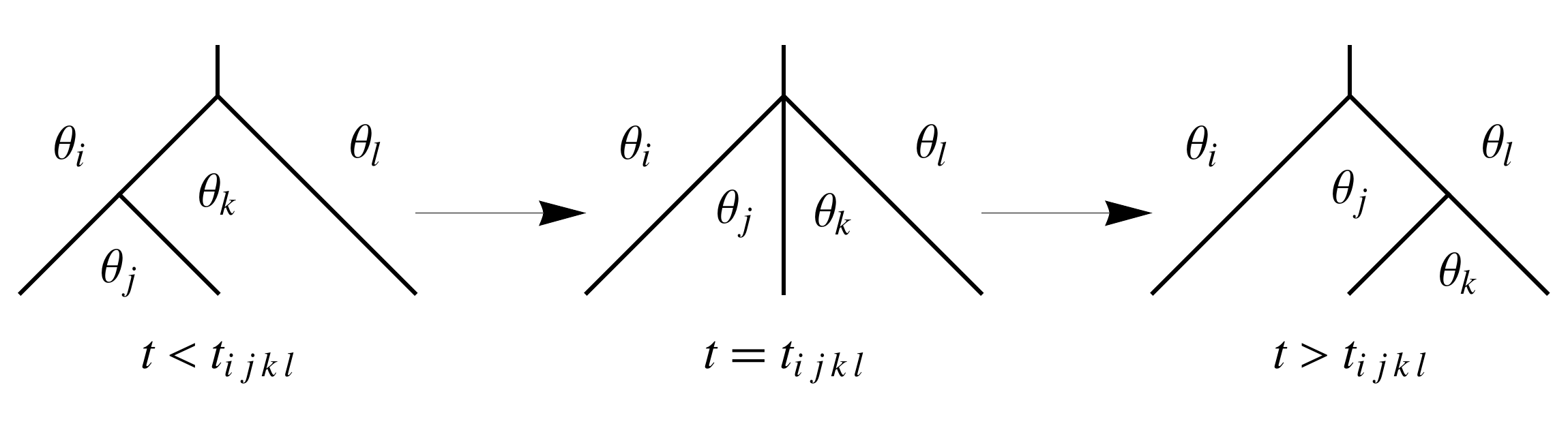}
}
\parbox{15cm}{
\caption{The evolution through a critical time $t_{ijkl}$ with visible critical point $P_{ijkl}$, 
where $p_i < p_j < p_k < p_l$. 
It amounts to a \emph{right-rotation} (see e.g. \cite{Knuth73}) applied to the first binary tree. 
This expresses a central feature of Tamari lattices, 
the rightward application of the associativity law mentioned in the introduction, in the language of 
binary trees (see also \cite{Pallo86,Pallo87,Pallo09,STT88,Loday02,Aguiar+Sottile06}). 
It has to be considered as a \emph{local} process, i.e. a binary tree displayed in this figure 
typically appears as a substructure of a bigger binary tree. 
\label{fig:t_ijkl}  }
}
\end{center}
\end{figure}

Together with Proposition~\ref{prop:P3_non-visible}, the next proposition describes what 
happens when time passes a critical value $t_{ijkl}$ with a 
visible critical point $P_{ijkl}$, see also Fig.~\ref{fig:t_ijkl}. 
In particular it follows that, disregarding the ``degenerate'' cases at a critical time, for $M>1$ 
the graphs have the structure of a \emph{rooted binary tree}.

\begin{proposition}
\label{prop:P3_visible}
Let $p_i < p_j < p_k < p_l$. If $P_{ijkl}$ is visible at $t=t_{ijkl}$ and not a meeting 
point of more than four phases, then \\
(1) $P_{ijk}(t)$ and $P_{ikl}(t)$ are visible for $t < t_{ijkl}$,  \\
(2) $P_{ijl}(t)$ and $P_{jkl}(t)$ are visible for $t > t_{ijkl}$. \\
Here $t$ is assumed to be close enough to $t_{ijkl}$ so that no other critical time with a 
visible critical point is in between. 
\end{proposition}
\noindent
\textit{Proof:} 
For $t$ close enough to $t_{ijkl}$,  
a dominating phase in the vicinity of $P_{ijkl}$ can only be one of the four phases $\theta_i,\theta_j,\theta_k,\theta_l$, as a consequence of the assumptions. 
As in the proof of Proposition~\ref{prop:P3_non-visible}, at $P_{ijl}(t)$ we have
\bez
    \theta_i - \theta_k = (p_k-p_i)(p_k-p_j)(p_l-p_k)(t-t_{ijkl}) \, ,
\eez
which is positive if $t > t_{ijkl}$. This excludes $\theta_k$ as a dominating phase. Since 
$\theta_i=\theta_j=\theta_l$ at $P_{ijl}(t)$, this critical point is visible. 
Clearly, $P_{ijl}(t)$ remains visible unless $t$ takes another critical value with a 
visible critical point. 
A similar argument applies to the other critical points. 
\hfill $\square$
\vskip.2cm

Let us recall that, disregarding critical time values, any line soliton solution from the 
class defined in section~\ref{sec:simplest_class} 
determines a time-ordered sequence of rooted binary trees (with the same number of leaves). Proposition~\ref{prop:P3_visible} tells us 
that the rule according to which the transition from a binary tree to the next takes place is 
precisely the characteristic property of a Tamari lattice (see also Fig.~\ref{fig:t_ijkl}). 
This leads to the following conclusion. 

\begin{theorem}
Each line soliton solution with $\tau$ of the form (\ref{tau_tree-class}), $M>1$, and without 
coincidences\footnote{This restriction ensures that at a critical time only a single ``rotation'' 
takes place. At a coincidence at least two rotations are applied simultaneously and that 
means a direct transition in the Tamari lattice to a more remote neighbor on a chain.}  
of critical times defines a sequence of rooted binary trees which is a maximal chain in a 
Tamari lattice.  \hfill $\square$
\end{theorem}
\vskip.2cm 

Up to $M=5$ we will show explicitly how every maximal chain in $\bbT_{M-1}$ 
is realized by line soliton solutions. 
Propositions~\ref{prop:y-order}, \ref{prop:P3_non-visible} and \ref{prop:P3_visible} 
have generalizations which are elaborated in Appendix~\ref{AppA} and which will be important 
in the following. 
In particular, $x_{ij}, y_{ijk}, t_{ijkl}$ are special cases of (\ref{t^(n-1)_k1...kn}). 

\begin{figure}[H] 
\begin{center} 
\resizebox{!}{2.cm}{
\includegraphics{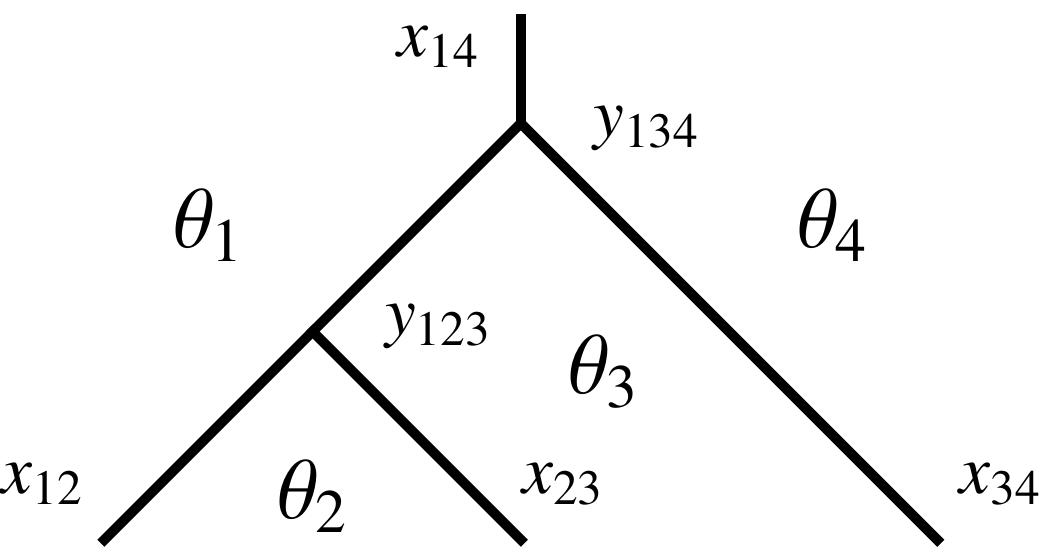}
}
\parbox{15cm}{
\caption{A binary tree constructed from the data $y_{134} > y_{123}$, starting with the root line 
$x_{14}$ and applying the branching rule (\ref{y->x_rule}) consecutively to the two nodes. 
\label{fig:construction_rule} }
}
\end{center} 
\end{figure}

Based on results of section~\ref{sec:simplest_class} (in particular Propositions~\ref{prop:x-order} and
\ref{prop:non-vis_halflines}), a simple recipe to construct soliton binary trees can be formulated. 
A line soliton binary tree at a fixed time is indeed easily constructed from the sequence 
of ordered coordinates $y_{ijk}$ of the visible critical points $P_{ijk}$ via
\be
     x_{ik} \quad \stackrel{y_{ijk}}{\longrightarrow} \quad (x_{ij}, x_{jk}) \, ,
          \label{y->x_rule}
\ee
to be applied in the top to bottom direction (assuming $p_i<p_j<p_k$). Here we understand 
momentarily $x_{ij}$ to represent only the \emph{visible} part of the line between (then dominating) 
phase regions $\theta_i$ and $\theta_j$. 
See Fig.~\ref{fig:construction_rule} and also Appendix~\ref{AppC} for further consequences. 
\noindent
The transition to another binary tree at the critical time $t_{ijkl}$, i.e. the ``rotation'' 
shown in Fig.~\ref{fig:t_ijkl}, can be expressed as 
\be
     (y_{ikl}, y_{ijk}) \quad \stackrel{t_{ijkl}}{\longrightarrow} \quad (y_{ijl}, y_{jkl}) \, ,
          \label{y_t-map}
\ee
assuming $p_i<p_j<p_k<p_l$. Here $(y_{ikl}, y_{ijk})$ is a pair of neighbors in the 
decreasingly ordered sequence of critical $y$-values that determines a rooted binary tree 
associated with a line soliton solution at some event. In order to apply this map, it may be 
necessary to first apply a permutation (see Example~\ref{ex:M=4} below and also Appendix~\ref{AppC}). 
The initial rooted binary tree, 
corresponding to a line soliton solution at large negative values of $t$ (cf. Proposition~\ref{prop:P3_asympt}), 
is determined by the sequence $(y_{1,M,M+1}, y_{1,M-1,M}, \ldots, y_{123})$. 
If we know the order of all critical times $t_{ijkl}$ that correspond to visible events, 
then (\ref{y_t-map}) generates a description of the line soliton evolution as a chain of 
rooted binary trees. 

\begin{remark}
\label{rem:y-poset-family}
In section~\ref{sec:simplest_class} we met a family of posets associated with simplexes, where the 
(directed) edges correspond to the critical values of $x$. There is a new family of posets where the 
nodes are given by the maximal chains in the corresponding poset of the first family. The (directed) 
edges are associated with the critical values of $y$, which are ordered increasingly from 
top to bottom along a chain. Now we note that the 
\emph{process} determined by propositions~\ref{prop:y-order}, \ref{prop:P3_non-visible} 
and \ref{prop:P3_visible}, hence (\ref{y_t-map}), can be expressed as the graph in 
Fig.~\ref{fig:tetragon}. 
\begin{figure}[H] 
\begin{center} 
\resizebox{!}{3.cm}{
\includegraphics{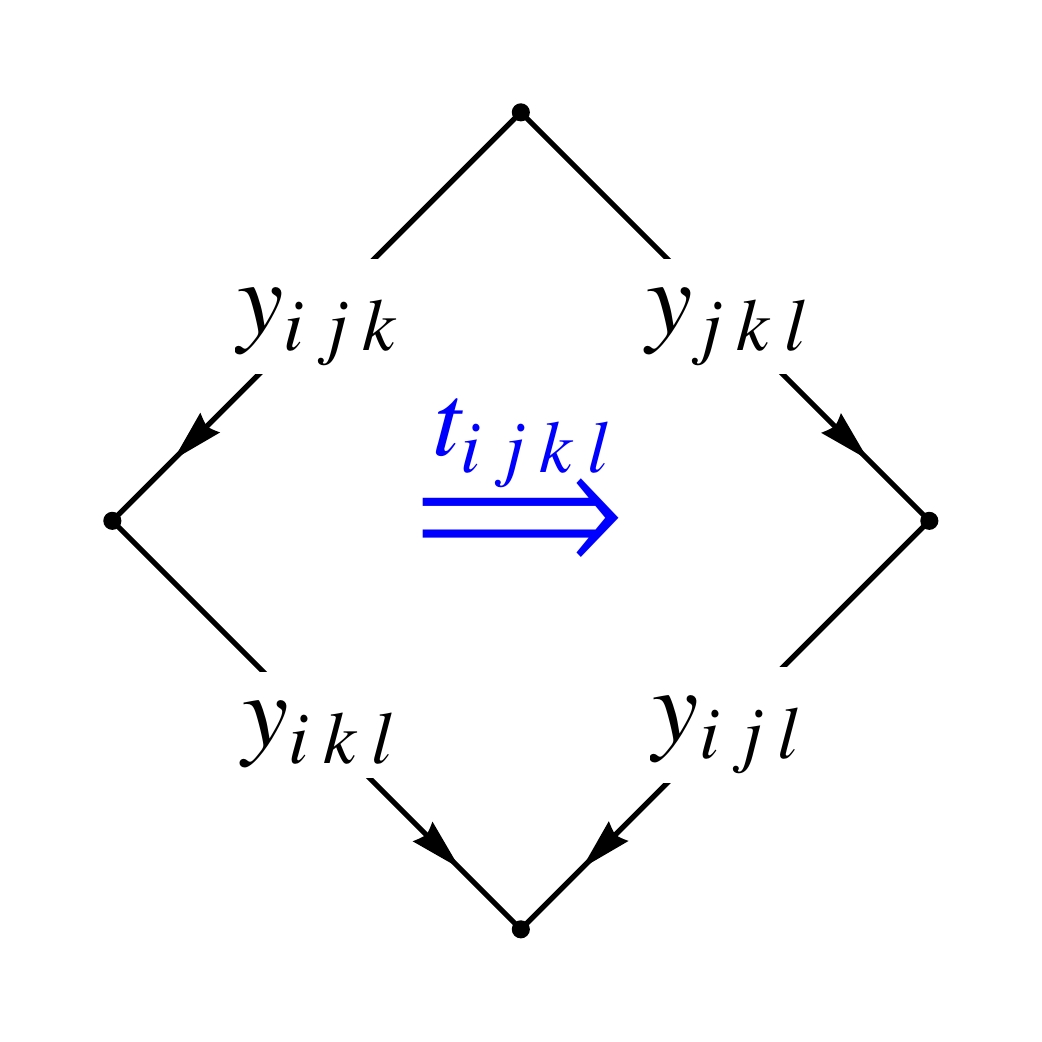}
}
\parbox{15cm}{
\caption{The process connected with the passage of $t$ through a visible critical value $t_{ijkl}$, 
where $p_i<p_j<p_k<p_l$. 
For $t < t_{ijkl}$ (left half of the figure), $y_{jkl}$ and $y_{ijl}$ are \emph{non}-visible, whereas 
$y_{ijk}$ and $y_{ikl}$ are visible (at least if $t$ is greater than all other visible critical 
values that are smaller than $t_{ijkl}$). For $t > t_{ijkl}$ (right half of the figure), the latter 
pair becomes \emph{non}-visible whereas $y_{jkl}$ and $y_{ijl}$ become visible (and remain visible 
at least until $t$ reaches the next critical value).  
  \label{fig:tetragon}  }
}
\end{center} 
\end{figure} 
\noindent
For $M=3$, Fig.~\ref{fig:tetragon} already displays the whole poset, which is thus a tetragon. 
The top node is given by the chain $x_{12} < x_{23} < x_{34}$, the left and right nodes by 
$x_{13}<x_{34}$ and $x_{12}<x_{24}$, respectively, and the bottom node by $x_{14}$. 
These data can be read off from the tetrahedron poset in Fig.~\ref{fig:trigon}. 
For $M=4$, we obtain the cube poset in Fig.~\ref{fig:cube}. The top node is given by the longest 
maximal chain in the $M=4$ simplex poset of the first family, which is 
$x_{12} < x_{23} < x_{34} < x_{45}$. Using the rule expressed by the left graph of Fig.~\ref{fig:trigon}, 
the nodes in the next row are $x_{13} < x_{34} < x_{45}$, $x_{12} < x_{24} < x_{45}$ 
and $x_{12} < x_{23} < x_{35}$, respectively.\footnote{The combinatorics is simpler described 
as follows. Assign the sequence $12345$ to the top node (which stands for the list of 
all phases $\theta_i$, $i=1,\ldots,5$). The next neighbor nodes are obtained by deleting the 
second, third and forth number, respectively. Hence we obtain $1345$, $1245$ and $1235$. 
Each of them has two next lower neighbors, obtained by deleting one of the two numbers in 
the middle. For example, $1345$ is connected with $145$ and $135$. Finally, from these we 
obtain $15$ to represent the bottom node.}
In the next lower row we have 
$x_{14} < x_{45}$, $x_{13} < x_{35}$, and $x_{12} < x_{25}$. The bottom node is given 
by $x_{15}$. For $M>4$ we obtain a hypercube.
\begin{figure}[H] 
\begin{center} 
\resizebox{!}{5.cm}{
\includegraphics{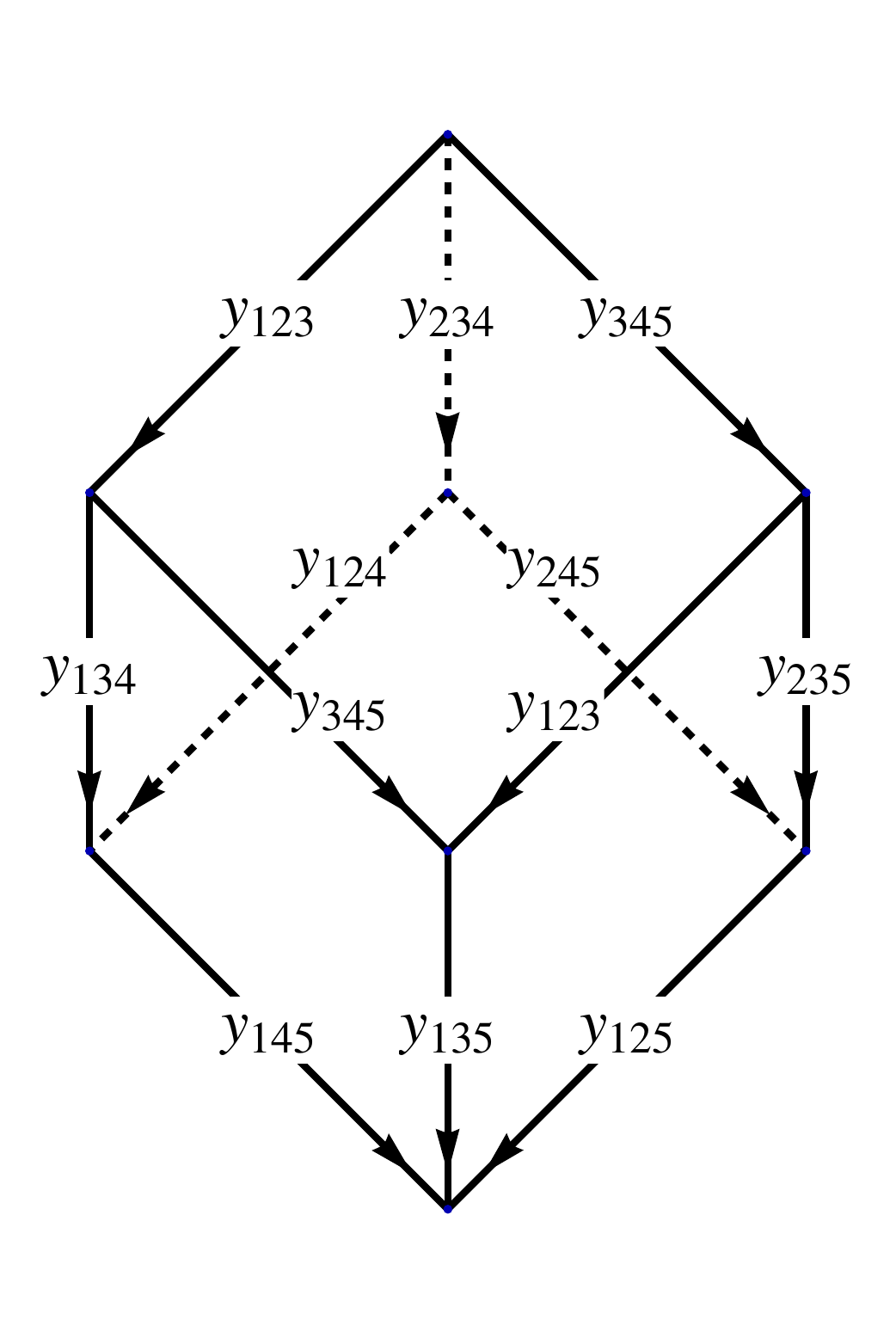}
}
\parbox{15cm}{
\caption{A poset with $M=4$. Five of its faces are associated with the critical values of $t$ 
according to the rule expressed in Fig.~\ref{fig:tetragon}. The top face is an exception. 
It is associated with an additional kind of ``critical value" of $t$, see Appendix~\ref{AppB}. 
 \label{fig:cube}  }
}
\end{center} 
\end{figure} 
\noindent
For $t < \min\{t_{ijkl}\}$ we read off from the cube in Fig.~\ref{fig:cube} the chain 
$y_{123} < y_{134} < y_{145}$, which is the initial (rooted binary tree) configuration. 
If the first critical time is $t_{1234}$, then a transition to the 
tree determined by $y_{234} < y_{124} < y_{145}$ takes place, and for the further 
time development the only possibility is via the critical time $t_{1245}$ to 
$y_{234} < y_{245} < y_{125}$, and afterwards via $t_{2345}$ to $y_{345} < y_{235} < y_{125}$, 
which is the configuration for $t>\max\{t_{ijkl}\}$. If the first critical time is $t_{1345}$, 
then we have a transition to $y_{123} < y_{345} < y_{135}$. As a rooted binary tree, this is 
equivalent to the tree given by $y_{345} < y_{123} < y_{135}$, a transition encoded by the top face 
in Fig.~\ref{fig:cube}. For the latter tree, the only possible further transition is via $t_{1235}$ 
to the unique final configuration for $t>\max\{t_{ijkl}\}$. All this results in the 
Tamari lattice $\bbT_3$ shown in Fig.~\ref{fig:T3} below. We will take a somewhat different route 
to it in order to be able to determine conditions under which the left or the right chain 
is realized, corresponding to which of the critical time values $t_{1234}$ and $t_{1345}$ 
is the smaller one. 
\end{remark}

\subsection{The second step}
\label{subsec:2ndstep}
To further classify the possible line soliton evolutions with $M>3$, we have to look at the 
cases where some of the critical times are equal. This corresponds to particular choices 
of the constants $c_k$. 
In order to analyze this, it turns out to be convenient to redefine the latter via
\bez
        c_k \mapsto p_k^4 \, t^{(4)} + c_k    \qquad \quad k=1,\ldots,M+1 \, ,
\eez
with a new parameter $t^{(4)}$. If $t^{(4)}$ is identified with the next to $t$ evolution variable of the 
KP hierarchy, then the function $u$ (see section~\ref{sec:simplest_class}) also solves the 
second KP hierarchy equation. It should not be a big surprise that the 
hierarchy structure plays a simplifying role in the classification problem of line soliton solutions. 
Let us introduce the complete homogeneous symmetric polynomials
\bez
   h_m(p_{i_1}, \ldots, p_{i_n}) 
  = \sum_{ \alpha_1 + \cdots + \alpha_n = m } p_{i_1}^{\alpha_1} \,  p_{i_2}^{\alpha_2} \cdots 
     \, p_{i_n}^{\alpha_n} \, ,
\eez
where $\alpha_k \in \bbN \cup \{0\}$. Then we have (see also (\ref{t^(n-1)_k1...kn}))
\bez
  x_{ij}(y,t,t^{(4)}) &=& -h_1(p_i,p_j) \, y - h_2(p_i,p_j) \, t 
           - h_3(p_i,p_j) \, t^{(4)} - c_{ij} \, , \\
  x_{ijk}(t,t^{(4)}) &=& x_{ij}(y_{ijk}(t,t^{(4)}),t,t^{(4)}) \, , \\
  y_{ijk}(t,t^{(4)}) &=& -h_1(p_i,p_j,p_k) \, t - h_2(p_i,p_j,p_k) \, t^{(4)} 
                     - c_{ijk} \, , \\
  t_{ijkl}(t^{(4)}) &=& - h_1(p_i,p_j,p_k,p_l) \, t^{(4)} - c_{ijkl} \, ,
\eez
with $c_{ij}, c_{ijk}$ and $c_{ijkl}$ given in terms of $c_i$ by the previous formulae.
We note that now a critical point $P_{ijk}$ depends on $t$ and $t^{(4)}$, hence it  
forms a surface in $\mathbb{R}^4$. The critical point $P_{ijkl}$ depends on 
$t^{(4)}$, hence it forms a line in $\mathbb{R}^4$, which is the intersection of the 
surfaces corresponding to $P_{ijk}, P_{ikl}, P_{ijl}, P_{jkl}$. 

 We find (see also Proposition~\ref{prop:t^(n-1)-diff})
\be
     t_{ijkl}(t^{(4)}) - t_{ijkm}(t^{(4)}) = (p_m-p_l)(t^{(4)}-t^{(4)}_{ijklm}) 
             \qquad \mbox{with} \quad
     t^{(4)}_{ijklm} = - c_{ijklm} \, , 
         \label{tdiff-t^{(4)}_ijklm}
\ee
where
\bez
     c_{ijklm} = \frac{ c_{ijkl} - c_{ijkm} }{ p_l - p_m }
     = \frac{c_i}{(p_i-p_j)(p_i-p_k)(p_i-p_l)(p_i-p_m)} + \mbox{cyclic permutations} \; .
\eez
If two critical times sharing three indices are equal, i.e. $t_{ijkl} = t_{ijkm}$, then 
it follows that (at least) five critical times are equal: $t_{ijkl} = t_{ijkm} = t_{ijlm} 
= t_{iklm} = t_{jklm}$. This happens when $t^{(4)} = t^{(4)}_{ijklm}$. 
At the critical event 
\bez
    (P_{ijklm},t_{ijkl}(t^{(4)}_{ijklm}),t^{(4)}_{ijklm}) \in \mathbb{R}^4 \, ,
\eez
with projection point
\bez
     P_{ijklm} := P_{ijkl}(t^{(4)}_{ijklm}) 
\eez
in the $xy$-plane, we thus have $\theta_i = \theta_j = \theta_k = \theta_l = \theta_m$. 
The following is a direct consequence of (\ref{tdiff-t^{(4)}_ijklm}) (see also 
Proposition~\ref{prop:t^(n-1)-ordering}).

\begin{proposition}
\label{prop:t_ijkl-ordering}
If $p_i < p_j < p_k < p_l < p_m$ we have
\bez
   \begin{array}{l}  t_{ijkl}(t^{(4)}) < t_{ijkm}(t^{(4)}) < t_{ijlm}(t^{(4)}) < t_{iklm}(t^{(4)}) < t_{jklm}(t^{(4)}) \\
     t_{ijkl}(t^{(4)}) > t_{ijkm}(t^{(4)}) > t_{ijlm}(t^{(4)}) > t_{iklm}(t^{(4)}) > t_{jklm}(t^{(4)})
   \end{array} 
                    \quad \mbox{for} \quad  
   \begin{array}{l}   t^{(4)} < t^{(4)}_{ijklm} \\ t^{(4)} > t^{(4)}_{ijklm} \end{array} \; .
\eez
\hfill $\square$
\end{proposition}

The next results are special cases of Propositions~\ref{prop:non-visible_P}
and \ref{prop:visible_P} in Appendix~\ref{AppA}.

\begin{proposition}
\label{prop:P4_non-visible}
Let $p_i < p_j < p_k < p_l < p_m$. Then \\ 
(1) $P_{iklm}(t^{(4)})$ and $P_{ijkm}(t^{(4)})$ are non-visible for $t^{(4)} < t^{(4)}_{ijklm}$.  \\
(2) $P_{ijkl}(t^{(4)}), P_{ijlm}(t^{(4)})$ and $P_{jklm}(t^{(4)})$ are non-visible for 
$t^{(4)} > t^{(4)}_{ijklm}$.
\hfill $\square$
\end{proposition}

\begin{proposition}
\label{prop:P4_visible}
Let $p_i < p_j < p_k < p_l < p_m$ and suppose $P_{ijklm}$ is visible at $t = t_{ijkl}$, 
$t^{(4)} = t^{(4)}_{ijklm}$, and not a meeting point of more than five phases. 
The following holds for values of $t^{(4)}$ that are close enough to $t^{(4)}_{ijklm}$, 
so that no other critical value of $t^{(4)}$ with visible projection point is in between. \\ 
(1) $P_{ijkl}(t^{(4)}), P_{ijlm}(t^{(4)})$ and $P_{jklm}(t^{(4)})$ 
are visible, at the respective critical time, if $t^{(4)} < t^{(4)}_{ijklm}$.\footnote{For example, 
for $t^{(4)} < t^{(4)}_{ijklm}$,  $P_{ijkl}(t^{(4)})$ is visible at $t=t_{ijkl}(t^{(4)})$. } \\
(2) $P_{iklm}(t^{(4)})$ and $P_{ijkm}(t^{(4)})$ are visible, at the respective critical time, 
 if $t^{(4)} > t^{(4)}_{ijklm}$. 
\hfill $\square$
\end{proposition}

\begin{figure}[H] 
\begin{center} 
\resizebox{!}{3.5cm}{
\includegraphics{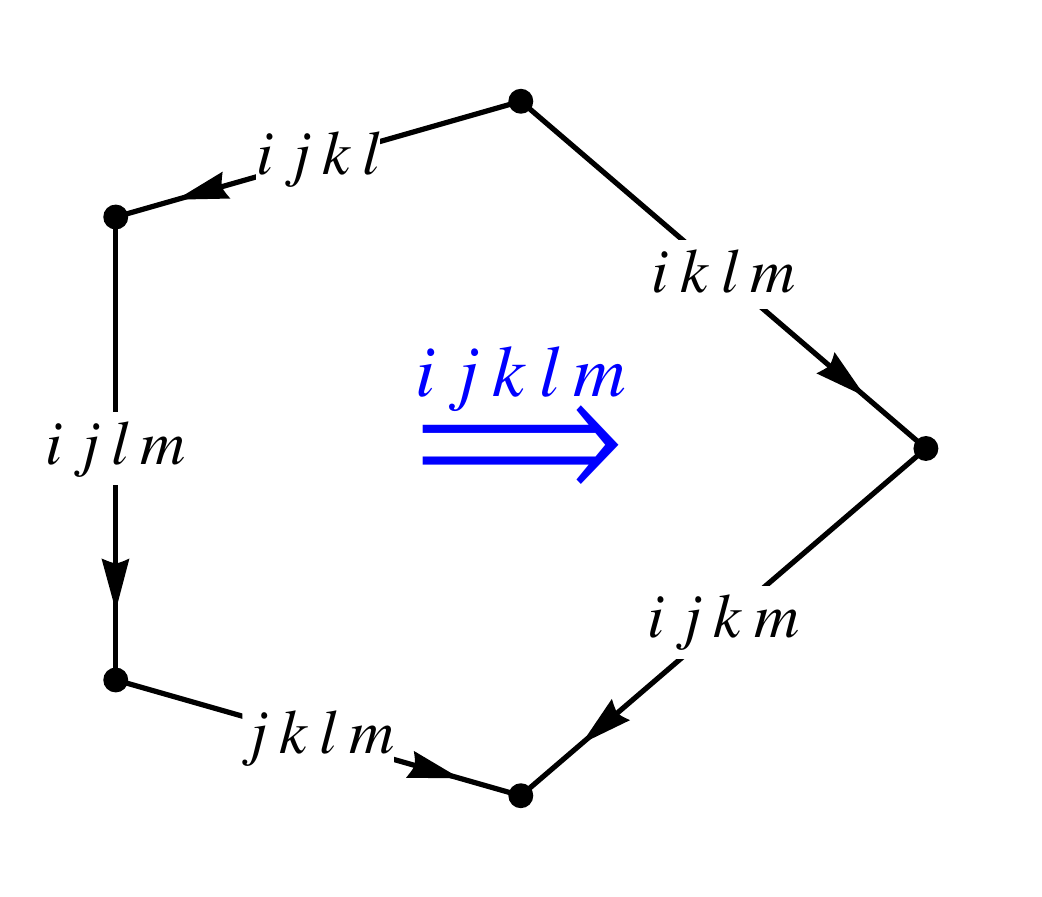}
}
\parbox{15cm}{
\caption{The process connected with the passage of $t^{(4)}$ through a critical value $t^{(4)}_{ijklm}$ 
with \emph{visible} critical point $P_{ijklm}$, where $p_i < p_j < p_k < p_l < p_m$.
Here $ijkl$ stands for $t_{ijkl}$ and $ijklm$ for $t^{(4)}_{ijklm}$. The left chain corresponds to 
$t^{(4)} < t^{(4)}_{ijklm}$, the right to $t^{(4)} > t^{(4)}_{ijklm}$. See also Fig.~\ref{fig:T3} 
for a special case.
\label{fig:pentagon}  }
}
\end{center} 
\end{figure}

Fig.~\ref{fig:pentagon} expresses the subgraph structure determined by  Propositions~\ref{prop:t_ijkl-ordering}, \ref{prop:P4_non-visible}
and \ref{prop:P4_visible} as a \emph{process}.

\begin{example}
\label{ex:M=4}
Let $M=4$. For any fixed $t^{(4)}$, we have \emph{five} critical times 
$t_{1234}(t^{(4)})$, $t_{1235}(t^{(4)})$, $t_{1245}(t^{(4)})$, $t_{1345}(t^{(4)})$, $t_{2345}(t^{(4)})$. 
The corresponding critical events have projection points
$P_{1234}(t^{(4)})$, $P_{1235}(t^{(4)})$, $P_{1245}(t^{(4)})$, $P_{1345}(t^{(4)})$, $P_{2345}(t^{(4)})$, 
at which four phases meet. 
All these critical events coincide for $t^{(4)} = t^{(4)}_{12345}$. At the associated projection 
point $P_{12345}$ all the five phases meet, and it is therefore visible at $t=t_{1234}$ and 
$t^{(4)} = t^{(4)}_{12345}$. 
A description of the evolution of the line soliton pattern thus has to distinguish 
the cases $t^{(4)} < t^{(4)}_{12345}$ and $t^{(4)} > t^{(4)}_{12345}$. 
\\
\textbf{(1)} $t^{(4)} < t^{(4)}_{12345}$. From Propositions~\ref{prop:t_ijkl-ordering}, 
\ref{prop:P4_non-visible} and \ref{prop:P4_visible}, we obtain all ``visible'' critical times and 
they satisfy $t_{1234} < t_{1245} < t_{2345}$. Via (\ref{y_t-map}) this yields
\bez
    (y_{145}, y_{134}, y_{123}) \stackrel{t_{1234}}{\longrightarrow} (y_{145}, y_{124}, y_{234})
   \stackrel{t_{1245}}{\longrightarrow} (y_{125}, y_{245}, y_{234})
   \stackrel{t_{2345}}{\longrightarrow} (y_{125}, y_{235}, y_{345}) \, ,
\eez
which translates into the first sequence of rooted binary trees in Fig.~\ref{fig:T3_chains}. 
\begin{figure}[H] 
\begin{center} 
\resizebox{!}{1.5cm}{
\includegraphics{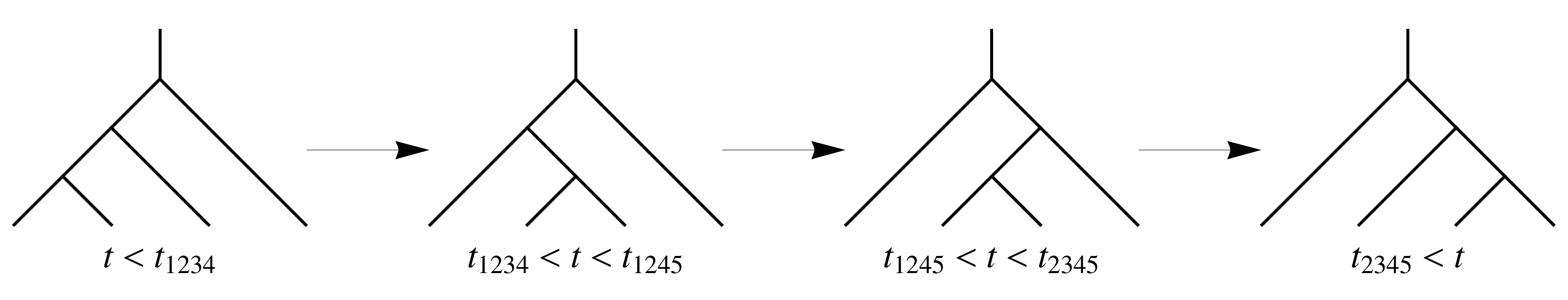} 
}
\hspace{1cm}
\resizebox{!}{1.5cm}{
\includegraphics{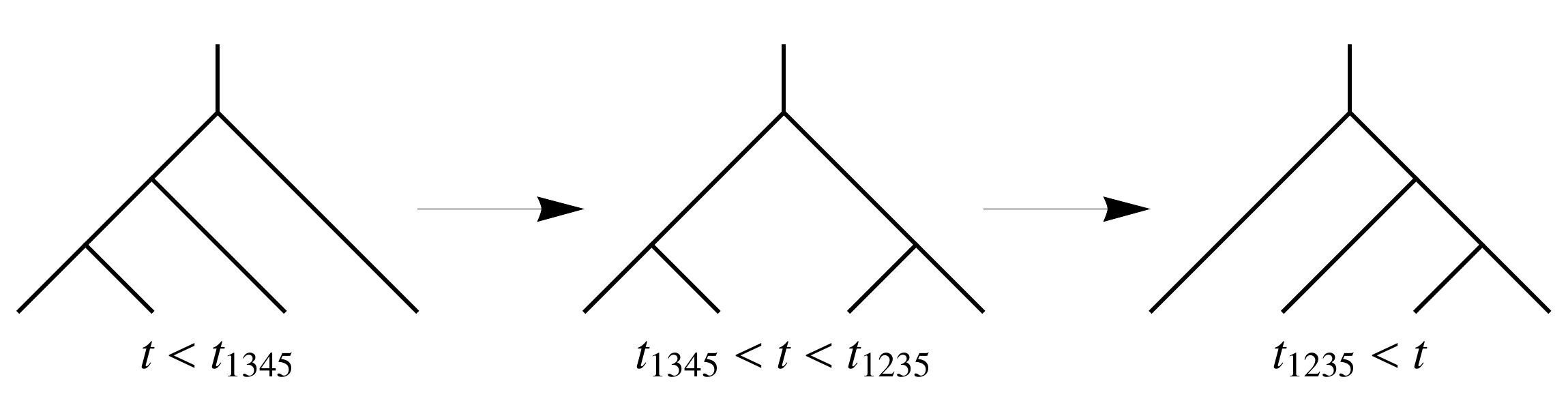} 
}
\parbox{15cm}{
\caption{Evolution of line soliton patterns with $M=4$ and $t^{(4)} < t^{(4)}_{12345}$ (first chain), 
respectively $t^{(4)} > t^{(4)}_{12345}$ (second chain). These are the two maximal chains in the 
Tamari lattice $\bbT_3$, which forms a pentagon (see Fig.~\ref{fig:T3}). 
Instead of assigning $t$-intervals to the trees, it is  
convenient to assign the corresponding critical values $t_{ijkl}$ to the arrows, i.e. the edges 
of the Tamari lattice (as in Figs.~\ref{fig:pentagon} and \ref{fig:T3}). 
\label{fig:T3_chains}  }
}
\end{center}
\end{figure} 
\noindent
\textbf{(2)} $t^{(4)} > t^{(4)}_{12345}$. Then the ``visible'' critical times satisfy 
$t_{1345} < t_{1235}$. This leads to 
\bez
    (y_{145}, y_{134}, y_{123}) \stackrel{t_{1345}}{\longrightarrow} (y_{135}, y_{345}, y_{123})
   \stackrel{\mathrm{permutation}}{\longrightarrow}
    (y_{135}, y_{123}, y_{345})
   \stackrel{t_{1235}}{\longrightarrow} (y_{125}, y_{235}, y_{345}) \, ,
\eez
which translates into the second chain in Fig.~\ref{fig:T3_chains}. 
The tree in the middle allows the two possibilities $y_{123} < y_{345}$ and $y_{345} < y_{123}$ 
(in accordance with Proposition~\ref{prop:y-order}).
A permutation is necessary in order to be able to apply (\ref{y_t-map}) 
with the second critical time to the respective pair of neighbors. This makes sense if we 
regard the two possibilities as equivalent (and this has been done in Fig.~\ref{fig:T3_chains}). 
Resolving the ``fine structure'', by determining the event where $y_{123} = y_{345}$, 
they can be distinguished in a setting of trees with levels \cite{Loday+Ronco98}, 
see Appendix~\ref{AppB}. 
\\
The two sequences of rooted binary trees obtained for $t^{(4)} < t^{(4)}_{12345}$, respectively 
$t^{(4)} > t^{(4)}_{12345}$, are the two maximal chains in the Tamari lattice $\bbT_3$
(see Fig.~\ref{fig:T3}). 
\begin{figure}[H] 
\begin{center}
\resizebox{!}{3.5cm}{
\includegraphics{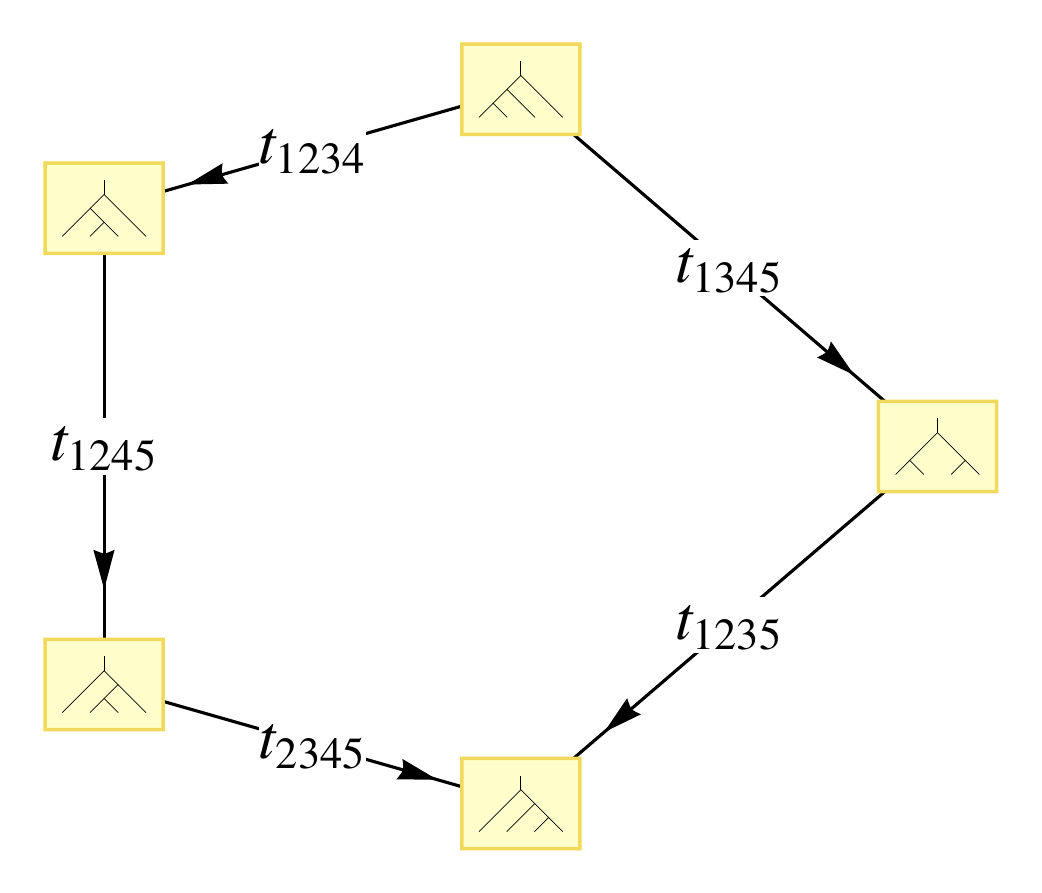} 
}
\parbox{15cm}{
\caption{Representation of the Tamari lattice $\bbT_3$ by line soliton graphs (which is a special 
case of Fig.~\ref{fig:pentagon}, see also Remark~\ref{rem:y-poset-family}). 
The left chain is realized if $t^{(4)} < t^{(4)}_{12345}$, 
the right chain if $t^{(4)} > t^{(4)}_{12345}$ (see also Fig.~\ref{fig:T3_chains}).
At $t=t_{1234}(t^{(4)}_{12345})$ and $t^{(4)} = t^{(4)}_{12345}$, a direct transition takes 
place from the uppermost to the lowermost tree.   \label{fig:T3}  }   
}
\end{center} 
\end{figure}
\end{example}

\subsection{The third step}
For $M>4$ we redefine the constants $c_k$ once more,
\bez
        c_k \mapsto p_k^5 \, t^{(5)} + c_k    \qquad \quad k=1,\ldots,M+1 \, ,
\eez
with a new parameter $t^{(5)}$. Then we have
\bez
  x_{ij}(y,t,t^{(4)},t^{(5)}) &=& -h_1(p_i,p_j) \, y - h_2(p_i,p_j) \, t 
           - h_3(p_i,p_j) \, t^{(4)} - h_4(p_i,p_j) \, t^{(5)} - c_{ij} \, , \\
  x_{ijk}(t,t^{(4)},t^{(5)}) &=& x_{ij}(y_{ijk}(t,t^{(4)},t^{(5)}),t,t^{(4)},t^{(5)}) \, , \\
  y_{ijk}(t,t^{(4)},t^{(5)}) &=& -h_1(p_i,p_j,p_k) \, t - h_2(p_i,p_j,p_k) \, t^{(4)} - h_3(p_i,p_j,p_k) \, t^{(5)}
                     - c_{ijk} \, , \\
  t_{ijkl}(t^{(4)},t^{(5)}) &=& h_1(p_i,p_j,p_k,p_l) \, t^{(4)} 
                   - h_2(p_i,p_j,p_k,p_l) \, t^{(5)} - c_{ijkl} \, , \\
  t^{(4)}_{ijklm}(t^{(5)}) &=& -h_1(p_i,p_j,p_k,p_l,p_m) \, t^{(5)} - c_{ijklm} \, ,
\eez
with $c_{ij}$, $c_{ijk}$, $c_{ijkl}$ and $c_{ijklm}$ as defined previously (see also (\ref{c_k1...kn_sum})).
Coincidences of critical values of $t^{(4)}$ can only occur at the following critical values 
of $t^{(5)}$,
\bez
     t^{(5)}_{ijklmn} = - c_{ijklmn} \, ,
\eez
where
\bez
     c_{ijklmn} 
   = \frac{ c_{ijklm} - c_{ijkln} }{p_m - p_n}
   = \frac{c_i}{(p_i-p_j)(p_i-p_k)(p_i-p_l)(p_i-p_m)(p_i-p_n)} + \mbox{cyclic permutations} \; .
\eez
This follows from the identity (see also Proposition~\ref{prop:t^(n-1)-diff})
\be
     t^{(4)}_{ijklm} - t^{(4)}_{ijkln} = (p_n - p_m)(t^{(5)} - t^{(5)}_{ijklmn})  \; .
           \label{t4diff-t^{(5)}_ijklmn}
\ee
Furthermore, at $t^{(5)}=t^{(5)}_{ijklmn}$ we have $t^{(4)}_{ijklm} = t^{(4)}_{ijkln} 
= t^{(4)}_{ijkmn} = t^{(4)}_{ijlmn} = t^{(4)}_{iklmn} = t^{(4)}_{jklmn}$. At this critical event 
(now a point in $\mathbb{R}^5$ with coordinates $x,y,t,t^{(4)},t^{(5)}$) having the projection
\bez
     P_{ijklmn} := P_{ijklm}(t^{(5)}_{ijklmn}) 
\eez
in the $xy$-plane, we have $\theta_i = \theta_j = \theta_k = \theta_l = \theta_m = \theta_n$. 
The following is a consequence of (\ref{t4diff-t^{(5)}_ijklmn}) 
(see also Proposition~\ref{prop:t^(n-1)-ordering}). 

\begin{proposition}
\label{prop:t4_order}
If $p_i < p_j < p_k < p_l < p_m < p_n$, then
\bez
   \begin{array}{l}  t^{(4)}_{ijklm} < t^{(4)}_{ijkln} < t^{(4)}_{ijkmn} < t^{(4)}_{ijlmn} 
     < t^{(4)}_{iklmn} < t^{(4)}_{jklmn}  \\
     t^{(4)}_{ijklm} > t^{(4)}_{ijkln} > t^{(4)}_{ijkmn} 
     > t^{(4)}_{ijlmn} > t^{(4)}_{iklmn} > t^{(4)}_{jklmn}
   \end{array} 
                    \quad \mbox{for} \quad
   \begin{array}{l}   t^{(5)} < t^{(5)}_{ijklmn} \\ t^{(5)} > t^{(5)}_{ijklmn} \end{array} \; .
\eez
\hfill $\square$
\end{proposition}

The next two propositions are special cases of Proposition~\ref{prop:non-visible_P} 
and \ref{prop:visible_P}, respectively.

\begin{proposition}
\label{prop:P5_non-visible}
Let $p_i < p_j < p_k < p_l < p_m < p_n$. Then \\ 
(1) $P_{ijkln}(t^{(5)}), P_{ijlmn}(t^{(5)})$ and $P_{jklmn}(t^{(5)})$ are non-visible for 
$t^{(5)} < t^{(5)}_{ijklmn}$. \\
(2) $P_{ijklm}(t^{(5)}), P_{ijkmn}(t^{(5)})$ and $P_{iklmn}(t^{(5)})$ are non-visible for 
$t^{(5)} > t^{(5)}_{ijklmn}$.
\hfill $\square$
\end{proposition}

\begin{proposition}
\label{prop:P5_visible}
Let $p_i < p_j < p_k < p_l < p_m < p_n$ and suppose that $P_{ijklmn}$ is visible at 
$t=t_{ijkl}$, $t^{(4)}=t^{(4)}_{ijklm}$, $t^{(5)} = t^{(5)}_{ijklmn}$, and not a meeting point 
of more than six phases. 
The following holds for values of $t^{(5)}$ that are close enough to $t^{(5)}_{ijklmn}$, so that no other 
critical value of $t^{(5)}$ with visible projection point is in between. \\ 
(1) $P_{ijklm}(t^{(5)})$, $P_{ijkmn}(t^{(5)})$ and $P_{iklmn}(t^{(5)})$ 
are visible, at the respective critical values of $t$ and $t^{(4)}$, if $t^{(5)} < t^{(5)}_{ijklmn}$.  \\
(2) $P_{jklmn}(t^{(5)}), P_{ijlmn}(t^{(5)})$ and $P_{ijkln}(t^{(5)})$ 
are visible, at the respective critical values of $t$ and $t^{(4)}$, if $t^{(5)} > t^{(5)}_{ijklmn}$.
\hfill $\square$
\end{proposition}

\begin{figure}[H] 
\begin{center} 
\resizebox{!}{4.cm}{
\includegraphics{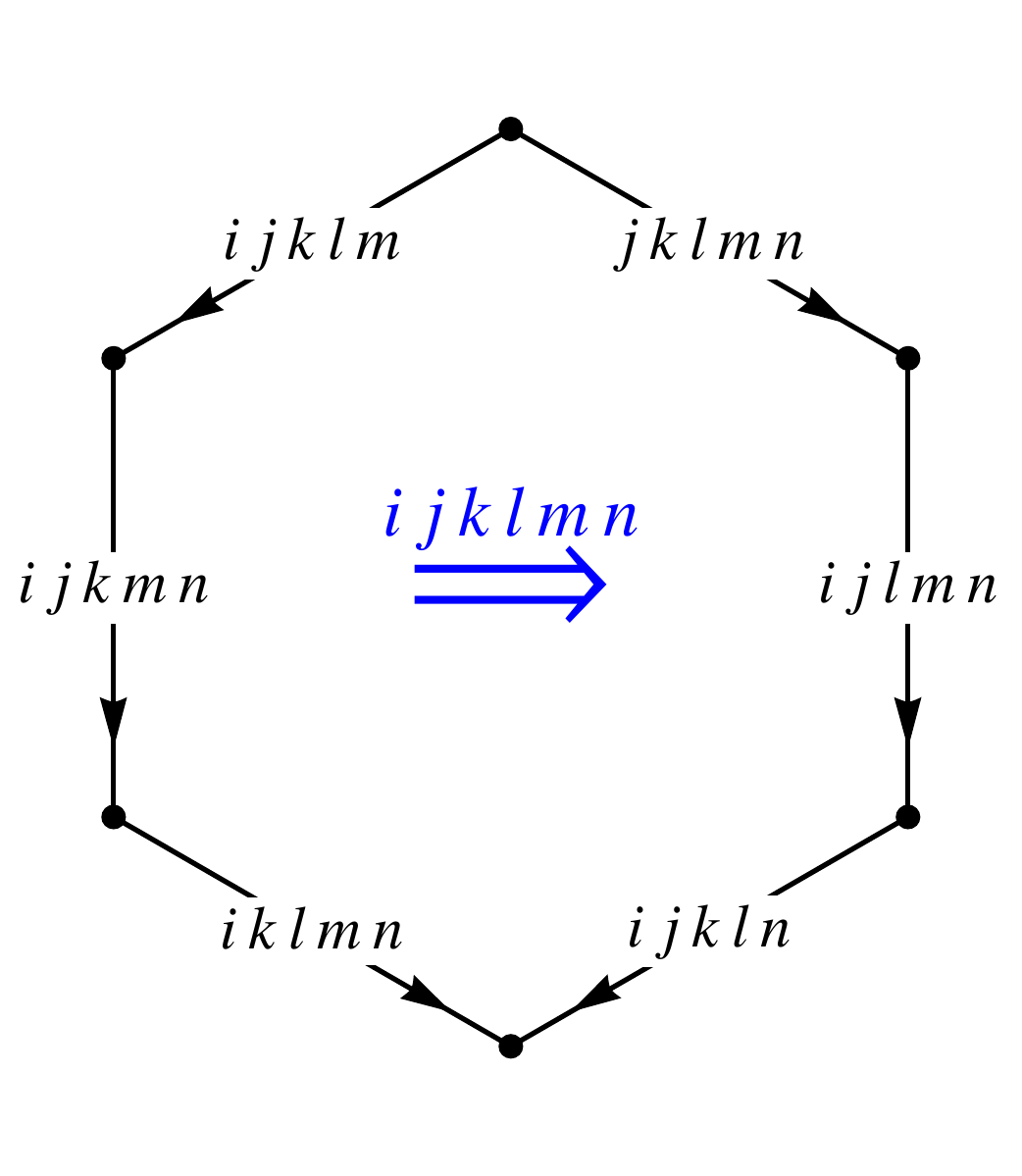}
}
\parbox{15cm}{ 
\caption{The process connected with the passage of $t^{(5)}$ through a critical value $t^{(5)}_{ijklmn}$ 
with a \emph{visible} critical event, where $p_i < p_j < p_k < p_l < p_m < p_n$. 
Here $ijklm$ stands for 
$t^{(4)}_{ijklm}$ and $ijklmn$ for $t^{(5)}_{ijklmn}$. The left chain corresponds to 
$t^{(5)} < t^{(5)}_{ijklmn}$, the right to $t^{(5)} > t^{(5)}_{ijklmn}$. The nodes  
are classes of chains in a Tamari lattice, see Example~\ref{ex:M5}. 
\label{fig:hexagon}  }
}
\end{center} 
\end{figure}

Fig.~\ref{fig:hexagon} expresses a consequence of Propositions~\ref{prop:t4_order}, 
\ref{prop:P5_non-visible} and \ref{prop:P5_visible} as a \emph{process}.

\begin{example}
\label{ex:M5}
For $M=5$ there is only a single critical value of $t^{(5)}$, namely $t^{(5)}_{123456}$, and 
$P_{123456}$ is visible at $t=t_{1234}$, $t^{(4)}=t^{(4)}_{12345}$ and $t^{(5)}= t^{(5)}_{123456}$, as 
a meeting point of all six phases. There are six critical values of $t^{(4)}$, namely 
$t^{(4)}_{12345}, t^{(4)}_{12346},t^{(4)}_{12356},t^{(4)}_{12456},t^{(4)}_{13456},t^{(4)}_{23456}$. 
For $t^{(5)} < t^{(5)}_{123456}$, according to Proposition~\ref{prop:t4_order}
we have to distinguish the cases where (1) $t^{(4)} < t^{(4)}_{12345}$, 
(2) $t^{(4)}_{12345} < t^{(4)} < t^{(4)}_{12356}$, (3) $t^{(4)}_{12356} < t^{(4)} < t^{(4)}_{13456}$, and 
(4) $t^{(4)}_{13456} < t^{(4)}$. \\
In case (1) we obtain from Proposition~\ref{prop:t_ijkl-ordering} the inequalities 
(a) $t_{1234} < t_{1235} < t_{1245} < t_{1345} < t_{2345}$, 
(b) $t_{1234} < t_{1236} < t_{1246} < t_{1346} < t_{2346}$,
(c) $t_{1235} < t_{1236} < t_{1256} < t_{1356} < t_{2356}$,
(d) $t_{1245} < t_{1246} < t_{1256} < t_{1456} < t_{2456}$,
(e) $t_{1345} < t_{1346} < t_{1356} < t_{1456} < t_{3456}$,
(f) $t_{2345} < t_{2346} < t_{2356} < t_{2456} < t_{3456}$. 
According to Proposition~\ref{prop:P4_non-visible}, the critical points appearing at the times
$t_{1235}$, $t_{1236}$, $t_{1246}$, $t_{1345}$, $t_{1346}$, $t_{1356}$, $t_{1456}$, $t_{2346}$, $t_{2456}$ 
are non-visible. Their elimination leads to 
(a) $t_{1234} < t_{1245} < t_{2345}$, 
(b) $t_{1234}$,
(c) $t_{1256} < t_{2356}$,
(d) $t_{1245} < t_{1256}$,
(e) $t_{3456}$,
(f) $t_{2345} < t_{2356} < t_{3456}$, 
and the union determines the second poset in Fig.~\ref{fig:M5_posets}.\footnote{We are not 
aware of a general argument why the union of sequences of ordered critical times, as in 
one of the cases (1)-(4) of Example~\ref{ex:M5} (see also 
Fig.~\ref{fig:M5_posets}), are posets. At least this turns out to be the case for $M \leq 6$. } 
Since Proposition~\ref{prop:vis_neigb-levels} does not identify any of the remaining critical 
times as ``non-visible" (note that $t^{(4)}_{12345}$, $t^{(4)}_{12356}$ and $t^{(4)}_{13456}$ 
correspond to visible events according to Proposition~\ref{prop:P5_visible}), we can refer to
Proposition~\ref{prop:exhaust} in order to conclude that they all correspond to \emph{visible} 
events only. 
\begin{figure}[H] 
\begin{center} 
\resizebox{!}{6.cm}{
\includegraphics{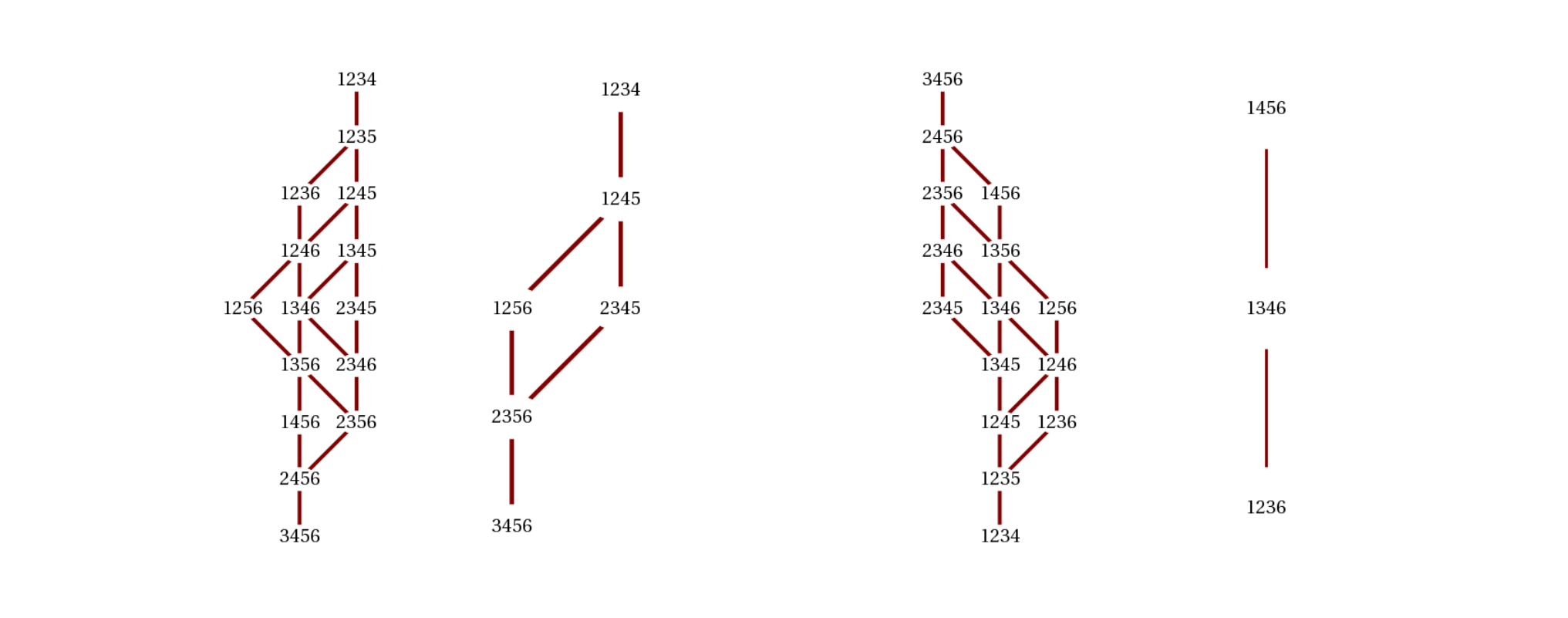} 
} 
\parbox{15cm}{
\caption{Intermediate step in the derivation of the possible line soliton evolutions  
for $M=5$ and $t^{(4)} < \min\{t^{(4)}_{ijklm}\}$ (pair of posets on the left), 
respectively $t^{(4)} > \max\{t^{(4)}_{ijklm}\}$ (pair of posets on the right). 
In both cases the first diagram is obtained as the union of all sequences of ordered 
critical times. The second diagram then results by dropping those critical times for 
which the critical event (or rather its projection in the $xy$-plane) is non-visible 
(and removing redundant edges). A four-digit number stands for the corresponding critical time.
\label{fig:M5_posets}  }
}
\end{center}
\end{figure} 
\noindent
Depending on the order of the critical time values $t_{1256}$ and $t_{2345}$, the evolution follows 
one of the two sequences of rooted binary trees in Fig.~\ref{fig:T4_chain1+2}, easily 
elaborated with the help of (\ref{y_t-map}). 
\begin{figure}[H] 
\begin{center} 
\resizebox{!}{3.cm}{
\includegraphics{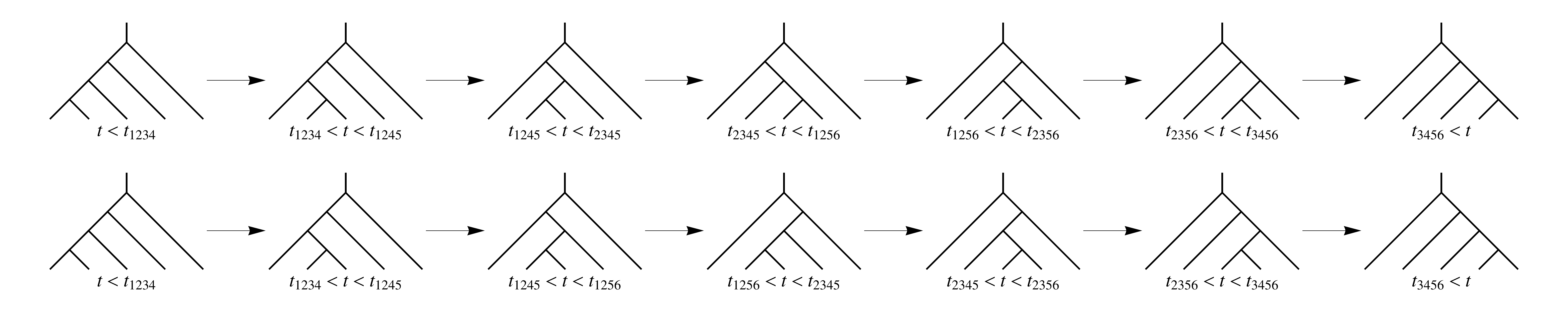} 
}
\parbox{15cm}{ 
\caption{The two possible evolutions for $M=5$ and $t^{(4)} < \min\{t^{(4)}_{ijklm}\}$. They correspond
to chains in the Tamari lattice $\bbT_4$.
\label{fig:T4_chain1+2}  }
}
\end{center}
\end{figure} 
\noindent
In case (4) we have the inequalities 
(a') $t_{1234} > t_{1235} > t_{1245} > t_{1345} > t_{2345}$,
(b') $t_{1234} > t_{1236} > t_{1246} > t_{1346} > t_{2346}$,
(c') $t_{1235} > t_{1236} > t_{1256} > t_{1356} > t_{2356}$,
(d') $t_{1245} > t_{1246} > t_{1256} > t_{1456} > t_{2456}$,
(e') $t_{1345} > t_{1346} > t_{1356} > t_{1456} > t_{3456}$,
(f') $t_{2345} > t_{2346} > t_{2356} > t_{2456} > t_{3456}$.
Now we have to eliminate
$t_{1234}$, $t_{1235}$, $t_{1245}$, $t_{1246}$, $t_{1256}$, $t_{1345}$, $t_{1356}$, $t_{2345}$, 
$t_{2346}$, $t_{2356}$, $t_{2456}$, $t_{3456}$ (see Fig.~\ref{fig:M5_posets}). 
The resulting sequence of rooted binary trees is shown in Fig.~\ref{fig:T4_chain3}.
\begin{figure}[H] 
\begin{center} 
\resizebox{!}{1.5cm}{
\includegraphics{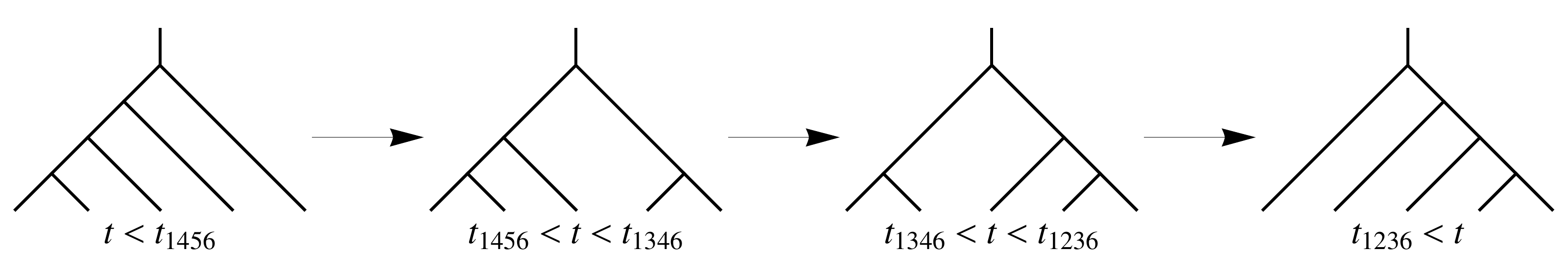} 
}
\parbox{15cm}{  
\caption{The evolution for $M=5$ and $t^{(4)} > \max\{t^{(4)}_{ijklm}\}$, another chain in the 
Tamari lattice $\bbT_4$. \label{fig:T4_chain3}  }
}
\end{center}
\end{figure} 
\noindent
Fig.~\ref{fig:T4_t4deform1} shows the resulting classes of chains in the cases (1)-(4). 
\begin{figure}[H] 
\begin{center} 
\resizebox{!}{6.5cm}{
\includegraphics{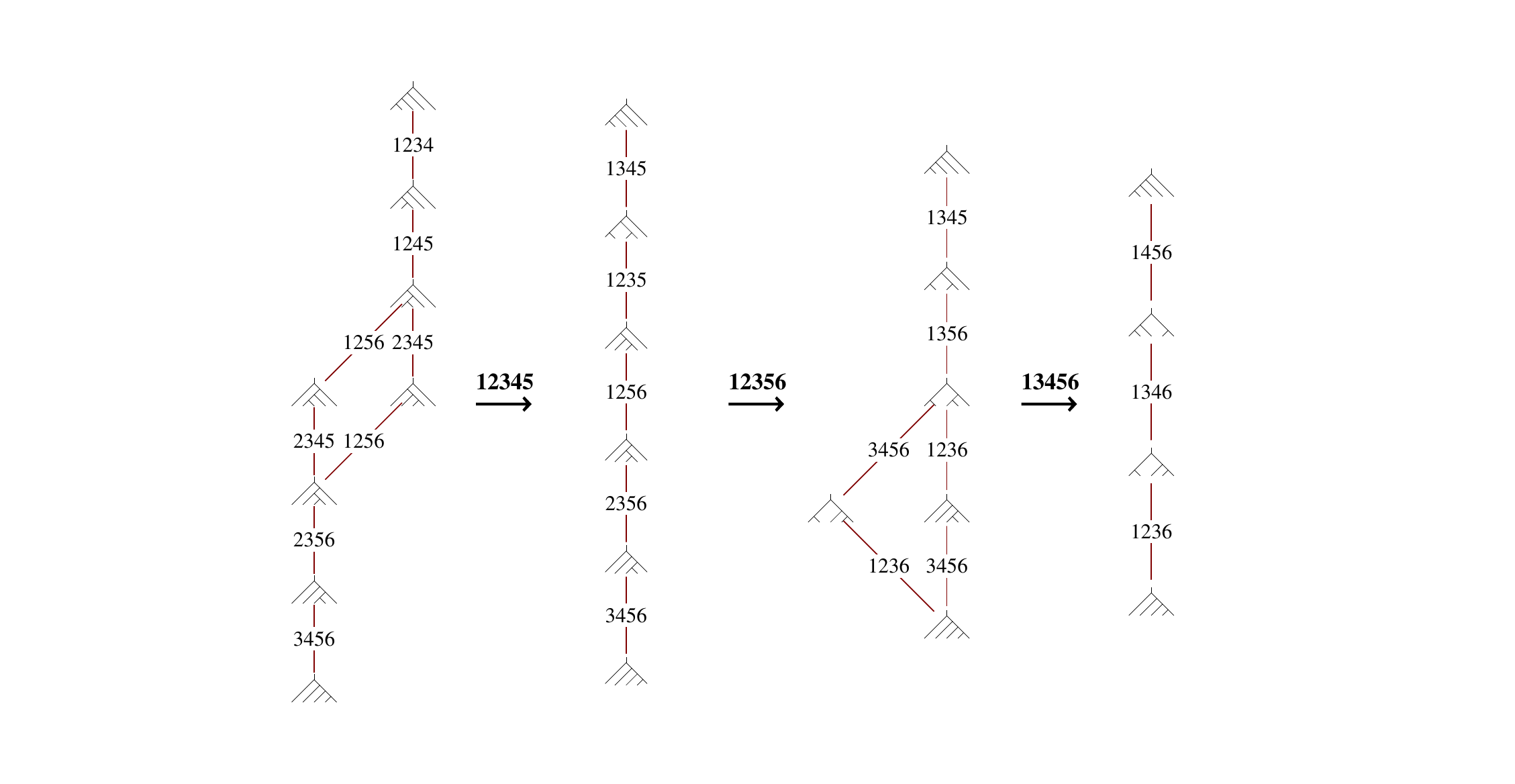}
}
\parbox{15cm}{ 
\caption{Classes of possible evolutions for $M=5$ and $t^{(5)} < t^{(5)}_{123456}$, corresponding 
to intervals of $t^{(4)}$ and transitions at critical values of $t^{(4)}$. 
This corresponds to the left chain in Fig.~\ref{fig:hexagon}. 
\label{fig:T4_t4deform1}  }
}
\end{center} 
\end{figure} 
\noindent
For $t^{(5)} > t^{(5)}_{123456}$, the corresponding (classes of) chains are displayed in 
Fig.~\ref{fig:T4_t4deform2}. Collecting all different chains, we obtain the representation of the 
Tamari lattice $\bbT_4$ in Fig.~\ref{fig:Tamari4}. 
\begin{figure}[H] 
\begin{center} 
\resizebox{!}{6.5cm}{
\includegraphics{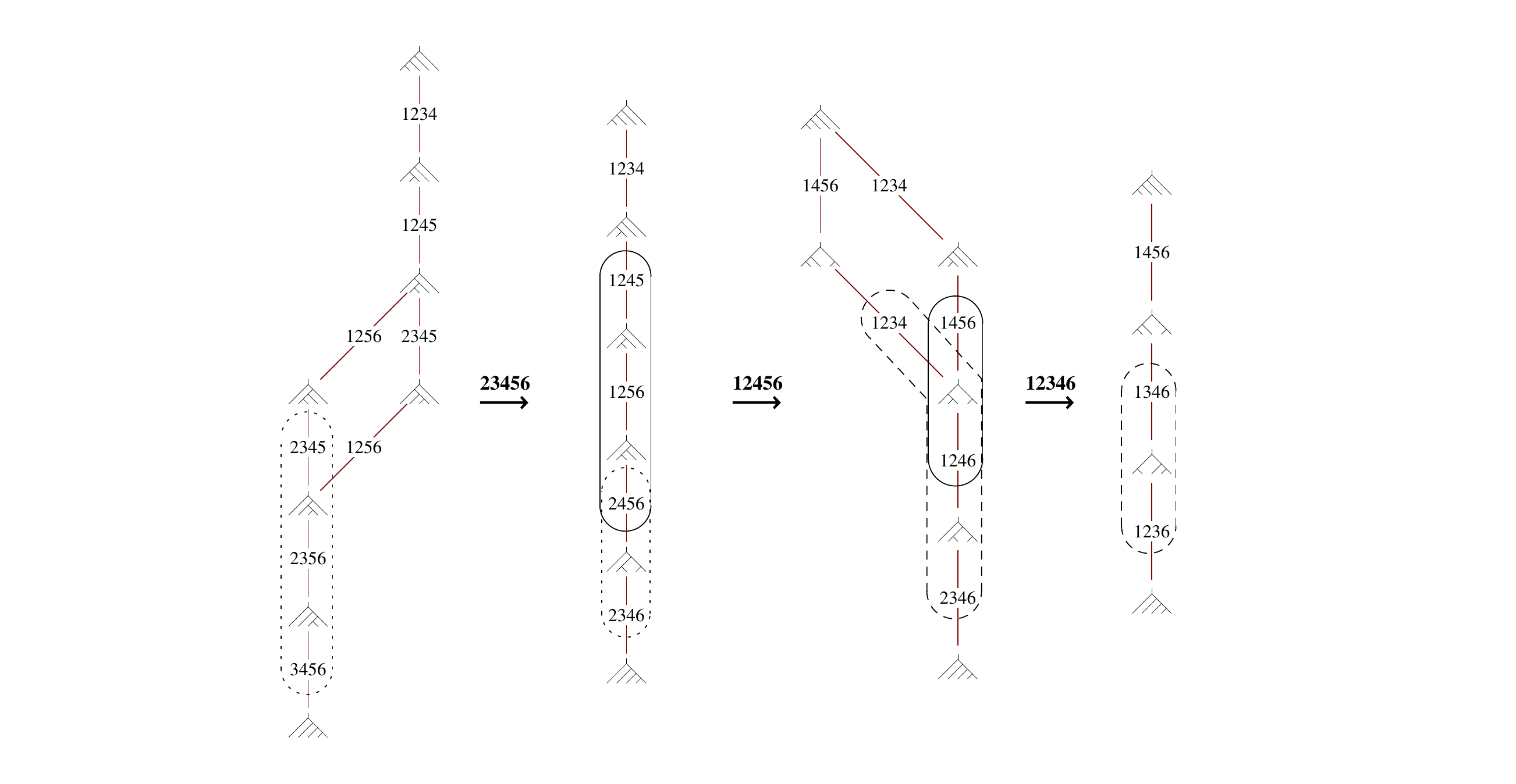}
}
\parbox{15cm}{
\caption{Classes of possible evolutions for $M=5$ and $t^{(5)} > t^{(5)}_{123456}$. 
This corresponds to the right chain in Fig.~\ref{fig:hexagon}. 
As indicated by equally encircled parts, the deformation of a class into the next, 
as $t^{(4)}$ passes through a critical value, proceeds according to the left to right half 
pentagon structure of $\bbT_3$. 
\label{fig:T4_t4deform2}  }
}
\end{center} 
\end{figure} 
\vspace{-1.cm}
\begin{figure}[H]
\begin{center}
\begin{minipage}[t]{5.cm}
\resizebox{!}{8.cm}{
\includegraphics{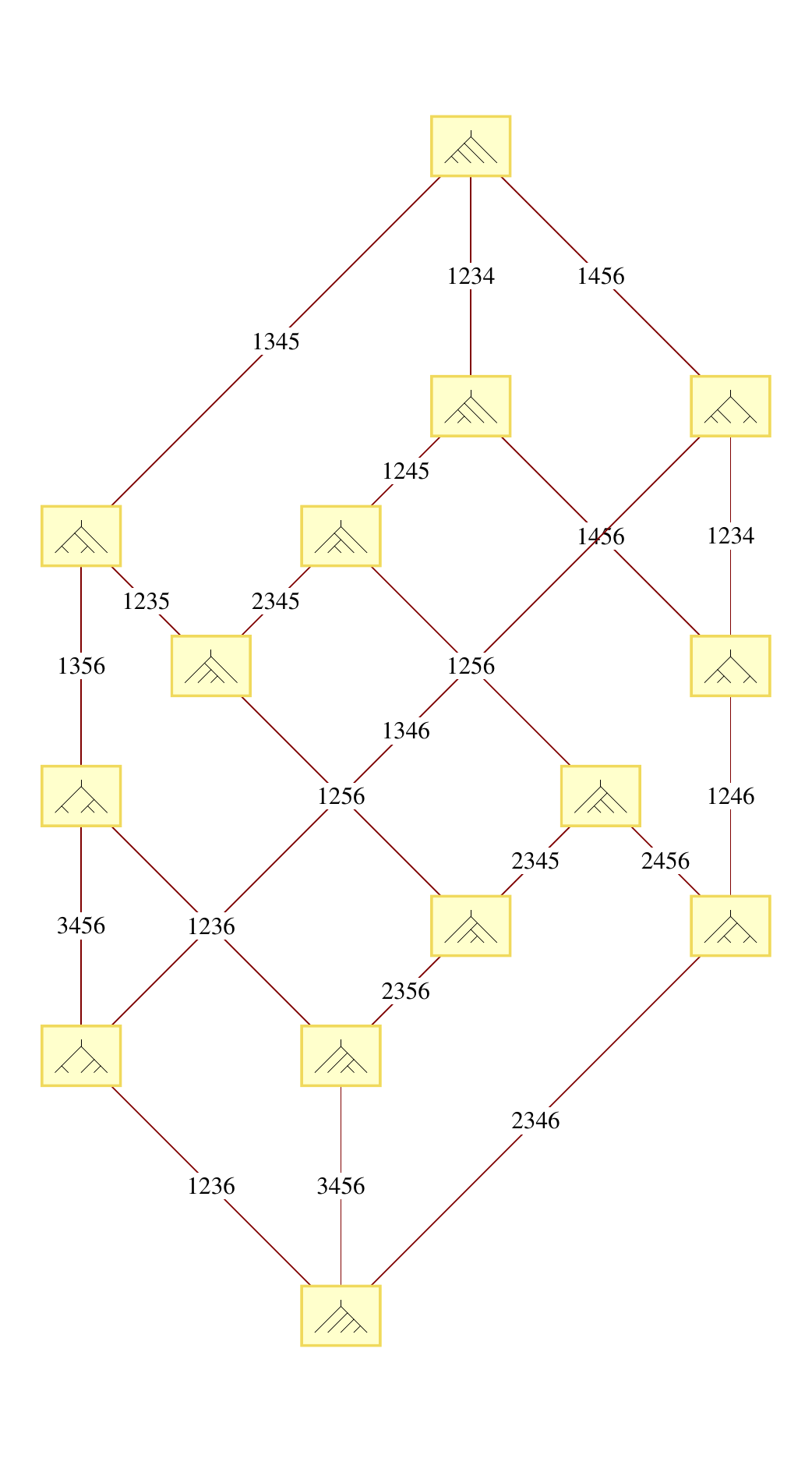}
}
\end{minipage}
\hspace{1.cm}
\begin{minipage}[t]{5cm}
\vspace{-6.cm}
\resizebox{!}{4.cm}{
\includegraphics{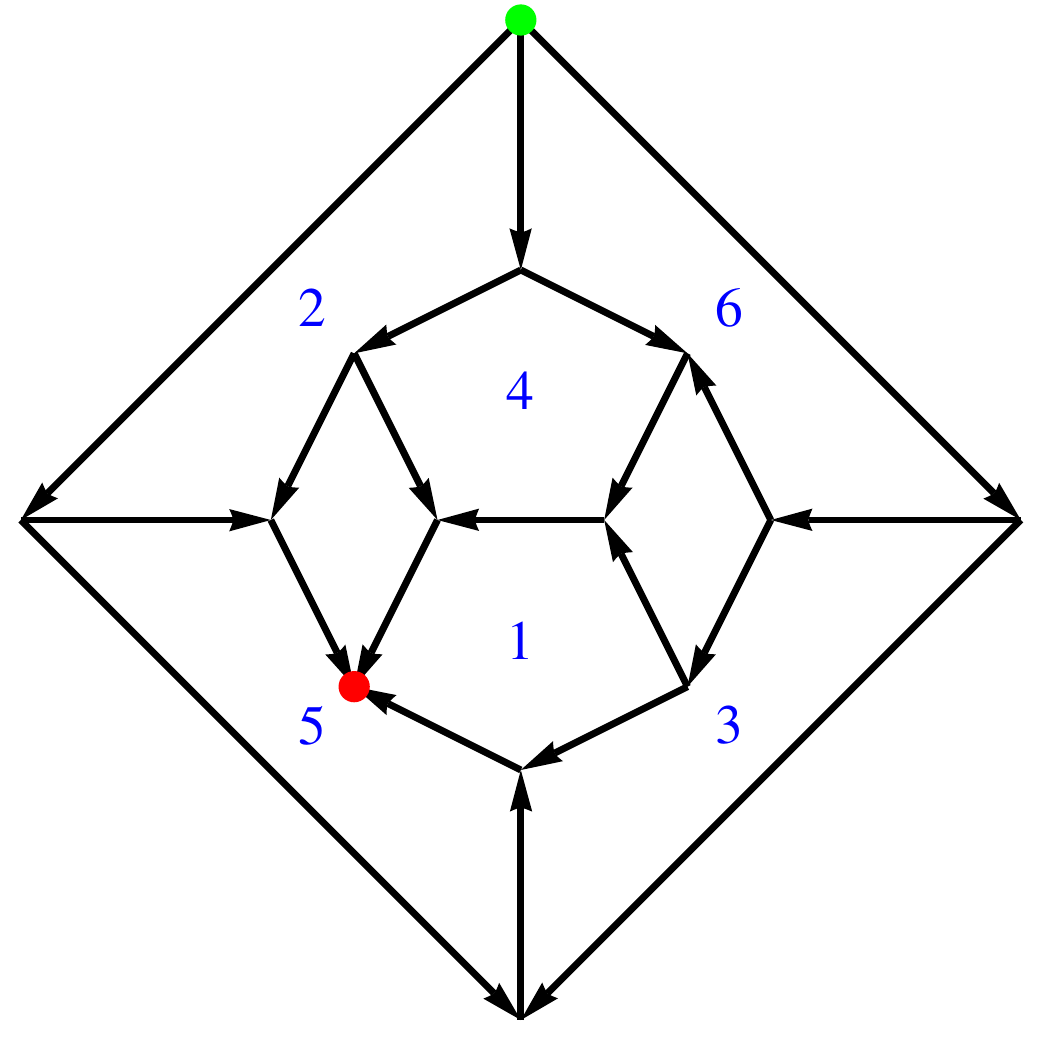}
}
\end{minipage} 
\parbox{15cm}{ \vspace{-.5cm}
\caption{The left graph shows the representation of the Tamari lattice $\bbT_4$ in terms of rooted binary 
trees which represent line soliton patterns. The digraph on the right describes $\bbT_4$ without 
overlapping edges and makes evident that it consists of six pentagons and three tetragons. (The 
2-faces of Tamari lattices are pentagons or tetragons, see e.g. \cite{Casp+LeConte04}.) 
Here the numbers assigned to the pentagons encode the critical times associated with the edges. 
For an edge between pentagon $i$ and pentagon $j$ we form the complement of $ij$ in $123456$, which 
then determines the associated critical time. Opposite edges of a tetragon have to be identified 
for this counting, so that e.g. $t_{1456}$ is assigned to the left uppermost arrow. 
Another familiar representation is as a poset structure on the \emph{associahedron} 
\cite{Stash98,Stash04,Loday04} in three dimensions. 
\label{fig:Tamari4}  }
}
\end{center}
\end{figure}
\noindent
In order to realize a certain chain in $\bbT_4$, the critical times appearing along it all 
have to be smaller than the critical times of neighboring branches. Solving the 
inequalities arising in this way, one obtains the conditions in Table~\ref{table:T4conditions}
(see also Tables~\ref{table:t4rels} and \ref{table:T4_t4conds} in Appendix~\ref{AppB}).
An example of a line soliton solution of type 1 in the table is displayed in Fig.~\ref{fig:T4_evolution_type1}.
\begin{table}[H]
\begin{center}
\begin{tabular}{|c|l|l|l|} 
\hline
 1 & $t_{1234}, t_{1245},t_{2345},t_{1256},t_{2356},t_{3456}$ & $\mu<0$ & 
           $\frac{p_3+p_4-p_1-p_6}{(p_3-p_6)(p_4-p_6)} < \lambda/\mu$  \\
\hline 
 2 & $t_{1234}, t_{1245},t_{1256},t_{2345},t_{2356},t_{3456}$ &  & 
           $\frac{1}{p_1-p_6} < \lambda/\mu < \frac{p_3+p_4-p_1-p_6}{(p_3-p_6)(p_4-p_6)} $  \\
\hline
 3 & $t_{1234}, t_{1245},t_{1256},t_{2456},t_{2346}$ &  & 
           $\frac{1}{p_3-p_6} < \lambda/\mu < \frac{1}{p_1-p_6}$  \\
\hline
 4 & $t_{1234}, t_{1456},t_{1246},t_{2346}$ &  & 
           $ \frac{p_2+p_3-p_5-p_6}{(p_2-p_6)(p_3-p_6)} < \lambda/\mu < \frac{1}{p_3-p_6}$  \\
\hline
 5 & $t_{1456}, t_{1234},t_{1246},t_{2346}$ &  & 
           $ \frac{1}{p_5-p_6} < \lambda/\mu < \frac{p_2+p_3-p_5-p_6}{(p_2-p_6)(p_3-p_6)}$  \\
\hline
 6 & $t_{1456}, t_{1346},t_{1236}$ & $\mu \leq 0$ & 
           $\frac{1}{p_5-p_6} \mu < \lambda$  \\
\hline
   &  & $\mu >0$  &  $\frac{1}{p_2-p_6} < \lambda/\mu $ \\
\hline
 7 & $t_{1345}, t_{1356},t_{3456},t_{1236}$ &  & 
      $\frac{p_1+p_2-p_4-p_5}{p_1p_2-p_4p_5 + (p_4+p_5-p_1-p_2)\, p_6} < \lambda/\mu < \frac{1}{p_2-p_6}$ \\
\hline
 8 & $t_{1345}, t_{1356},t_{1236},t_{3456}$ &  & 
      $\frac{1}{p_4-p_6} < \lambda/\mu < \frac{p_1+p_2-p_4-p_5}{p_1p_2-p_4p_5 + (p_4+p_5-p_1-p_2)\, p_6}$ \\
\hline
 9 & $t_{1345}, t_{1235},t_{1256},t_{2356},t_{3456}$ &  & $\lambda/\mu < \frac{1}{p_4-p_6}$ \\
\hline
\end{tabular}
\parbox{15cm}{
\caption{The nine maximal chains in the Tamari lattice $\bbT_4$, here described by the corresponding 
sequences of critical times $t_{ijkl}$, and the parameter conditions under which they occur. 
We set $\mu = t^{(4)} - t^{(4)}_{12345}$ and $\lambda = t^{(5)} - t^{(5)}_{123456}$. 
Recall that $p_1 < p_2 < p_3 < p_4 < p_5 < p_6$. 
\label{table:T4conditions} }
}
\end{center}
\end{table}
\begin{figure}[H] 
\begin{center} 
\resizebox{16.cm}{!}{
\includegraphics{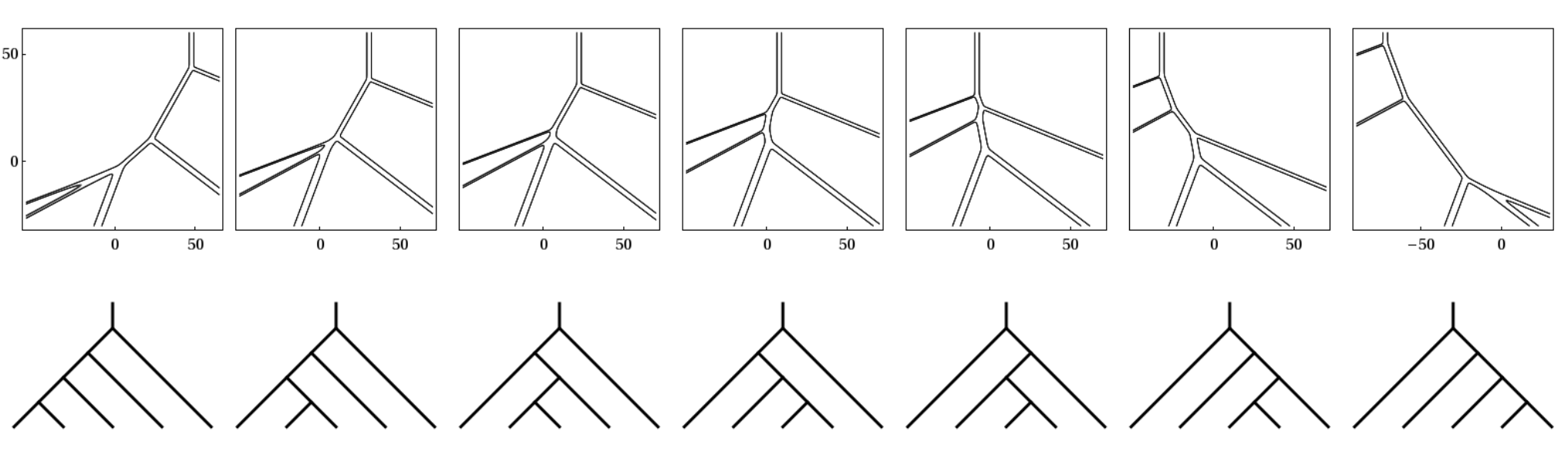} 
}
\parbox{15cm}{
\caption{The evolution of a line soliton pattern of type 1 in Table~\ref{table:T4conditions}.
Here we have $\theta_i = p_i x + p_i^2 y + p_i^3 t + p_i^4 t^{(4)} + p_i^5 t^{(5)} + c_i$ 
with the choice $p_1=-2, p_2=-3/2, p_3=-1,p_4=1/2, p_5=5/4,p_6=2$ and $c_1=10, c_2=c_3=c_4=c_5=0, c_6=-10$. 
Furthermore, we set $\mu=-2$ and $\lambda=-1$. The line soliton plots are taken at 
$t=-10, -5.7, -3.6, 0, 4, 10, 20$. 
\label{fig:T4_evolution_type1}  }
}
\end{center} 
\end{figure}
\end{example}

The further steps needed to treat the cases $M>5$ should now be obvious and we refer to 
Appendix~\ref{AppA} for corresponding general results.

\section{On the general class of KP line soliton solutions}
\label{sec:general}
\setcounter{equation}{0}
In the preceding sections (and Appendix~\ref{AppA}) we restricted our considerations to a special 
class of line soliton solutions and achieved a complete description (in the tropical approximation) 
of their evolution. In this section we argue that more general solutions can actually be 
understood fairly well as superimpositions of solutions from the special class, with 
rather simple modifications. The somewhat qualitative picture layed out in this section 
still has to be elaborated in more detail, however. 

The general class of line soliton solutions of the KP-II equation (and more generally its hierarchy) 
in Hirota form is well-known to be given by
\bez
     \tau = f_1 \wedge f_2 \wedge \cdots \wedge f_n \, ,
\eez
where
\bez
    f_i = \sum_{j=1}^{M+1} \epsilon_{ij} \, e_j \, , \qquad
    e_j = e^{\theta_j} \, , \qquad
    \theta_j = \sum_{r = 1}^M p_j^r \, t^{(r)} + c_j  \, ,
\eez
and the exterior product on the space of functions generated by the exponential functions 
$e_j$, $j=1,\ldots,M+1$, is defined by
\bez
    e_{i_1} \wedge \cdots \wedge e_{i_m} 
 = \Delta(p_{i_1},\ldots,p_{i_m}) \, e_{i_1} \cdots e_{i_m} \, ,
\eez
with the Vandermonde determinant (\ref{Vandermonde_det}). Hence
\bez
    \tau = \sum_{1 \leq i_1 < \cdots < i_n \leq M+1} A_{i_1 \ldots i_n} \, e_{i_1} \cdots e_{i_n} 
   \quad \mbox{where} \quad
   A_{i_1 \ldots i_n} = \epsilon_{1i_1} \cdots \epsilon_{ni_n} \, \Delta(p_{i_1},\ldots,p_{i_n}) \; .
\eez

\begin{example}
\label{ex:y_to_-y}
A subclass of the above class of solutions is given by 
\bez
    \tau &=& (e_1 + e_2) \wedge (e_2 + e_3) \wedge \cdots \wedge (e_M + e_{M+1}) \\
         &=& \sum_{i=1}^{M+1} e_1 \wedge \cdots \wedge \widehat{e_{i}} \wedge e_{i+1} \wedge 
             \cdots \wedge e_{M+1} \\
         &=& \Delta(p_1,\ldots,p_{M+1}) \, e^{\theta_1 + \cdots + \theta_{M+1}} \, 
             \sum_{i=1}^{M+1} a_i \, e^{-\theta_i} \, , 
\eez
where a hat indicates an omission, and 
$a_i = 1/[(p_i - p_1) \cdots (p_i - p_{i-1}) (p_{i+1}-p_i) \cdots (p_{M+1}-p_i)]$. 
Assuming $p_1 < p_2 < \cdots < p_{M+1}$, $a_i$ is positive, hence it can be absorbed into 
the constant $c_i$. Moreover, the factor in 
front of the sum drops out in the expression $u = 2 \log(\tau)_{xx}$ for the KP soliton 
solution, so that an equivalent $\tau$-function is given by
\bez
     \tilde{\tau} = \sum_{i=1}^{M+1} e^{-\theta_i} \; .
\eez
Via $p_i \mapsto -p_i$ and $c_i \mapsto -c_i$ (and with a renumbering of the $p$'s), this is 
the class of solutions treated in the main part of this work, up to the reflection 
$t^{(2r)} \mapsto -t^{(2r)}$, $r=1,2,\ldots$, which includes $y \mapsto -y$. 
The corresponding rooted binary trees are hence given by those of our simple class, 
but drawn upside down. 
\end{example}

\begin{remark}
 Since the above expression for $\tau$ determines a KP solution, this also holds for
\bez 
  \tau' =  e^{-\theta_1 - \cdots - \theta_{M+1}} \, \tau 
        = \sum_{1 \leq i_1 < \cdots < i_n \leq M+1} A_{i_1 \ldots i_n} \, e^{-\theta_{k_1}- 
            \cdots - \theta_{k_{n'}}} \, ,
\eez
where $\{ k_1,\ldots, k_{n'} \}=\{1,\ldots , M+1\} \setminus \{i_1,\ldots, i_n\}$. 
Since the reflection $t^{(r)} \mapsto -t^{(r)}$, $r=1,2,\ldots$, is a symmetry of the KP equation, 
and since $c_i \mapsto -c_i$ preserves the above class of solutions, we conclude that also 
\bez
    \tau_\star = \sum_{1 \leq i_1 < \cdots < i_n \leq M+1} A_{i_1 \ldots i_n} 
       \, e^{\theta_{k_1} + \cdots + \theta_{k_{n'}}} 
   = \sum_{1 \leq i_1 < \cdots < i_n \leq M+1} A_{i_1 \ldots i_n} 
       \, e_{k_1} \cdots e_{k_{n'}} 
\eez
is a solution, which we call the \emph{dual} of $\tau$. 
\end{remark}

Let us order the constants $p_i$ such that $p_1 < \cdots < p_{M+1}$ and let us assume that 
no pair of the functions $f_i$, $i=1,\ldots,n$, has an $e_j$ in common. 
The cases excluded by this assumption can be recovered by taking a limit where pairs of neighboring 
$p$'s coincide. 
By absorbing the modulus of a nonvanishing constant $\epsilon_{ij}$ via a redefinition of the 
constant $c_j$, without restriction of generality we can assume that
\bez
      \epsilon_{ij} \in \{ 0, \pm 1 \} \; .
\eez
By demanding that the coefficients $A_{i_1 \ldots i_n}$ are all non-negative, and at least one 
of them different from zero, we ensure that $\tau$ is positive and the KP solution is then regular. 
Then we obtain 
\bez
  \tau = \sum_{1 \leq i_1 < \cdots < i_n \leq M+1} 
         |\epsilon_{1 i_1} \cdots \epsilon_{n i_n}| \; e^{\theta_{i_1 \ldots i_n}} \, ,
\eez
where
\bez
  \theta_{i_1 \ldots i_n} = \theta_{i_1} + \cdots + \theta_{i_n} + \log \Delta(p_{i_1}, \ldots, p_{i_n}) \; .
\eez
In particular, 
\bez
       \theta_{ij} = \theta_i + \theta_j + \log(p_j-p_i) \; .  
\eez 

For fixed values of the parameters $t^{(3)}=t,t^{(4)},\ldots$, the $xy$-plane is divided into 
regions where one of the phases $\theta_{i_1 \ldots i_n}$ dominates all others. 
The line soliton segments are given by the visible boundaries of these regions. 
The tropical approximation now reads 
\bez
   \log(\tau) \simeq \max\{ \theta_{i_1 \ldots i_n} \, | \, 1 \leq i_1 < \cdots < i_n \leq M+1, \, 
      \epsilon_{1 i_1} \cdots \epsilon_{n i_n} \neq 0 \} \; .
\eez
In principle one can approach a classification of solutions in a similar way as done for the 
special class in the main part of this work. In the following we set $t^{(n)}=0$ for 
$n>3$ (more precisely, we absorb these variables into the constants $c_i$).
Assuming $p_i < p_j$, $p_k < p_l$ and $p_i + p_j \neq p_k + p_l$, we have
\be
     \theta_{ij} - \theta_{kl} = (p_i + p_j - p_k - p_l)( x - x_{ij,kl}(y,t) ) \, ,
               \label{theta_ij-theta_kl}
\ee
where
\be
  x_{ij,kl}(y,t) &=& \frac{1}{p_i + p_j - p_k - p_l} \Big( -(p_i^2 + p_j^2 - p_k^2 - p_l^2) \, y 
                - (p_i^3 + p_j^3  - p_k^3 - p_l^3) \, t   \nonumber \\
  && - c_i - c_j + c_k + c_l + \log\Big(\frac{p_l-p_k}{p_j-p_i}\Big) \Big)  \; .
          \label{x_ij,kl}
\ee
The boundary between the regions associated with the two phases $\theta_{ij}$ and $\theta_{kl}$ 
is therefore given by $x = x_{ij,kl}(y,t)$. In particular, we find that
\be
     x_{ik,jk} = x_{ij} + \frac{1}{p_j - p_i} \, \log\Big(\frac{p_k-p_i}{p_k-p_j}\Big) \, ,
               \label{x_ik,jk}
\ee
with 
\bez
    x_{ij}(y,t) = -(p_i+p_j) \, y - (p_i^2 + p_i p_j + p_j^2) \, t - c_{ij} 
\eez
(an expression that already appeared in section~\ref{subsec:first_step}). This in turn implies
\bez
     x_{il,jl} - x_{ik,jk} = - \frac{1}{p_j - p_i} \, \ell(p_i,p_j,p_k,p_l)  \, ,
\eez
where
\bez
     \ell(p_i,p_j,p_k,p_l) = \log\Big(\frac{(p_k-p_i)(p_l-p_j)}{(p_l-p_i)(p_k-p_j)} \Big) 
\eez
is the logarithm of the \emph{cross ratio} of the constants $p_i,p_j,p_k,p_l$. Hence the 
boundary lines $x = x_{ik,jk}$ and $x= x_{il,jl}$, $k \neq l$, are always parallel with 
a constant (i.e. $y$- and $t$-independent) separation on the $x$-axis. We note that these 
``shifts'' also do not depend on the parameters $c_i$ (hence also not on $t^{(n)}$, $n>3$). 
In particular, they coincide with the asymptotic phase shifts (difference of phase values for 
$x \to \pm \infty$) given in \cite{Koda10}.

Furthermore, the boundary lines $x_{ij,kl}$, $x_{kl,mn}$ meet at the point with $y$-coordinate
\bez
    y_{ij,kl,mn} &=& - \Big( \frac{p_i^2 + p_j^2 - p_k^2 - p_l^2}{p_i+p_j-p_k-p_l} 
               - \frac{p_k^2 + p_l^2 - p_m^2 - p_n^2}{p_k+p_l-p_m-p_n} \Big)^{-1} \Big(  \nonumber \\ 
      &&  \frac{1}{p_i+p_j-p_k-p_l}  
          \Big[ (p_i^3 + p_j^3 - p_k^3 -p_l^3) \, t + c_i + c_j - c_k - c_l
              + \log\Big( \frac{p_j-p_i}{p_l-p_k}\Big) \Big]  \nonumber \\
      && - \frac{1}{p_k+p_l-p_m-p_n} \Big[ (p_k^3 + p_l^3 - p_m^3 -p_n^3) \, t + c_k + c_l - c_m - c_n
              + \log\Big( \frac{p_l-p_k}{p_n-p_m}\Big) \Big] \Big), 
                     \label{y_ijklmn} 
\eez
provided that the inverses exist. Moreover, we have the identities
\be
   \theta_{ij} - \theta_{mn} &=& (p_i+p_j-p_m-p_n)(x - x_{ij,kl}) 
    + \frac{1}{p_i+p_j-p_k-p_l} \Big( (p_i^2+p_j^2)(p_m+p_n-p_k-p_l) \nonumber \\
   && + (p_k^2+p_l^2)(p_i+p_j-p_m-p_n) + (p_m^2+p_n^2)(p_k+p_l-p_i-p_j) \Big) (y - y_{ij,kl,mn})  
    , \quad       \label{theta_ij-theta_mn} 
\ee
and
\be
   \theta_{ij}-\theta_{kl} &=& (p_i + p_j - p_k - p_l)(x-x_{ij,ik})
   + (p_i-p_l)(p_i-p_j-p_k+p_l)(y-y_{ij,ik,jl}) \nonumber \\
   && + \ell(p_i,p_l,p_k,p_j) \; . \label{theta_diff_x_ijik}
\ee
They do \emph{not explicitly} depend on $t$, nor on the constants $c_i$ and the $\log \Delta$ terms. 
The further analysis turns out to be quite involved, though. 
A fair qualitative understanding can be reached without a deeper analysis, however, as outlined 
in the following. 

According to our assumptions, 
$\mathfrak{U}_i = \{ \theta_j \, | \, \epsilon_{ij} \neq 0, \, j=1,\ldots, M+1 \}$, $i=1,\ldots,n$, 
are disjoint sets. 
If we can \emph{neglect the effect of all the terms} $\log \Delta(p_{i_1}, \ldots, p_{i_n})$, then the  
tropical approximation is given by\footnote{Whereas in section~\ref{sec:simplest_class} we 
only used the tropical binary operation $a \oplus b = \max\{a,b\}$, here also  
the complementary one shows up, i.e. $a \odot b = a + b$.  }
\bez 
      \log(\tau) \simeq \sum_{i=1}^n \max(\mathfrak{U}_i) \, ,
\eez
which unveils the line soliton configuration as a \emph{superimposition} of the line soliton configurations 
corresponding to the constituents $f_i$, $i=1,\ldots,n$.\footnote{In the case of the solution $\tau_P$
below, we have $f_1 = e_1 - e_4$, which leads to a \emph{singular} solution. However, we note 
that this strong approximation does \emph{not} depend on the sign of the coefficients $\epsilon_{ij}$. 
As a consequence, in this approximation $f_1$ gets replaced by $e_1 + e_4$, which determines 
a line soliton.}

Superimposing two line soliton configurations, due to the locality of the KP equation there can 
only be an interaction between them at points where a branch of one of them crosses a branch 
of the other. This is locally an interaction between two line solitons, 
where now we should switch on the $\log \Delta$ term. We shall see in the next example what 
this brings about. 

For $M=3$, i.e. four phases, the regularity condition only allows the two 2-forms
\bez
      \tau_O = (e_1 + e_2) \wedge (e_3 + e_4) \qquad \mbox{and} \qquad
      \tau_P = (e_1 - e_4) \wedge (e_2 + e_3) \, ,
\eez
which belong to classes called ``O-type" and ``P-type" by some authors 
(see e.g. \cite{CLM10,Koda10}). We will consider the O-type solution in  
detail in Example~\ref{ex:O-type}. The analysis of the P-type solution is very much 
the same. In addition to the 2-form solutions, further regular solutions for $M=3$ 
are given by $\tau = e_1 + e_2 + e_3 + e_4$, belonging to our special class, and its dual 
$\tau_\star = e_2 \wedge e_3 \wedge e_4 + e_1 \wedge e_3 \wedge e_4 
+ e_1 \wedge e_2 \wedge e_4 + e_1 \wedge e_2 \wedge e_3$ (cf. Example~\ref{ex:y_to_-y}). 
Further regular solutions are obtained from solutions with $M>3$ by taking limits 
where pairs of neighboring $p$'s coincide, see Example~\ref{ex:2-forms_as_limits} below.

\begin{example}
\label{ex:O-type}
Assuming $p_1 < p_2 < p_3 < p_4$, we have
\bez
      \tau_O = (p_3 - p_1) \, e_1 e_3 + (p_4 - p_1) \, e_1 e_4 + (p_3 - p_2) \, e_2 e_3 
             + (p_4 - p_2) \, e_2 e_4 
           = e^{\theta_{13}} + e^{\theta_{14}} + e^{\theta_{23}} + e^{\theta_{24}} \; .
\eez
Our tropical approximation is given by 
\bez
    \log(\tau) \simeq \max\{ \theta_{13}, \theta_{14}, \theta_{23}, \theta_{24} \} \; .
\eez 
If the constants $\log(p_j-p_i)$ are negligible, then $\log(\tau) \simeq \max\{ \theta_{1},\theta_2\} + 
\max\{\theta_{3},\theta_{4}\}$ and a plot of $\log(\tau)_{xx}$ is simply the result of superimposing 
the plots of $\log(e_1+e_2)_{xx}$ and $\log(e_3+e_4)_{xx}$, hence displaying two crossing lines, corresponding 
to $\theta_1=\theta_2$, respectively $\theta_3=\theta_4$ (see Fig.~\ref{fig:contplot_wedge}). 
In general, however, the constants $\log(p_j-p_i)$ are 
\emph{not} negligible, of course, and the situation is more complicated (see the right plot in 
Fig.~\ref{fig:contplot_wedge}). 
\begin{figure}[H] 
\begin{center} 
\resizebox{!}{3.5cm}{
\includegraphics{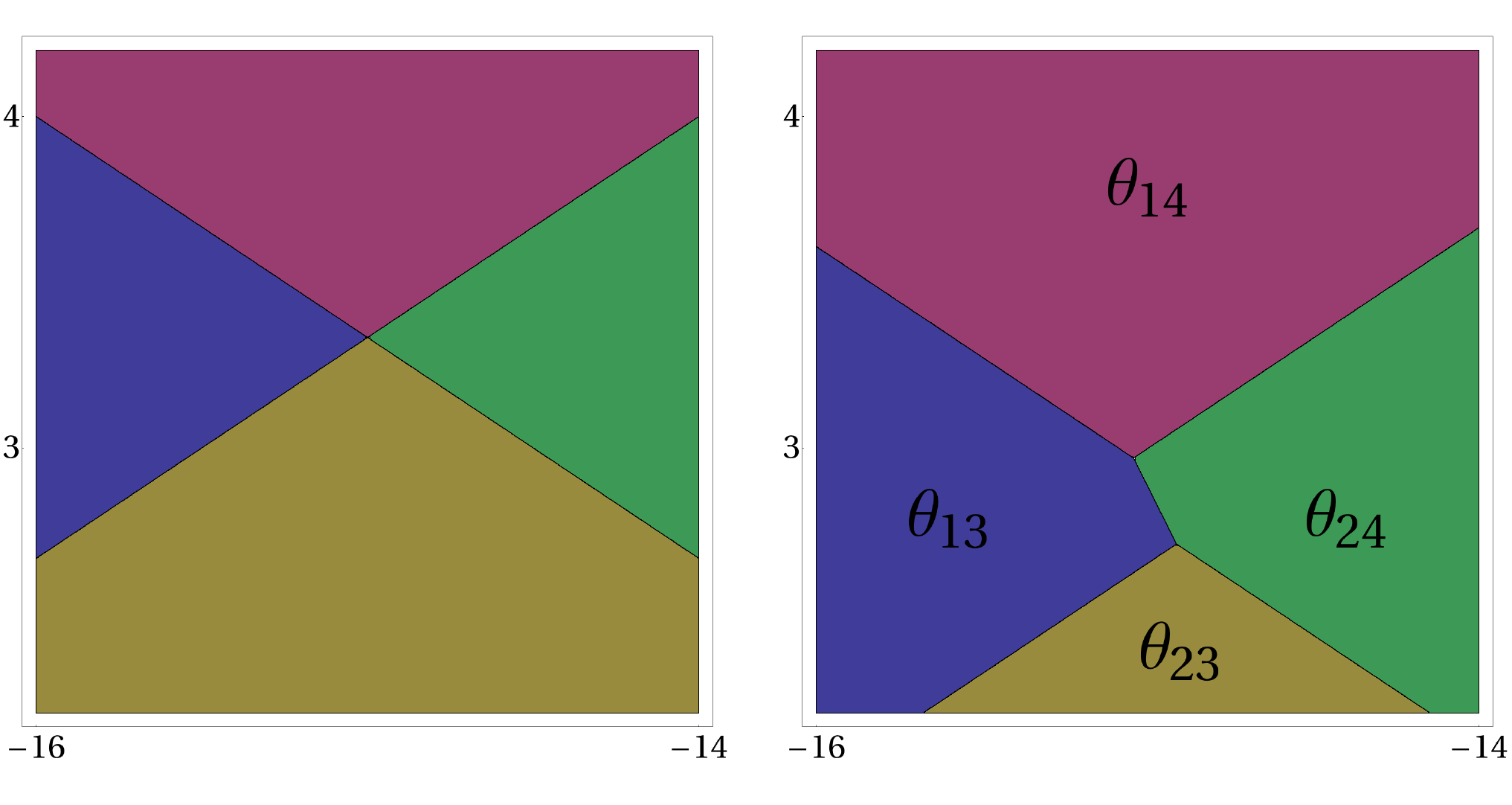}
}
\parbox{15cm}{
\caption{The left phase region plot shows $\max\{\theta_1,\theta_2\} + \max\{\theta_3,\theta_4\}$ 
as a function of $x$ (horizontal axis) and $y$, the right one shows 
the full tropical approximation $\max\{\theta_{13},\theta_{14},\theta_{23},\theta_{24}\}$, at $t=0$ (and 
all higher $t^{(r)}$ also set to zero). Here we chose the solution of Example~\ref{ex:O-type} 
with $p_1=-1, p_2 = -1/2, p_3=1/4, p_4=5/4$ and $c_1=c_4=0, c_2=-c_3=-10$.
\label{fig:contplot_wedge} }
}
\end{center} 
\end{figure}
\noindent
(\ref{theta_ij-theta_kl}) together with (\ref{x_ik,jk}) yields
\bez
    \theta_{13} = \theta_{14} \quad \Longleftrightarrow 
     && x = x_{34}(y,t) + \frac{1}{p_4 - p_3} \, \log\Big(\frac{p_3-p_1}{p_4-p_1}\Big) 
       =: x_{13,14}(y,t) \, , \\
    \theta_{23} = \theta_{24} \quad \Longleftrightarrow 
     && x = x_{34}(y,t) + \frac{1}{p_4 - p_3} \, \log\Big(\frac{p_3-p_2}{p_4-p_2}\Big) 
       =: x_{23,24}(y,t)  \, , \\
     \theta_{13} = \theta_{23} \quad \Longleftrightarrow 
     && x = x_{12}(y,t) + \frac{1}{p_2 - p_1} \, \log\Big(\frac{p_3-p_1}{p_3-p_2}\Big) 
       =: x_{13,23}(y,t)   \, , \\
     \theta_{14} = \theta_{24} \quad \Longleftrightarrow 
     && x = x_{12}(y,t) + \frac{1}{p_2 - p_1} \, \log\Big(\frac{p_4-p_1}{p_4-p_2}\Big) 
       =: x_{14,24}(y,t)   \; . 
\eez
Moreover, according to (\ref{theta_ij-theta_kl}) and (\ref{x_ij,kl}) we have
\bez
    \theta_{13} = \theta_{24} \quad \Longleftrightarrow 
     && x = - \frac{1}{p_1+p_3-p_2-p_4} \Big( (p_1^2+p_3^2-p_2^2-p_4^2) \, y
         + (p_1^3+p_3^3-p_2^3-p_4^3) \, t \\
     && \qquad + c_1 + c_3 - c_2 - c_4 
        - \log\Big(\frac{p_4-p_2}{p_3-p_1}\Big) \Big) 
         =: x_{13,24}(y,t) \, , \\
    \theta_{14} = \theta_{23} \quad \Longleftrightarrow 
     && x = - \frac{1}{p_1+p_4-p_2-p_3} \Big( (p_1^2+p_4^2-p_2^2-p_3^2) \, y
         + (p_1^3+p_4^3-p_2^3-p_3^3) \, t \\
     && \qquad + c_1 + c_4 - c_2 - c_3
        - \log\Big(\frac{p_3-p_2}{p_4-p_1}\Big) \Big) 
        =: x_{14,23}(y,t) \; .
\eez
The boundary $\theta_{14} = \theta_{23}$ cannot be expressed in the form $x=x_{14,23}(y,t)$ if 
$p_1+p_4-p_2-p_3$ vanishes, it is then parallel to the $x$-axis. 
The other boundaries can always be expressed in this form (as a consequence of 
$p_1 < p_2 < p_3 < p_4$). The two boundary lines given by 
$\theta_{13} = \theta_{14}$ and $\theta_{23} = \theta_{24}$, respectively, and also those 
given by $\theta_{13} = \theta_{23}$ and $\theta_{14} = \theta_{24}$, respectively, 
are always \emph{parallel}, with a constant separation on the $x$-axis given by 
\bez
   x_{13,14}(y,t) - x_{23,24}(y,t)
                   &=& \frac{1}{p_4 - p_3} \, \ell(p_1,p_2,p_3,p_4) \, , \\
   x_{14,24}(y,t) - x_{13,23}(y,t)
                   &=& - \frac{1}{p_2 - p_1} \, \ell(p_1,p_2,p_3,p_4) \; .
\eez
The point in which the boundary lines $x=x_{13,14}(y,t)$ and $x=x_{14,24}(y,t)$ 
intersect (at time $t$), and thus the three phases $\theta_{13}, \theta_{14}, \theta_{24}$ meet, 
has the $y$-coordinate
\bez
     y_{13,14,24} 
 &=& - \frac{1}{p_1+p_2-p_3-p_4} \Big( (p_1^2 + p_1 p_2 + p_2^2 - p_3^2 - p_3 p_4 -p_4^2)  \, t
     + c_{12} - c_{34} \\
  && + \frac{1}{p_4-p_3} \log\Big( \frac{p_3-p_1}{p_4-p_1} \Big)
     + \frac{1}{p_2-p_1} \log\Big( \frac{p_4-p_2}{p_4-p_1} \Big) \Big) \; .
\eez 
Similarly, the intersection point of the lines $x=x_{23,24}(y,t)$ and $x=x_{13,23}(y,t)$, 
where the three phases $\theta_{13}, \theta_{23}, \theta_{24}$ meet, 
has the $y$-coordinate 
\bez
     y_{13,23,24} 
 &=& - \frac{1}{p_3+p_4-p_1-p_2} \Big( (p_3^2 + p_3 p_4 + p_4^2 - p_1^2 - p_1 p_2 -p_2^2)  \, t
     + c_{34} - c_{12} \\
  && + \frac{1}{p_4-p_3} \log\Big( \frac{p_4-p_2}{p_3-p_2} \Big)
     + \frac{1}{p_2-p_1} \log\Big( \frac{p_3-p_1}{p_3-p_2} \Big) \Big) \; .
\eez 
The difference is 
\bez
    \delta y_{\mathrm{shift}} = \frac{1}{p_3-p_1+p_4-p_2} \Big( \frac{1}{p_4-p_3} + \frac{1}{p_2-p_1} \Big) 
            \, \ell(p_1,p_2,p_3,p_4)  \, ,
\eez
which is constant and moreover independent of the constants $c_i$. 
The difference of the corresponding $x$-coordinates is given by 
\bez
  \delta x_{\mathrm{shift}} = \frac{p_2^2 - p_1^2 + p_4^2 - p_3^2}{(p_3 -p_1 +p_4 -p_2)
           (p_2-p_1)(p_4-p_3)} \, \ell(p_1,p_2,p_3,p_4) \; .
\eez
The slope of the corresponding soliton line segment is
\bez
     \delta y_{\mathrm{shift}} / \delta x_{\mathrm{shift}}
  = -\frac{p_2-p_1 + p_4-p_3}{p_2^2 - p_1^2 + p_4^2 - p_3^2}  \; .
\eez
We note that the shift becomes infinite in the limit $p_2 \to p_3$ (since $\ell(p_1,p_2,p_3,p_4)$ 
then becomes infinite), so that the phase region $\theta_{23}$ in Fig.~\ref{fig:contplot_wedge} 
disappears towards $y=-\infty$. Hence we end up with a Miles resonance in this limit. 

As special cases of (\ref{theta_ij-theta_mn}), we obtain
\bez
    \theta_{13}-\theta_{24} &=& -(p_4-p_3+p_2-p_1)(x-x_{13,23}) 
             - (p_4-p_3)(p_3 +p_4-p_1-p_2)(y-y_{13,23,24}) \, , \\
    \theta_{23}-\theta_{13} &=& (p_2-p_1)(x-x_{23,24}) 
             - (p_2-p_1)(p_3 +p_4-p_1-p_2)(y-y_{13,23,24}) \, , \\
    \theta_{13}-\theta_{23} &=& -(p_2-p_1)(x-x_{13,24}) 
             + (p_2-p_1)(p_4-p_3) \frac{p_3 +p_4-p_1-p_2}{p_4-p_3+p_2-p_1} \, (y-y_{13,23,24}) \; .
\eez 
As a consequence, (for fixed $t$) the half lines $\{ x=x_{13,23}(y,t) \, | \, y > y_{13,23,24} \}$,
$\{ x=x_{23,24}(y,t) \, | \, y > y_{13,23,24} \}$, and 
$\{ x_{13,24}(y,t) \, | \, y < y_{13,23,24} \}$ are non-visible. Furthermore, we find
\bez
    \theta_{13}-\theta_{24} &=& -(p_4-p_3+p_2-p_1)(x-x_{13,14}) 
             + (p_2-p_1)(p_3 +p_4-p_1-p_2)(y-y_{13,14,24}) \, , \\
    \theta_{14}-\theta_{13} &=& (p_4-p_3)(x-x_{14,24}) 
             + (p_4-p_3)(p_3 +p_4-p_1-p_2)(y-y_{13,14,24}) \, , \\
    \theta_{13}-\theta_{14} &=& -(p_4-p_3)(x-x_{13,24}) 
             - (p_2-p_1)(p_4-p_3) \frac{p_3 +p_4-p_1-p_2}{p_4-p_3+p_2-p_1} \, (y-y_{13,14,24}) \; .
\eez 
This shows that the half lines $\{ x=x_{13,14}(y,t) \, | \, y < y_{13,14,24} \}$,
$\{ x=x_{14,24}(y,t) \, | \, y < y_{13,14,24} \}$, and 
$\{ x_{13,24}(y,t) \, | \, y > y_{13,14,24} \}$ are non-visible. Moreover, one can show 
that the whole line given by $x=x_{14,23}$ is non-visible. 
All this is compatible with the right plot in Fig.~\ref{fig:contplot_wedge}, of course.
We know that the complementary half lines are visible in the approximation where we 
neglect the phase shift terms $\log \Delta$ (and the two triple phase coincidences merge). 
Since (\ref{theta_ij-theta_mn}) does not explicitly 
depend on these terms, we can conclude that they remain visible when switching the phase 
shifts on. Of course, we can confirm this by further explicit computations. For example, 
at the three-phase coincidence with $y$-coordinate $y_{13,23,24}$, (\ref{theta_diff_x_ijik}) 
implies $\theta_{23} - \theta_{14} = \ell(p_2,p_1,p_4,p_3) >0$, hence this point is visible. 
\end{example}

\begin{example}
In case of $\tau_P$, the tropical approximation is
$\log(\tau_P) \simeq \max\{ \theta_{12}, \theta_{13}, \theta_{24}, \theta_{34} \}$. 
We can proceed as in Example~\ref{ex:O-type}. The line determined by 
$\theta_{12}=\theta_{34}$ can always be solved for $x$ (as a consequence of $p_1<p_2<p_3<p_4$) 
and turns out to be non-visible (see also Fig.~\ref{fig:P-type}). 
The slope of the line given by $\theta_{13}=\theta_{24}$ is 
$-(p_1 - p_2 + p_3 - p_4)/(p_1^2 - p_2^2 + p_3^2 - p_4^2)$. Furthermore, we obtain
\bez
    x_{12,13} - x_{24,34} 
  = \frac{1}{p_3 - p_2} \, \ell(p_2,p_3,p_1,p_4) \, , \qquad
    x_{13,34} - x_{12,24} 
  = - \frac{1}{p_4 - p_1} \, \ell(p_2,p_3,p_1,p_4) \; .
\eez
In contrast to the case treated in Example~\ref{ex:O-type}, we need an additional 
condition, namely $p_1 + p_4 \neq p_2 + p_3$, in order to ensure the existence of 
(then visible) three-phase coincidences, here with $y$-coordinate $y_{12,13,24}$, 
respectively $y_{13,24,34}$. Their distance along the $y$-axis is
\bez
    y_{13,24,34} - y_{12,13,24} 
  = - \frac{p_1 - p_2 + p_3 - p_4}{(p_3 - p_2)(p_4 - p_1)(p_1 + p_4 - p_2 -p_3)}
       \, \ell(p_2,p_3,p_1,p_4) \; .
\eez
The excluded case where $p_1 + p_4 = p_2 + p_3$ is further considered in Example~\ref{ex:parallel}.
\begin{figure}[H] 
\begin{center} 
\resizebox{!}{3.5cm}{
\includegraphics{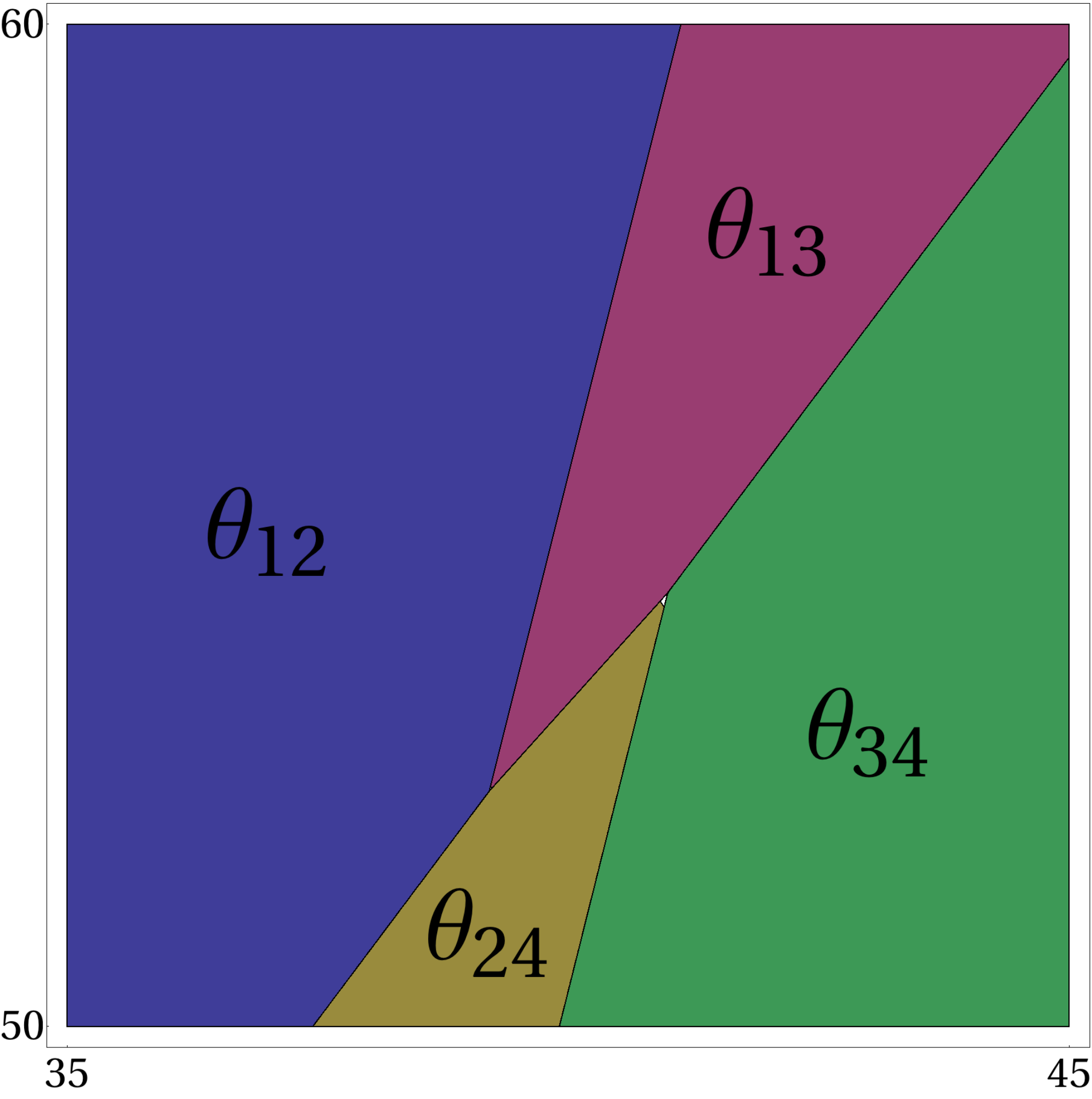}
}
\parbox{15cm}{
\caption{Phase region plot of $\max\{ \theta_{12}, \theta_{13}, \theta_{24}, \theta_{34} \}$
for the P-type solution at $t=0$ and with $p_1=-2, p_2 = -1/2, p_3=1/4, p_4=5/4$, 
$c_1=c_4=0$, $c_2=-c_3=-10$.   \label{fig:P-type} }
}
\end{center} 
\end{figure}
\end{example} 

\begin{example}
\label{ex:parallel}
Here we consider $\tau_P$ with $p_1 + p_4 = p_2 + p_3$. Writing
\bez
    p_1 = \frac{1}{2} (q - a - b) \, , \quad
    p_2 = \frac{1}{2} (q - a) \, , \quad
    p_3 = \frac{1}{2} (q + a) \, , \quad
    p_4 = \frac{1}{2} (q + a + b) \, , 
\eez
which real constants $a,b >0$ and $q$, we find
\bez
    x_{12,13} &=& - q \, y - \frac{1}{4} (a^2 + 3 q^2) \, t 
                  + \frac{1}{a} \Big( c_2 - c_3 - \log(1+ 2 a/b) \Big) \, , \\
    x_{12,24} &=& - q \, y - \frac{1}{4} [ (a+b)^2 + 3 q^2 ] \, t 
                  + \frac{1}{a+b} \Big( c_1 - c_4 - \log(1+ 2 a/b) \Big) \, , \\
    x_{12,34} &=& - q \, y - \frac{1}{4} (a^2 + ab + b^2 + 3 q^2) \, t 
                   + \frac{1}{2a+b} ( c_1 + c_2 - c_3 -c_4 ) \, , \\ 
    x_{13,24} &=& - q \, y - \frac{1}{4} (3 a^2 + 3ab + b^2 + 3 q^2) \, t 
                   + \frac{1}{b} ( c_1 - c_2 + c_3 -c_4 )  \, , \\
    x_{13,34} &=& - q \, y - \frac{1}{4} [ (a+b)^2 + 3 q^2 ] \, t 
                  + \frac{1}{a+b} \Big( c_1 - c_4 + \log(1+ 2 a/b) \Big) \, , \\
    x_{24,34} &=& - q \, y - \frac{1}{4} (a^2 + 3 q^2) \, t 
                  + \frac{1}{a} \Big( c_2 - c_3 + \log(1+ 2 a/b) \Big) \; .
\eez
Hence all these lines are parallel with slope $-1/q$. 
The two boundary lines $x=x_{12,13}$ and $x=x_{24,34}$ move with the same speed, and the same 
holds for $x=x_{12,24}$ and $x=x_{13,34}$. 
We note that $ x_{12,13} < x_{24,34}$ and $x_{12,24} < x_{13,34}$. 
Furthermore, we find the following coincidence events:
\bez
    x_{13,24} = x_{13,34} = x_{24,34} 
       \quad &\hspace{.2cm}& \quad t = t_0 - \Delta t =: t_- \\
    x_{12,34} = x_{13,24}  \quad &\mbox{at}& \quad t = t_0 \\
    x_{12,13} = x_{12,24} = x_{13,24} \quad &\hspace{.2cm}& \quad t = t_0 + \Delta t =: t_+ \, , 
\eez
where
\bez
    t_0 = 4 \frac{a(c_1-c_4) - (a+b)(c_2-c_3)}{ab (a+b)(2a+b)} \, , \qquad
    \Delta t = \frac{4 \log(1+2a/b)}{a(a+b)(2a+b)} > 0 \; .
\eez
Since $\theta_{12} - \theta_{13} = \frac{1}{4} ab (a+b)(t-t_0) - \log(1+2a/b)$ 
and $\theta_{12} - \theta_{24} = -\frac{1}{4} ab (a+b)(t-t_0) - \log(1+2a/b)$ on $x= x_{12,34}$, 
we conclude that this line is never visible. Hence also the event at $t_0$ is non-visible.
Along $x=x_{12,13}$ we find $\theta_{12}- \theta_{24} = - (q - 2 p_1)(q - p_1 - p_2)(p_2 - p_1)(t - t_+)$ and 
$\theta_{12}- \theta_{34} = - (q - 2 p_1)(q - p_1 - p_2)(p_2 - p_1)(t - t_+) + 2 \log[(q-p_2-p_1)/(p_2-p_1)]$, 
which are both positive for $t < t_+$, and the first expression is negative for $t > t_+$. 
Hence $x=x_{12,13}$ is visible for $t < t_+$ and non-visible for $t > t_+$. 
In the same way we find that $x=x_{13,34}$ is visible for $t < t_-$ and non-visible for $t > t_-$, 
$x=x_{12,24}$ is visible for $t > t_+$ and non-visible for $t < t_+$, $x=x_{24,34}$ is visible 
for $t > t_-$ and non-visible for $t < t_-$, and $x=x_{13,24}$ is visible for $t_- < t < t_+$ and 
non-visible otherwise. There are no further visible lines. Hence, for $t < t_-$ and $t > t_+$  
there are two visible boundary lines corresponding to \emph{two parallel} line solitons\footnote{The 
constants $a$ and $b$ determine the amplitudes of these line solitons, see Appendix~\ref{AppD}. }  
(oblique to the $x$-axis). But for $t_- < t < t_+$ there are \emph{three} parallel visible boundary lines, 
see also Fig.~\ref{fig:parallel}. 
This means that for $t < t_-$ and $t > t_+$ only three of the four phases are visible, and all four are 
visible only for $t_- < t < t_+$. 
\begin{figure}[H] 
\begin{center} 
\resizebox{!}{2.cm}{
\includegraphics{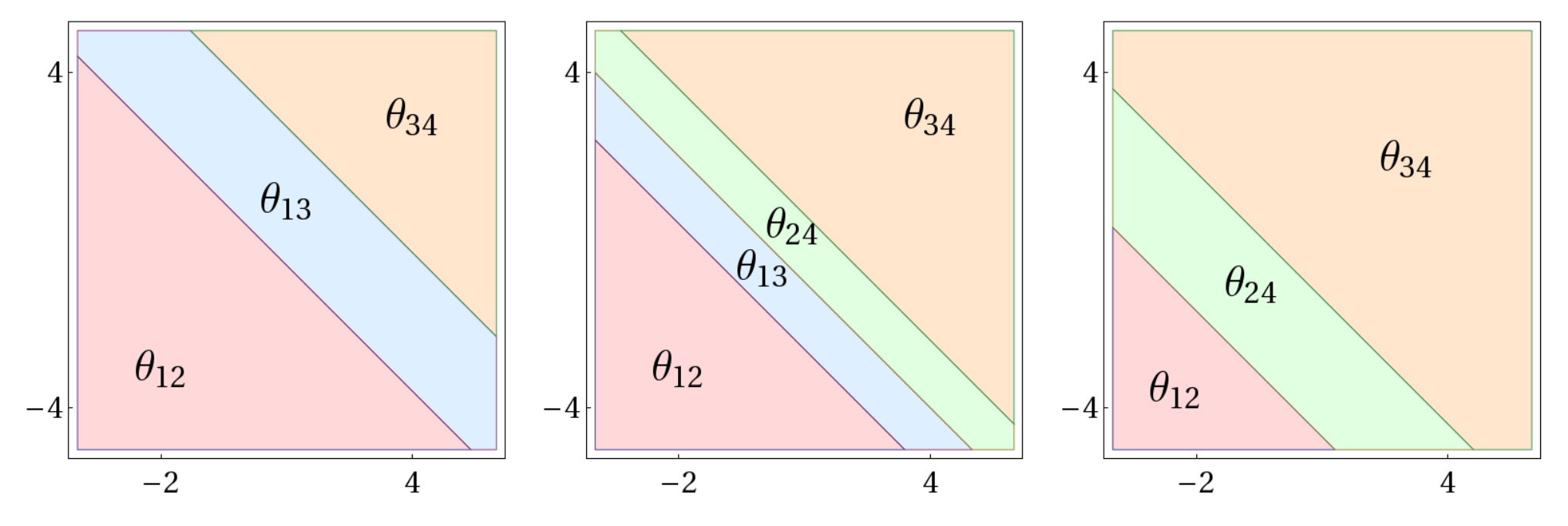}
}
\hspace{1cm}
\resizebox{!}{2.cm}{
\includegraphics{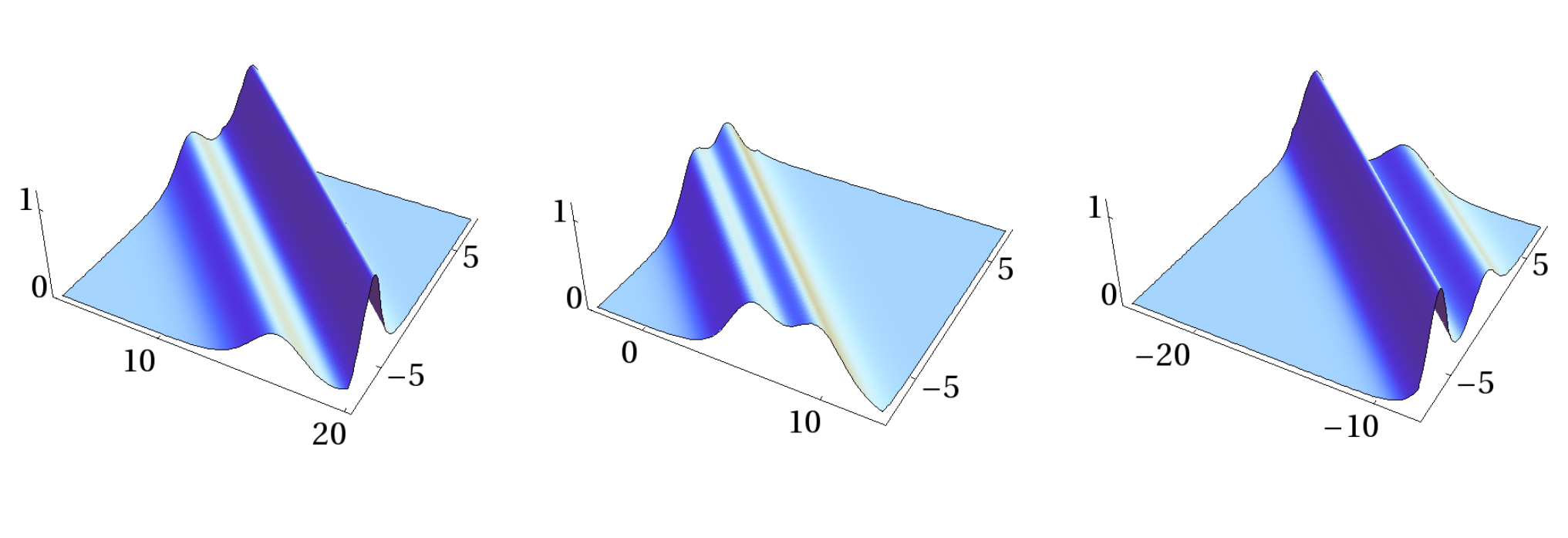}
}
\parbox{15cm}{
\caption{Phase region plot of a solution with two parallel line solitons, 
as described in Example~\ref{ex:parallel}, at times $t=-2,0,2$. They exchange a ``virtual line soliton''.
Here we chose $q=a=1$, $b=1/2$ and $c_i=0$, $i=1,\ldots,4$. To the right are plots of the exact solution 
at $t=-10,0,10$.    \label{fig:parallel} }
}
\end{center} 
\end{figure}
\noindent
The tropical description provides us with an interpretation of the soliton interaction process. 
For $t < t_-$ (left of the three region plots in Fig.~\ref{fig:parallel}) the lines $x_{12,13} \simeq x_{23}$ 
and $x_{13,34} \simeq x_{14}$ represent two line solitons moving from right to left, where the latter is 
faster than the first. At $t_-$ the faster soliton sends off a virtual line soliton (corresponding to 
$x_{13,24}$) and thereby mutates to $x_{24,34} \simeq x_{23}$, which is a new manifestation of the slower 
line soliton. At $t_+$ the original slower soliton swallows the virtual one and mutates to 
$x_{12,24} \simeq x_{14}$, which is a new manifestation of the original faster soliton. 
A generalization of this solution, now with $n$ \emph{parallel} line solitons\footnote{Actually, 
this case can be reduced to a discussion of the \emph{KdV} equation, in the tropical approximation. }, 
is given by
\bez
    \tau = (e_1 - (-1)^n e_{2n}) \wedge (e_2 - (-1)^{n-1} e_{2n-1}) \wedge \cdots \wedge
           (e_{n-1} - e_{n+2}) \wedge (e_n + e_{n+1}) \, , 
\eez  
where $p_1 < p_2 < \cdots < p_{2n}$ and $p_1+p_{2n} = p_2 + p_{2n-1} = \cdots = p_{n-1} + p_{n+2} 
= p_n + p_{n+1}$. 
Moreover, by taking the wedge product of two such functions, we can generate 
grid-like structures. For example, let
\bez
    \tau = (e_{-6} + e_{-1}) \wedge (e_{-5} - e_{-2}) \wedge (e_{-4} + e_{-3}) 
           \wedge (e_1 + e_6) \wedge (e_2 - e_5) \wedge (e_3 + e_4) \, ,
\eez
where $p_{-6} < p_{-5} < \cdots < p_{-1} < p_1 < \cdots < p_6$, 
$p_{-6} + p_{-1} = p_{-5} + p_{-2} = p_{-4} + p_{-3}$ and $p_1+p_6=p_2+p_5=p_3+p_4$. 
Fig.~\ref{fig:grid} shows a plot of such a solution. 
\begin{figure}[H] 
\begin{center} 
\resizebox{!}{3.5cm}{
\includegraphics{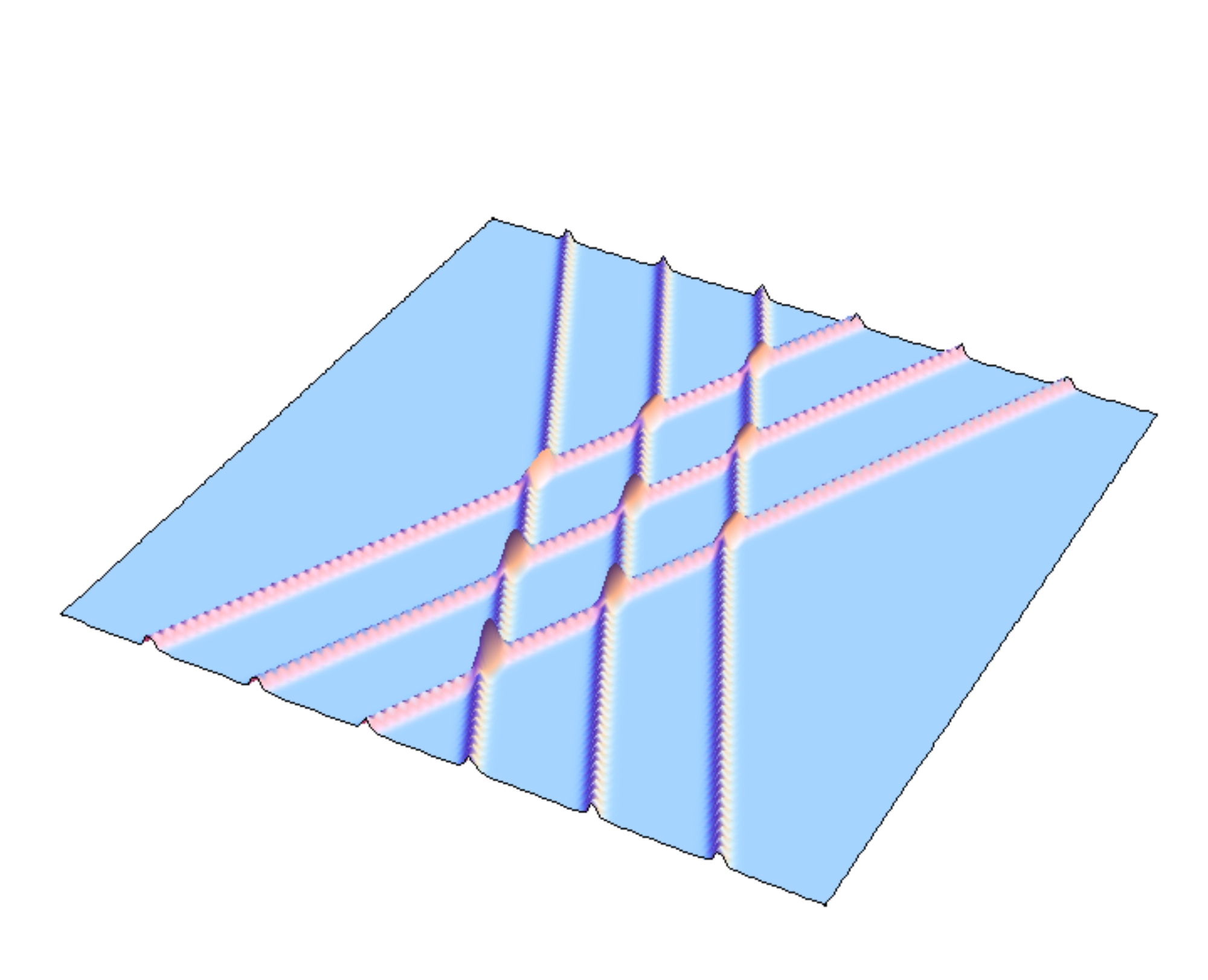}
}
\parbox{15cm}{
\caption{Plot of a grid-like solution at a fixed time, built from two families of parallel line solitons, 
as described in Example~\ref{ex:parallel}.
   \label{fig:grid} }
}
\end{center} 
\end{figure}
\end{example}

The above results suggest that the line soliton solution is generically obtained as a superimposition 
of the constituents (i.e. the factors in the wedge product, modulo conversion of negative to positive signs) 
and in addition with the creation of new line segments of constant length and slope 
due to the $\log \Delta$ phase shift terms, as in Example~\ref{ex:O-type}. Typically these 
new line segments will \emph{not} be visible in a line soliton plot, with the exception of 
the extremal cases considered next.

We explore what happens when two neighboring constants, say $p_i,p_{i+1}$, in the sequence 
of $p$'s approach each other. 
Writing $p_{i+1} - p_i = e^{-\alpha_i}$ and $b_i = e^{c_{i+1} - c_i}$, 
we have $e_{i+1} \simeq b_i \, e_i$ for large positive $\alpha_i$. 
If $i,i+1 \in \{k_1,\ldots, k_n \}$, we find
\bez
    \log \Delta(p_{k_1}, \ldots, p_{k_n}) \simeq 
    \log \Big( \frac{\pa}{\pa p_{i+1}} \Delta(p_{k_1}, \ldots, p_{k_n}) \Big)_{p_{i+1} = p_i} - \alpha_i \; .
\eez
As a consequence, the region dominated by $\theta_{k_1, \ldots,i,i+1, \ldots, k_n}$  
disappears in the limit $\alpha_i \to \infty$. 

If $i+1 \in \{k_1,\ldots, k_n \}$, but $i \not\in \{k_1,\ldots, k_n \}$, then
\bez
   \theta_{k_1 \ldots k_n} 
 &=& \theta_{k_1} + \cdots + \theta_{i+1} + \cdots + \theta_{k_n} 
       + \log \Delta(p_{k_1}, \ldots,p_{i+1}, \ldots, p_{k_n}) \\
 &\simeq& \theta_{k_1} + \cdots + \theta_i + \cdots + \theta_{k_n} + \log b_i
       + \log \Delta(p_{k_1}, \ldots,p_i, \ldots, p_{k_n})   \\
 &=& \theta_{k_1 \ldots i \ldots k_n} + \log b_i =: \tilde{\theta}_{k_1 \ldots i \ldots k_n}  \; .
\eez
Hence the $\theta_{k_1 \ldots i+1 \ldots k_n}$-region passes into a 
$\tilde{\theta}_{k_1 \ldots i \ldots k_n}$-region. Boundary lines between regions that 
do not carry an index $i+1$ remain unchanged. 

We conclude that, as $p_{i+1} \to p_i$, each region with dominating phase of the form 
$\theta_{k_1, \ldots,i,i+1, \ldots, k_n}$ is shifted away, the phase regions to its left 
and to its right meet, a corresponding boundary line is created.

\begin{example}
\label{ex:2-forms_as_limits}
We consider regular 2-form solutions with five phases (i.e. $M=4$), $p_1 < p_2 < p_3 < p_4 < p_5$, 
and limits where two neighboring constants coincide. \\
(1) $\tau = (e_1 + e_2) \wedge (e_3 + e_4 + e_5)$. Setting $p_3 = p_2$, we have $e_3 = a \, e_2$
with a constant $a >0$. Recalling that a constant overall factor 
of $\tau$ does not change the respective KP soliton solution, after a redefinition of $c_4$ and $c_5$, 
and a renumbering, we obtain $\tau' = (e_1 + e_2) \wedge (e_2 + e_3 + e_4)$. 
Fig.~\ref{fig:plimit_plots} shows an example for what happens as $p_3 \to p_2$. The phase region 
associated with this pair is shifted away to infinity in this limit. 
\begin{figure}[H] 
\begin{center} 
\resizebox{12.cm}{!}{
\includegraphics{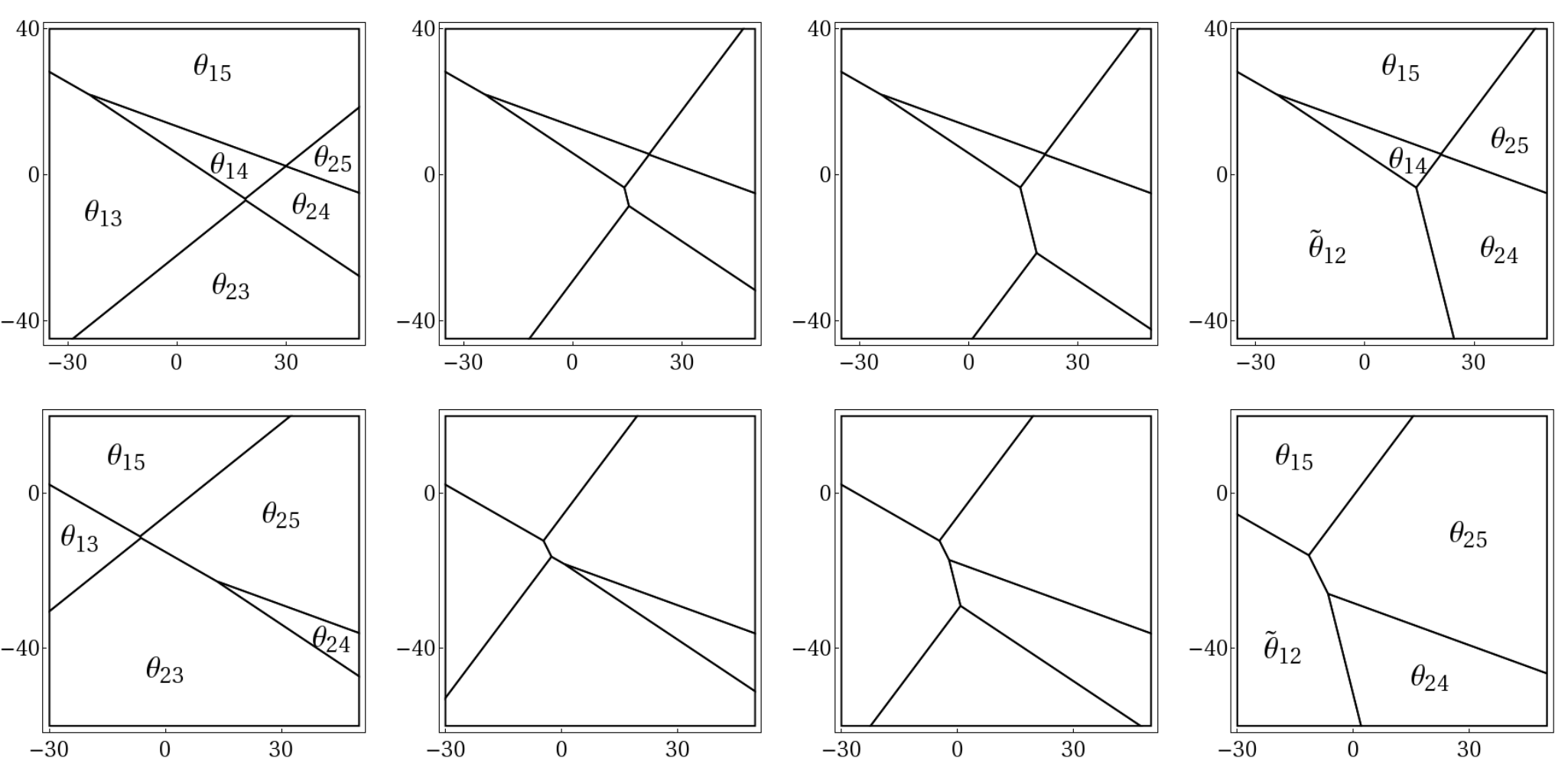}
}
\parbox{15cm}{
\caption{Tropical approximation of an exact solution of the form 
$\tau = (e_1 + e_2) \wedge (e_3 + e_4 + e_5)$, i.e. a superposition of a single 
line soliton and a Miles resonance, at two different times. 
From left to right, $p_3$ approaches $p_2$. In the rightmost plots we have $p_3=p_2$. 
 \label{fig:plimit_plots} }
}
\end{center} 
\end{figure}
\noindent
(2) $\tau = (e_1 + e_2 - e_5) \wedge (e_3 + e_4)$. 
Setting $p_5 = p_4$, after a redefinition of $c_1$ 
and $c_2$ we end up with $\tau' = (e_1 + e_2 - e_4) \wedge (e_3 + e_4)$. \\
(3) $\tau = (e_1 - e_5) \wedge (e_2 + e_3 + e_4)$. Setting $p_5 = p_4$, after a redefinition of $c_1$ 
we obtain $\tau' = (e_1 - e_4) \wedge (e_2 + e_3 + e_4)$. \\
(4) $\tau = (e_1 - e_4 - e_5) \wedge (e_2 + e_3)$. Setting $p_4 = p_3$, redefining $c_1$ 
and $c_5$, and finally renaming $p_5$ to $p_4$, we find $\tau' = (e_1 - e_3 - e_4) \wedge (e_2 + e_3)$. \\
Starting with a regular six phase solution, via \emph{two} limits we obtain a four phase solution: \\
(5)  $\tau = (e_1 + e_2 - e_6) \wedge (e_3 + e_4 + e_5)$. We set $p_6 = p_5$ and $p_3=p_2$ to obtain
$(e_1 + e_2 - a \, e_5) \wedge (b \, e_2 + e_4 + e_5)$. We can achieve $a=1$ with a redefinition of 
$c_1$ and $c_2$, or $b=1$ with a redefinition of $c_4$ and $c_5$, but not both simultaneously. 
After a renumbering we obtain $\tau' = (e_1 + e_2 - e_4) \wedge (a \, e_2 + e_3 + e_4)$, $a >0$. 
Fig.~\ref{fig:case7} shows a structure appearing in the tropical approximation that is not present 
in the full solution. But there are parameter values where the two bounded regions in the left plot 
in Fig.~\ref{fig:case7} indeed become visible (cf. Fig.~4 in \cite{Bion07}).
\begin{figure}[H] 
\begin{center} 
\resizebox{!}{2.5cm}{
\includegraphics{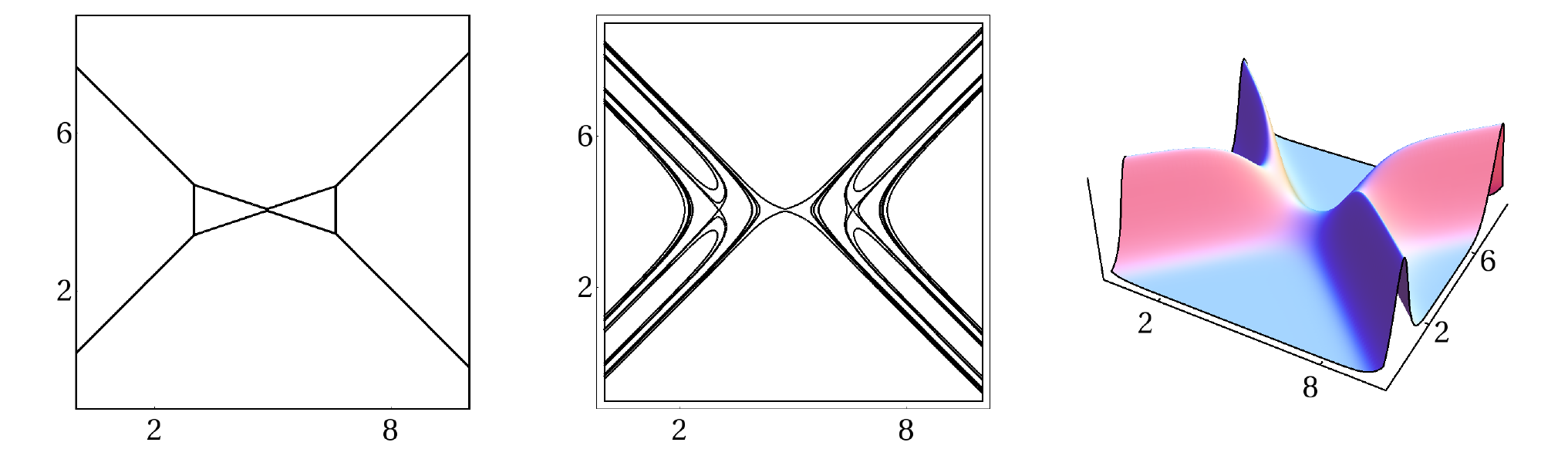}
}
\parbox{15cm}{
\caption{The left plot shows the tropical approximation of a solution of type (5) in 
Example~\ref{ex:2-forms_as_limits} at a fixed time in a region of size comparable with the 
width of a line soliton. The other plots show the full solution in the same region as 
a contour plot and a three-dimensional plot over the $xy$-plane. 
 \label{fig:case7} }
}
\end{center} 
\end{figure}
\noindent
Together with $\tau_O$ and $\tau_P$, we have seven types of four-phase 2-form solutions (cf. 
the seven cases of $(2,2)$-solutions in \cite{Koda10}). 
Modulo redefinitions of the constants $c_i$, $\tau'$ in (2) is the dual of 
that in (1), and also (3) and (4) are related in this way. $\tau_O, \tau_P$ and $\tau'$ 
in (5) are self-dual.
\end{example}

\section{Summary of further results and conclusions}
\label{sec:conclusions}
For the simplest class of KP-II line soliton solutions, we have shown that the time evolution 
can be described as a time-ordered sequence of rooted binary trees and that this constitutes a 
maximal chain in a Tamari lattice. 

Moreover, we derived general results (in particular in Appendix~\ref{AppA}) that allow to compute the data 
corresponding to transition events (where a rooted binary tree evolves into another). 
The fact that the soliton solutions extend to solutions of the KP \emph{hierarchy} plays 
a crucial role in the derivation of these results. 

Tamari lattices are related to quite a number of mathematical structures and our work adds to it by 
establishing a bridge to an integrable PDE, the KP equation (and moreover its hierarchy).\footnote{See
also \cite{Buch+Kori07} for a relation between integrable PDEs and polytopes.} 
The latter is well-known for other deep connections with various areas of mathematics. 

The family of Tamari lattices is actually not the only family of posets (or lattices) showing up 
in the line soliton classification problem. We already met in section~\ref{sec:simplest_class} a family 
where the nodes are the phases $\theta_i$ and the edges correspond to critical values of $x$. 
The underlying polytopes are a triangle ($M=2$), a tetrahedron ($M=3$), and their higher-dimensional 
analogs ($M>3$). 
Another family appeared in section~\ref{subsec:first_step}. Its nodes consist of chains of critical $x$-values 
and the edges correspond to critical $y$-values. The underlying polytopes are a tetragon ($M=3$), 
a cube ($M=4$), and hypercubes for $M>4$. 

\begin{figure}[H] 
\begin{center} 
\begin{minipage}[t]{5.5cm}
\resizebox{!}{8.cm}{
\includegraphics{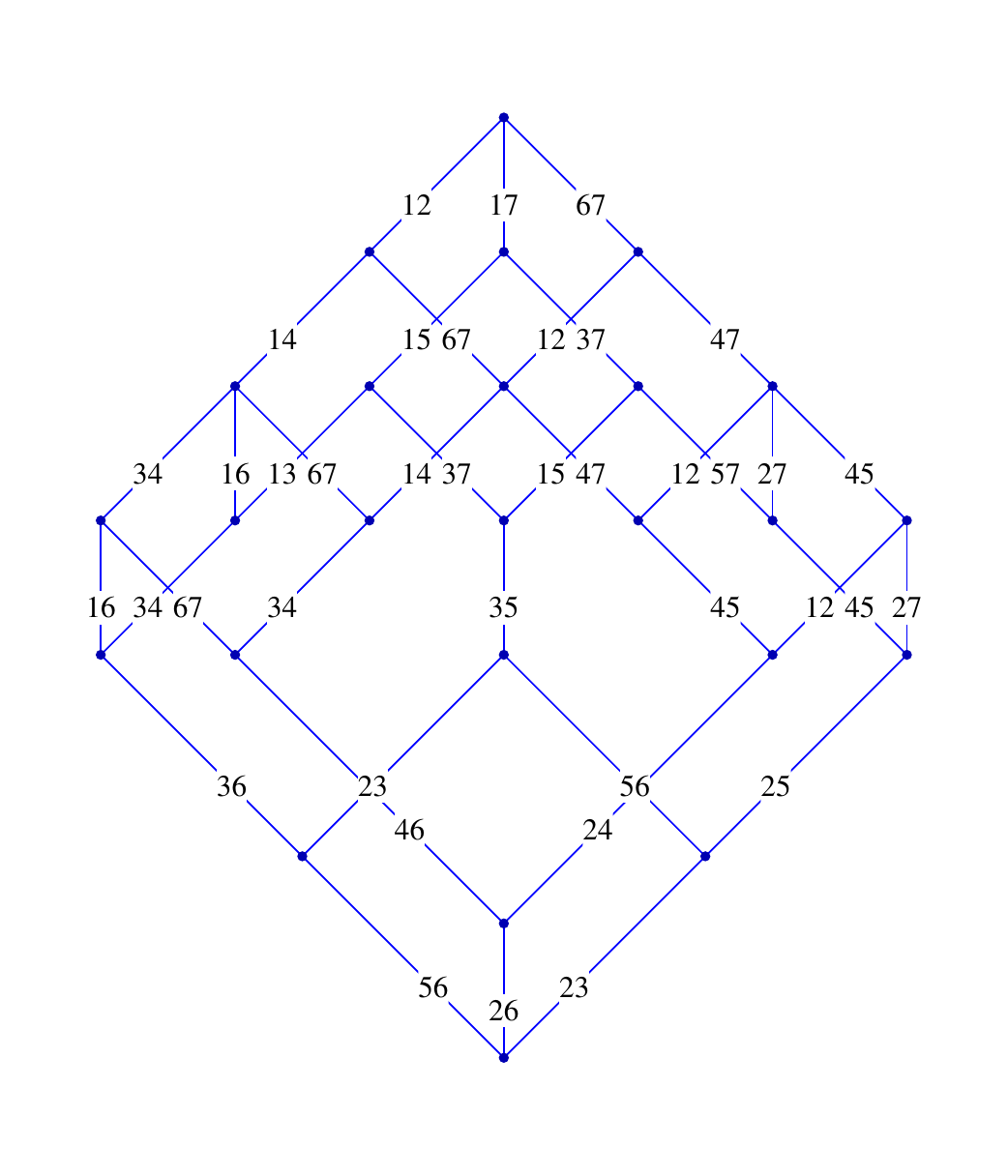}
}
\end{minipage}
\hspace{2cm}
\begin{minipage}[t]{4.5cm}
\vspace{-6.5cm}
\resizebox{!}{4.cm}{
\includegraphics{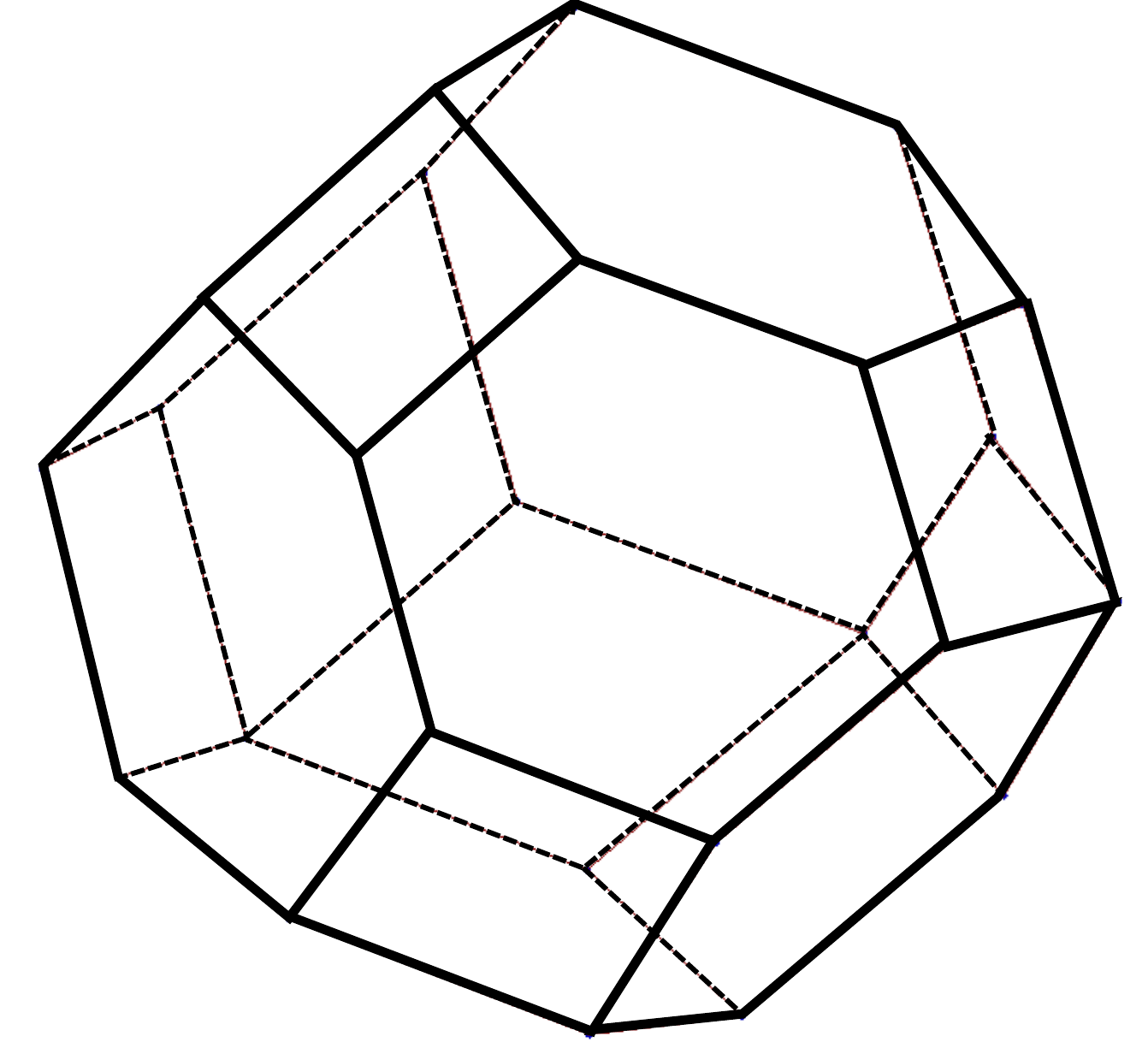}
}
\end{minipage}
\parbox{15cm}{
\caption{The left figure shows a new lattice. Its nodes are classes 
of maximal chains in $\bbT_5$ and the edges correspond to critical values of $t^{(4)}$. 
A two-digit number stands for its complement in $1234567$. 
The right figure represents this lattice as a polyhedron. It has 25 nodes, 39 edges and 16 faces, 
and it consists of seven hexagons and nine tetragons.  \label{fig:superT4} }
}
\end{center} 
\end{figure}

According to Figs.~\ref{fig:T4_t4deform1} and \ref{fig:T4_t4deform2} (see also
Fig.~\ref{fig:hexagon}) there is a new lattice of \emph{hexagon} form. 
Its six nodes are given by the six classes built from the nine maximal chains of $\bbT_4$ 
(see Figs.~\ref{fig:T4_t4deform1} and \ref{fig:T4_t4deform2}, and also Appendix~\ref{AppC}), 
and its edges correspond to the six critical values of $t^{(4)}$. This lattice is an analog of 
the pentagon Tamari lattice $\bbT_3$ and belongs to a new family. Its next member  
is obtained for $M=6$. It has 25 nodes, which consist of classes of maximal chains in $\bbT_5$ 
(see Appendix~\ref{AppC}), and its (directed) edges are again determined by the critical values 
of $t^{(4)}$, see Fig.~\ref{fig:superT4}. 
Moreover, we expect a hierarchy of families of lattices. We already mentioned the two families 
associated with the critical values of $x$ and $y$. The Tamari lattices correspond to the 
critical values of $t=t^{(3)}$, the next family is associated with the critical values of $t^{(4)}$. 
More generally, there is a family associated with the critical values of $t^{(n)}$, $n \in \mathbb{N}$. 
Comparison with the algebra of oriented simplexes formulated in \cite{Street87} (see also \cite{Cheng+Lauda04}) 
in terms of higher-dimensional categories shows striking relations which should be further elaborated.  
An exploration of more general classes of line solitons might exhibit relations 
with other posets (or lattices) and polytopes.

In this work we solved the classification problem for the simplest class of KP-II line soliton solutions, 
corresponding to rooted trees. 
Our classification rests upon the exploration of events where phases coincide. At such an event 
the tree that describes the line soliton configuration changes its form. A finer description 
is obtained by taking also events into account at which a transition between two trees \emph{with 
levels} (associated with the same rooted binary tree) takes place. Our exposition made contact 
with such a refinement at various places. A nice example is the ``missing face'' in the 
cube poset in Fig.~\ref{fig:cube}. We elaborated this refinement in Appendix~\ref{AppB} and explained 
in Appendix~\ref{AppC} how it lifts the Tamari lattices (or associahedra) to permutohedra. 

At first sight the classification for the simple class of line soliton solutions appears to 
be only a small step towards 
the classification of the whole set of line soliton solutions, which exhibit a much more 
complicated behavior. But this is not quite so, as outlined in section~\ref{sec:general}. 
Any line soliton solution can be written as a (suitably defined) 
exterior product of $\tau$-functions from the simple class. Generically such a product corresponds 
to superimposing the soliton graphs associated with the constituents. Since the interaction is local, 
there can only be a change in a neighborhood of a point where a soliton branch of one constituent 
meets a branch of another. At such a point a new line soliton segment (due to a phase shift) is 
created and its length does \emph{not} depend on time and \emph{not} on the constants $c_i$, but only on 
the values of the constants $p_i$. For generic parameter values, this effect is hardly visible. 
It becomes significant, however, in cases where some of the 
(a priori assumed to be different) constants $p_i$ coincide. These are the more complicated cases 
which should still be explored in more detail.  

We expect that our tropical approximations of KP line soliton solutions have a place 
in the tropical (totally positive) Grassmannian \cite{Stan+Pitm02,Spey+Will05}.
For other approaches to the KP line soliton classification problem we refer in particular to the 
review \cite{Koda10} and the references therein. 

Finally, we would like to stress that the tropical approximation allows to zoom into the interaction 
structure of solitons and enriches it with an underlying quantum particle-like picture (see 
Example~\ref{ex:parallel}). We expect that this tropical approach will also be useful 
in case of other (in particular soliton) equations.

\renewcommand{\theequation} {\Alph{section}.\arabic{equation}}
\renewcommand{\thesection} {\Alph{section}}

\makeatletter
\newcommand\appendix@section[1]{%
  \refstepcounter{section}%
  \orig@section*{Appendix \@Alph\c@section: #1}%
  \addcontentsline{toc}{section}{Appendix \@Alph\c@section: #1}%
}
\let\orig@section\section
\g@addto@macro\appendix{\let\section\appendix@section}
\makeatother

\begin{appendix}

\section{Some general results}
\label{AppA}
\setcounter{equation}{0}

\subsection{Preparations}
The phases $\theta_k$ appearing in the expression for the function $\tau$ have the form
\bez
    \theta_k = \sum_{r=1}^{n-1} p_k^r \, t^{(r)} + c_k \, , 
\eez
where $p_k,c_k \in \mathbb{R}$ and $t^{(r)}$, $r=1,\ldots,n-1$, are real variables. 
The constants $p_i$ are assumed to be pairwise different. 
In previous sections we wrote $t^{(1)}=x, t^{(2)}=y, t^{(3)}=t$. 
In order to find the values of $t^{(r)}$ for which 
$\theta_{k_1} = \theta_{k_2} = \cdots = \theta_{k_n} =:-t^{(0)}$, 
we have to solve the linear system
\bez
     p_{k_i}^{n-1} \, t^{(n-1)} + p_{k_i}^{n-2} \, t^{(n-2)} + \cdots + p_{k_i} \, t^{(1)} + t^{(0)} 
  = - c_{k_i} \qquad \quad i=1,\ldots,n \, ,
\eez
which is done with the help of Cramer's rule. In particular, for $t^{(n-1)}$ we obtain
the solution
\be
     t^{(n-1)}_{k_1\ldots k_n} = - c_{k_1\ldots k_n} \, ,   \label{t=-c}
\ee
where
\be
     c_{k_1\ldots k_n} = \frac{\kappa(p_{k_1},\ldots,p_{k_n})}{\Delta(p_{k_1},\ldots,p_{k_n})}
          \label{c=kappa/Vandermonde}
\ee
with the Vandermonde determinant
\be
     \Delta(p_{k_1},\ldots,p_{k_n}) 
   = \left| \begin{array}{cccc} 1 & p_{k_1} & \cdots & p_{k_1}^{n-1} \\
                   1 & p_{k_2} & \cdots & p_{k_2}^{n-1}  \\
                   \vdots & \vdots &  & \vdots \\
                   1 & p_{k_n} & \cdots & p_{k_n}^{n-1} \end{array} \right|
   = \prod_{1 \leq i < j \leq n} ( p_{k_j} - p_{k_i} )    \label{Vandermonde_det}
\ee
and
\bez
     \kappa(p_{k_1},\ldots,p_{k_n})
 &=& \left| \begin{array}{ccccc} 1 & p_{k_1} & \cdots & p_{k_1}^{n-2} & c_{k_1} \\
                   1 & p_{k_2} & \cdots & p_{k_2}^{n-2} & c_{k_2} \\
                   \vdots & \vdots &  & \vdots & \vdots \\
                   1 & p_{k_n} & \cdots & p_{k_n}^{n-2} & c_{k_n} \end{array} \right|
 = (-1)^{n-1} \sum_{i=1}^n (-1)^{i-1} c_{k_i} 
   \left| \begin{array}{cccc} 1 & p_{k_1} & \cdots & p_{k_1}^{n-2} \\
                   \vdots & \vdots & & \vdots \\
                   1 & p_{k_{i-1}} & \cdots & p_{k_{i-1}}^{n-2} \\
                   1 & p_{k_{i+1}} & \cdots & p_{k_{i+1}}^{n-2} \\
                   \vdots & \vdots & & \vdots \\
                   1 & p_{k_n} & \cdots & p_{k_n}^{n-2} \end{array} \right| \\
 &=& \sum_{i=1}^n (-1)^{n-i} c_{k_i} \, \Delta(p_{k_1},\ldots,\widehat{p_{k_i}},\ldots,p_{k_n}) \; .
\eez
Here a hat indicates an omission. Now (\ref{c=kappa/Vandermonde}) implies
\be
    c_{k_1\ldots k_n} = \sum_{i=1}^n \frac{c_{k_i}}{(p_{k_i}-p_{k_1}) \cdots (p_{k_i}-p_{k_{i-1}})
                        (p_{k_i}-p_{k_{i+1}}) \cdots (p_{k_i}-p_{k_n})} \; .  \label{c_k1...kn_sum}
\ee

\begin{proposition}
\label{prop:c_formula}
\bez
     c_{k_1\ldots k_{n+1}} 
 = \frac{c_{k_1\ldots \widehat{k_i} \ldots k_{n+1}} - c_{k_1\ldots \widehat{k_j} \ldots k_{n+1}}}{p_{k_j}-p_{k_i}}
   \qquad \quad (i \neq j) \; .
\eez
\end{proposition}
\noindent
\textit{Proof:} Since $c_{k_1\ldots k_n}$ is totally symmetric, it suffices to prove the formula 
for $i=1$ and $j=n+1$. Using (\ref{c_k1...kn_sum}) we have
\bez
 &&    c_{k_2 \ldots k_{n+1}} - c_{k_1 \ldots k_n}
 = \frac{c_{k_{n+1}}}{(p_{k_{n+1}} -p_{k_2}) \cdots (p_{k_{n+1}} -p_{k_n}) }
     - \frac{c_{k_1}}{(p_{k_1} -p_{k_2}) \cdots (p_{k_1} -p_{k_n}) }  \\
 &&    + \sum_{r=2}^n \Big( \frac{1}{(p_{k_r} - p_{k_{n+1}})} - \frac{1}{(p_{k_r} - p_{k_1})} \Big) 
       \frac{c_{k_r}}{(p_{k_r} - p_{k_2}) \cdots (p_{k_r} - p_{k_{r-1}}) (p_{k_r} - p_{k_{r+1}})
       \cdots (p_{k_r} - p_{k_n}) } \, , 
\eez
hence
\bez
 &&    \frac{c_{k_2 \ldots k_{n+1}} - c_{k_1 \ldots k_n}}{p_{k_{n+1}} - p_{k_1}}
  =  \frac{c_{k_1}}{(p_{k_1} -p_{k_2}) \cdots (p_{k_1} - p_{k_{n+1}})}
   + \frac{c_{k_{n+1}}}{(p_{k_{n+1}} - p_{k_1}) \cdots (p_{k_{n+1}} -p_{k_n}) }  \\
 &&  + \sum_{r=2}^n \frac{c_{k_r}}{(p_{k_r} - p_{k_1}) \cdots (p_{k_r} - p_{k_{r-1}}) (p_{k_r} - p_{k_{r+1}})
       \cdots (p_{k_r} - p_{k_{n+1}}) }  
 = c_{k_1\ldots k_{n+1}} \; .
\eez
\hfill $\square$

\begin{proposition}
\label{prop:c_redef}
The substitution $c_k \mapsto c_k + p_k^r \, t^{(r)}$, with a variable $t^{(r)}$, has the following effect,
\bez
     c_{k_1 \ldots k_n} \mapsto c_{k_1 \ldots k_n} + h_{r-n+1}(p_{k_1},\ldots,p_{k_n}) \, t^{(r)} \, ,
\eez
where $h_m$, $m=1,2,\ldots$, are the complete symmetric polynomials, and $h_m = 0$ if $m < 0$, $h_0 =1$. 
\end{proposition}
\noindent
\textit{Proof:} By linearity of the determinant, the substitution effects 
$\kappa(p_{k_1},\ldots,p_{k_n})$ as follows,
\bez
    \kappa(p_{k_1},\ldots,p_{k_n}) \mapsto \kappa(p_{k_1},\ldots,p_{k_n}) 
  + \left| \begin{array}{ccccc} 1 & p_{k_1} & \cdots & p_{k_1}^{n-2} & p_{k_1}^r \\
                   \vdots & \vdots & & \vdots & \vdots \\
                    1 & p_{k_n} & \cdots & p_{k_n}^{n-2} & p_{k_n}^r \end{array} \right| \, t^{(r)} \; .
\eez
The latter determinant equals $h_{r-n+1}(p_{k_1},\ldots,p_{k_n}) \, \Delta(p_{k_1},\ldots,p_{k_n})$ 
(see e.g. \cite{Macd95}). Now the assertion follows from the expression (\ref{c=kappa/Vandermonde}) 
for $c_{k_1\ldots k_n}$.
\hfill $\square$
\vskip.2cm

A redefinition $c_k \mapsto c_k + \sum_{r=n}^N p_k^r \, t^{(r)}$ changes the expression for 
the phases to 
\bez
      \theta_k = \sum_{r=1}^N p_k^r \, t^{(r)} + c_k \, ,
\eez
and, by application of the last proposition to (\ref{t=-c}), the critical value 
for the $n$-phase coincidence $\theta_{k_1} = \theta_{k_2} = \cdots = \theta_{k_n}$ 
now reads\footnote{Despite of our notation, $t^{(n-1)}_{k_1\ldots k_n}$ depends on the 
choice of $N$, of course. Note that it is a function of $t^{(n)},\ldots,t^{(N)}$. }
\be
     t^{(n-1)}_{k_1\ldots k_n} = - \sum_{r=1}^{N+1-n} h_r(p_{k_1},\ldots,p_{k_n}) \, t^{(n+r-1)} 
        - c_{k_1 \ldots k_n}  \, ,    \label{t^(n-1)_k1...kn}
\ee
where $n=2,3,\ldots,N+1$. In particular, $t^{(N)}_{k_1\ldots k_{N+1}} = - c_{k_1 \ldots k_{N+1}}$.
(\ref{t^(n-1)_k1...kn}) also makes sense for $n=1$, where $\theta_k = -t^{(0)}_k$. 
The following proposition presents identities that have the same form irrespective of 
the value of $N$, i.e. their form is not affected by the redefinitions expressed in the last proposition. 
Of course, the ingredients (\ref{t^(n-1)_k1...kn}) do depend on $N$. 

\begin{proposition}
\label{prop:t^(n-1)-diff}
For $n=1,\ldots,N+1$ we have
\be
    t^{(n-1)}_{k_1 \ldots \widehat{k_i} \ldots k_{n+1}}
    - t^{(n-1)}_{k_1 \ldots \widehat{k_j} \ldots k_{n+1}}
  = (p_{k_i} - p_{k_j})(t^{(n)} - t^{(n)}_{k_1 \ldots k_{n+1}})   \qquad \qquad
    i,j=1,\ldots,n+1 \; .    \label{t-identities}
\ee
\end{proposition}
\noindent
\textit{Proof:} Using (\ref{t^(n-1)_k1...kn}) for fixed $N$, and Proposition~\ref{prop:c_formula},
we obtain 
\bez
 &&   t^{(n-1)}_{k_1 \ldots \widehat{k_i} \ldots k_{n+1}}
    - t^{(n-1)}_{k_1 \ldots \widehat{k_j} \ldots k_{n+1}} 
  =  (p_{k_i} - p_{k_j}) \, c_{k_1 \ldots k_{n+1}}                 \\
 && + \sum_{r=1}^{N+1-n} \Big( h_r(p_{k_1},\ldots,\widehat{p_{k_j}},\ldots,p_{k_{n+1}}) 
      - h_r(p_{k_1},\ldots,\widehat{p_{k_i}},\ldots,p_{k_{n+1}}) \Big) \, t^{(n+r-1)} \; .
\eez
Eliminating $c_{k_1 \ldots k_{n+1}}$ with the help of (\ref{t^(n-1)_k1...kn}) 
(with $n$ replaced by $n+1$), we obtain
\bez
 &&  t^{(n-1)}_{k_1 \ldots \widehat{k_i} \ldots k_{n+1}}
    - t^{(n-1)}_{k_1 \ldots \widehat{k_j} \ldots k_{n+1}}
 = \underbrace{[h_1(p_{k_1},\ldots,\widehat{p_{k_j}},\ldots,p_{k_{n+1}}) 
   - h_1(p_{k_1},\ldots,\widehat{p_{k_i}},\ldots,p_{k_{n+1}})]}_{\mbox{$=p_{k_i}-p_{k_j}$}}
     \, t^{(n)}  \\
 && - (p_{k_i} - p_{k_j}) \, t^{(n)}_{k_1\ldots k_{n+1}} 
    + \sum_{r=1}^{N+1-r} \Big( h_r(p_{k_1},\ldots,\widehat{p_{k_j}},\ldots,p_{k_{n+1}})
      - h_r(p_{k_1},\ldots,\widehat{p_{k_i}},\ldots,p_{k_{n+1}}) \\
 && - (p_{k_i} - p_{k_j}) \, h_{r-1}(p_{k_1}, \ldots, p_{k_{n+1}}) \Big) \, t^{(n+r-1)} \; .
\eez
But the last sum vanishes as a consequence of the identities\footnote{A proof of these identities 
is obtained via the substitution $c_k \mapsto p_k^r$ (cf. Proposition~\ref{prop:c_redef}) 
in the formula in Proposition~\ref{prop:c_formula}. }
\bez
    h_r(p_{k_1},\ldots,\widehat{p_{k_j}},\ldots,p_{k_{n+1}})
      - h_r(p_{k_1},\ldots,\widehat{p_{k_i}},\ldots,p_{k_{n+1}}) 
 = (p_{k_i} - p_{k_j}) \, h_{r-1}(p_{k_1}, \ldots, p_{k_{n+1}})  \; .
\eez
\hfill $\square$

\subsection{Main results}
For fixed $M$, we have $M+1$ phases
\bez
      \theta_i = \sum_{r=1}^M p_i^r \, t^{(r)} + c_i  \qquad \quad i=1,\ldots, M+1 \; .
\eez 
In the following we regard the variables $t^{(r)}$, $r=1,\ldots,M$, as Cartesian coordinates 
on $\mathbb{R}^M$. 
The region in $\mathbb{R}^M$ where $\theta_i$ dominates is given by
\bez
     \mathcal{U}_i = \{ \mathbf{t} \in \mathbb{R}^M \, | \, \max\{\theta_1(\mathbf{t}), \ldots,
       \theta_{M+1}(\mathbf{t})\} = \theta_i(\mathbf{t}) \} \; .
\eez
Associated with any set $\{k_1,\ldots, k_{n+1} \} \subset \{1,\ldots,M+1\}$, $n>0$, there is a  
\emph{critical plane}, 
\bez
  \mathcal{P}_{k_1\ldots k_{n+1}} = \{ \mathbf{t} \in \mathbb{R}^M \, | \, \theta_{k_1}(\mathbf{t}) =
       \cdots = \theta_{k_{n+1}}(\mathbf{t}) \} \, ,
\eez
which is an affine plane of dimension $M-n$. 
Since the $p_i$ are pairwise different, no pair of hyperplanes $\mathcal{P}_{ij}$, $1 \leq i <j \leq M+1$, 
can be parallel. In particular, they cannot coincide and thus $\mathcal{U}_i \neq \emptyset$, 
$i=1,\ldots,M+1$. We also note that $\bigcup_{1 \leq i \leq M+1} \mathcal{U}_i = \mathbb{R}^M$. 
Some obvious relations are
\bez
    \mathcal{P}_{k_1\ldots k_{n+1}} \subset \mathcal{P}_{k_1\ldots k_{m+1}}  \qquad \mbox{for} \qquad
    m < n \, , 
\eez
and
\bez
    \mathcal{P}_{k_1\ldots k_{n+1}} = \mathcal{P}_{k_1\ldots \widehat{k_r} \ldots k_{n+1}}
      \cap \mathcal{P}_{k_1\ldots \widehat{k_s} \ldots k_{n+1}} \qquad \mbox{for} \qquad
    r \neq s \, , 
\eez
where a hat again indicates an omission, hence also
\bez
    \mathcal{P}_{k_1\ldots k_{n+1}} = \bigcap_{r=1}^{n+1} \mathcal{P}_{k_1\ldots \widehat{k_r} \ldots k_{n+1}} \; .
\eez
We can use $t^{(n+1)}, \ldots, t^{(M)}$ as coordinates on $\mathcal{P}_{k_1\ldots k_{n+1}}$, since 
on this subset of $\mathbb{R}^M$ the remaining coordinates are fixed as solutions of the system 
$\theta_{k_1} = \cdots = \theta_{k_{n+1}} =: -t^{(0)}$, i.e. 
\bez
      \sum_{r=0}^n p_{k_j}^r \, t^{(r)} = - c_{k_j} - \sum_{r=n+1}^M p_{k_j}^r \, t^{(r)} 
      \qquad \quad j=1,\ldots,n+1 \; .
\eez
We solve this system for $t^{(r)}$, $r=1,\ldots,n$, and denote the solutions as 
$t^{(r)}_{k_1\ldots k_{n+1}}(t^{(n+1)}, \ldots, t^{(M)})$, $r=1,\ldots,n$. They depend linearly 
on the parameters $c_i$. For the highest we already found
\bez
     t^{(n)}_{k_1\ldots k_{n+1}}(t^{(n+1)}, \ldots, t^{(M)}) 
   = - \sum_{r=1}^{M-n} h_r(p_{k_1},\ldots,p_{k_{n+1}}) \, t^{(n+r)} - c_{k_1\ldots k_{n+1}} \, ,
\eez
which is totally symmetric in the lower indices.
These are called \emph{critical values} of $t^{(n)}$. 
The values of $t^{(r)}_{k_1\ldots k_{n+1}}(t^{(n+1)}, \ldots, t^{(M)})$, $r=1,\ldots,n-1$, are then 
determined iteratively as functions of $t^{(n+1)}, \ldots, t^{(M)}$. 
Hence the points of $\mathcal{P}_{k_1\ldots k_{n+1}}$ are given by
\bez
  \mathbf{t}_{k_1\ldots k_{n+1}}(t^{(n+1)}, \ldots, t^{(M)}) 
 := \mathbf{t}(t^{(1)}_{k_1\ldots k_{n+1}}, \ldots, t^{(n)}_{k_1\ldots k_{n+1}}, t^{(n+1)}, \ldots, t^{(M)})
  \, ,
\eez
where we suppressed the arguments of $t^{(r)}_{k_1\ldots k_{n+1}}$.

\begin{proposition}
\label{prop:t^(n-1)-ordering}
Let $\{k_1,\ldots,k_{n+1}\} \subset \{1,\ldots, M+1\}$ and $p_{k_i} < p_{k_j}$. Then we have
\bez
   t^{(n-1)}_{k_1\ldots \widehat{k_j} \ldots k_{n+1}} \lessgtr
   t^{(n-1)}_{k_1 \ldots \widehat{k_i} \ldots k_{n+1}} 
                        \qquad \mbox{for} \qquad  
    t^{(n)} \lessgtr  t^{(n)}_{k_1\ldots k_{n+1}}  \; .
\eez
\end{proposition}
\noindent
\textit{Proof:} This is an immediate consequence of (\ref{t-identities}).
\hfill $\square$

\begin{proposition}
Let $\{k_1,\ldots,k_{n+1}\} \subset \{1,\ldots,M+1\}$. Then
\be
    \theta_{k_1} - \theta_{k_{n+1}} 
 = -\sum_{r=1}^n \Big( \prod_{j=1}^r (p_{k_{n+1}} - p_{k_j}) \Big) \, 
     (t^{(r)} - t^{(r)}_{k_1\ldots k_{r+1}})    \; . 
\ee
\end{proposition}
\noindent
\textit{Proof:} (\ref{t-identities}) with $n=1$ reads
\bez
    \theta_{k_1} - \theta_{k_2} = -(p_{k_2} - p_{k_1}) (x - x_{k_1k_2}) \, ,
\eez 
which is the above formula for $n=1$. Assuming that the assertion holds for $n$, we can apply 
it with $\{k_1,\ldots,k_n,k_{n+2}\}$ to obtain
\bez
     \theta_{k_1} - \theta_{k_{n+2}} 
 &=& -\sum_{r=1}^{n-1} \Big( \prod_{j=1}^r (p_{k_{n+2}} - p_{k_j}) \Big) \, 
     (t^{(r)} - t^{(r)}_{k_1\ldots k_{r+1}})  \\
 &&  - \Big( \prod_{j=1}^n (p_{k_{n+2}} - p_{k_j}) \Big) \, 
     (t^{(n)} - t^{(n)}_{k_1\ldots k_{n+1}} 
        + t^{(n)}_{k_1\ldots k_{n+1}} - t^{(n)}_{k_1\ldots k_n k_{n+2}}) \\
 &=& -\sum_{r=1}^n \Big( \prod_{j=1}^r (p_{k_{n+2}} - p_{k_j}) \Big) \, 
     (t^{(r)} - t^{(r)}_{k_1\ldots k_{r+1}})   \\
 &&  - \Big( \prod_{j=1}^n (p_{k_{n+2}} - p_{k_j}) \Big) \, 
     (t^{(n)}_{k_1\ldots k_{n+1}} - t^{(n)}_{k_1\ldots k_n k_{n+2}}) \; .
\eez
Using (\ref{t-identities}) in the last factor, this becomes the asserted formula for $n+1$,
which thus completes the induction step.
\hfill $\square$
\vskip.2cm

\begin{corollary}
\label{cor:theta-diff}
Let $\{k_1,\ldots,k_{n+1}\} \subset \{1,\ldots,M+1\}$. On $\mathcal{P}_{k_1\ldots k_n}$, we have
\be
    \theta_{k_1} - \theta_{k_{n+1}} 
 = -(p_{k_{n+1}} - p_{k_1}) (p_{k_{n+1}} - p_{k_2}) \cdots (p_{k_{n+1}} - p_{k_n}) 
       (t^{(n)} - t^{(n)}_{k_1\ldots k_{n+1}}) 
   \; .    \label{theta-diff_general}
\ee
\hfill $\square$
\end{corollary}

A point $\mathbf{t}_0 \in \mathcal{P}_{k_1\ldots k_{n+1}}$ will be called \emph{non-visible} 
if there is an $m \not\in \{k_1,\ldots,k_{n+1}\}$ such that 
$\theta_m(\mathbf{t}_0) > \theta_{k_1}(\mathbf{t}_0)$.\footnote{In 
this case $m$ can be chosen such that $\theta_m$ is a dominating phase at $\mathbf{t}_0$.} 
Otherwise it will be called \emph{visible}. 
In previous sections we considered the projection into the $xy$-plane, 
$P_{k_1\ldots k_{n+1}}(t^{(n+1)}, \ldots, t^{(M)})$,  
of a point $\mathbf{t}_{k_1\ldots k_{n+1}}(t^{(n+1)}, \ldots, t^{(M)}) \in \mathcal{P}_{k_1\ldots k_{n+1}}$. 
Our previous notion of visibility of $P_{k_1\ldots k_{n+1}}(t^{(n+1)}, \ldots, t^{(M)})$, which means ordinary 
visibility in a plot of $\max\{\theta_1,\ldots,\theta_{M+1}\}$, is in fact equivalent to 
visibility of the latter point in $\mathbb{R}^M$. 
In the following, $\lceil n/2 \rceil$ denotes the smallest integer greater than or equal to $n/2$, 
and $\lfloor n/2 \rfloor$ the largest integer smaller than or equal to $n/2$.

\begin{proposition} 
\label{prop:non-visible_P}
For $n =1,2,\ldots$, let $\{k_1,\ldots,k_{n+1}\} \subset \{1,\ldots, M+1\}$,  
$p_{k_1} < p_{k_2} < \cdots < p_{k_{n+1}}$, and $t^{(n+1)}_0, \ldots, t^{(M)}_0 \in \mathbb{R}$. 
The following half-lines are non-visible: \\
(1) $\{ \mathbf{t}_{k_1 \ldots \,\widehat{k_{n-2r}} \, \ldots k_{n+1}}(t^{(n)},t^{(n+1)}_0, \ldots, t^{(M)}_0) 
 \, | \,  t^{(n)} < t^{(n)}_{k_1 \ldots k_{n+1}} \} \subset \mathcal{P}_{k_1 \ldots \,\widehat{k_{n-2r}} 
 \, \ldots k_{n+1}}$, 
$r=0,\ldots,\lceil n/2 \rceil -1$, \\
(2) $\{ \mathbf{t}_{k_1 \ldots \,\widehat{k_{n+1-2r}}\, \ldots k_{n+1}}(t^{(n)},t^{(n+1)}_0, \ldots, t^{(M)}_0) 
 \, | \, t^{(n)} > t^{(n)}_{k_1 \ldots k_{n+1}} \} \subset \mathcal{P}_{k_1 \ldots \,\widehat{k_{n+1-2r}}
 \, \ldots k_{n+1}}$, 
$r=0,\ldots,\lfloor n/2 \rfloor$.\\
Here $t^{(n)}_{k_1 \ldots k_{n+1}}$ stands for $t^{(n)}_{k_1 \ldots k_{n+1}}(t^{(n+1)}_0, \ldots, t^{(M)}_0)$.
\end{proposition}
\noindent
\textit{Proof:} 
On $\mathcal{P}_{k_1\ldots \widehat{k_m} \ldots k_n}$, (\ref{theta-diff_general}) can 
be written in the form
\bez
   \theta_{k_1} - \theta_{k_m} 
 = - \Big( \prod_{j=1,\ldots,n+1 \atop j \neq m} (p_{k_m} - p_{k_j}) \Big) \, 
    (t^{(n)} - t^{(n)}_{k_1\ldots k_{n+1}})  \qquad \quad
    m = 2,\ldots,n+1 \; .
\eez
We actually consider this equation on $\mathcal{P}_{k_1\ldots\widehat{k_m} \ldots k_n} \cap \mathcal{E}$,
where $\mathcal{E}$ is the plane in $\mathbb{R}^M$ 
determined by fixing the values of $t^{(n+1)},\ldots,t^{(M)}$ to $t^{(n+1)}_0,\ldots,t^{(M)}_0$.
As a consequence of our assumption $p_{k_1} < \cdots < p_{k_{n+1}}$, for $m=n+1$ the above expression 
is negative if $t^{(n)} > t^{(n)}_{k_1 \ldots k_{n+1}}$, hence 
$\mathcal{P}_{k_1 \ldots k_n} \cap \mathcal{E}$ is 
then non-visible. For $m=n$ the expression is negative if $t^{(n)} < t^{(n)}_{k_1 \ldots k_{n+1}}$, 
hence $\mathcal{P}_{k_1 \ldots k_{n-1} k_{n+1}} \cap \mathcal{E}$ is then non-visible. 
For $m=n-1$, the expression 
is negative if $t^{(n)} > t^{(n)}_{k_1 \ldots k_{n+1}}$, hence 
$\mathcal{P}_{k_1 \ldots k_{n-2} k_n k_{n+1}} \cap \mathcal{E}$ is then non-visible. 
This argument can be continued as long as $m>1$. 
On the remaining critical plane $\mathcal{P}_{k_2 \ldots k_{n+1}}$, which appears in case (1) 
for odd $n$ and in case (2) for even $n$, we can write the above equation as
\bez
   \theta_{k_{n+1}} - \theta_{k_1} 
 = (-1)^{n+1} \Big( \prod_{j=2,\ldots,n+1} (p_{k_j} - p_{k_1}) \Big) \,  
    (t^{(n)} - t^{(n)}_{k_1\ldots k_{n+1}}) \; .
\eez
This is negative if either $n$ is odd and $t^{(n)} < t^{(n)}_{k_1 \ldots k_{n+1}}$, 
or if $n$ is even and $t^{(n)} > t^{(n)}_{k_1 \ldots k_{n+1}}$. As a 
consequence, $\mathcal{P}_{k_2 \ldots k_{n+1}} \cap \mathcal{E}$ is then non-visible. 
\hfill $\square$
\vskip.2cm

We note that the set of critical planes in part 1 and part 2 of Proposition~\ref{prop:non-visible_P}
are complementary. In the following we call a critical point $\mathbf{t}_0 \in \mathcal{P}_{k_1 \ldots k_{n+1}}$ 
\emph{generic} if it is \emph{not} also a higher order critical point, i.e. if 
$\mathbf{t}_0 \not\in \mathcal{P}_{k_1 \ldots k_{n+2}}$ with any $k_{n+2} \not\in \{k_1, \ldots, k_{n+1} \}$.

\begin{proposition} 
\label{prop:visible_P}
Let $\{k_1,\ldots,k_{n+1}\} \subset \{1,\ldots, M+1\}$ and $p_{k_1} < p_{k_2} < \cdots < p_{k_{n+1}}$.
Let $\alpha, \beta$ be such that in the open interval $(\alpha,t^{(n)}_{k_1 \ldots k_{n+1}})$, 
respectively $(t^{(n)}_{k_1 \ldots k_{n+1}},\beta)$, there is no critical value 
of $t^{(n)}$ corresponding to a visible critical point. 
If $\mathbf{t}_{k_1 \ldots k_{n+1}}(t^{(n+1)}_0, \ldots, t^{(M)}_0)$ 
is generic and visible, then the following line segments are visible: \\
(1) $\{ \mathbf{t}_{k_1 \ldots \, \widehat{k_{n+1-2r}}\, \ldots k_{n+1}}(t^{(n)},t^{(n+1)}_0, \ldots, t^{(M)}_0) 
    \, | \, 
    \alpha \leq t^{(n)} \leq t^{(n)}_{k_1 \ldots k_{n+1}} \} 
    \subset \mathcal{P}_{k_1 \ldots \, \widehat{k_{n+1-2r}}\, \ldots k_{n+1}}$, where
    $r=0,\ldots,\lfloor n/2 \rfloor$, \\
(2) $\{ \mathbf{t}_{k_1 \ldots \,\widehat{k_{n-2r}} \, \ldots k_{n+1}}(t^{(n)},t^{(n+1)}_0, \ldots, t^{(M)}_0) 
     \, | \, 
    t^{(n)}_{k_1 \ldots k_{n+1}} \leq t^{(n)} \leq \beta \} 
    \subset \mathcal{P}_{k_1 \ldots \,\widehat{k_{n-2r}} \, \ldots k_{n+1}}$, where
    $r=0,\ldots,\lceil n/2 \rceil -1$. \\ 
Here we set $t^{(n)}_{k_1 \ldots k_{n+1}} = t^{(n)}_{k_1 \ldots k_{n+1}}(t^{(n+1)}_0, \ldots, t^{(M)}_0)$.
\end{proposition}
\noindent
\textit{Proof:} In the following we use $\mathcal{E}$ as defined in the proof of 
Proposition~\ref{prop:non-visible_P}.
Since we assume $\mathbf{t}_0 := \mathbf{t}_{k_1 \ldots k_{n+1}}(t^{(n+1)}_0, \ldots, t^{(M)}_0)
\in \mathcal{P}_{k_1 \ldots k_{n+1}}$ to be visible, at 
$\mathbf{t}_0$ the phases $\theta_{k_1},\ldots,\theta_{k_{n+1}}$ coincide and dominate. 
Since $\mathbf{t}_0$ is assumed to be generic, there is a neighborhood of
$\mathbf{t}_0$ in $\mathcal{E}$ which is covered by the polyhedral cones $\mathcal{U}_{k_j} 
\cap \mathcal{E}$, $j=1,\ldots,n+1$. 
Since each line $\mathcal{P}_{k_1 \ldots \widehat{k_{n-2r}} \ldots k_{n+1}} \cap \mathcal{E}$, 
$r \in \{1,\ldots,n+1\}$, contains $\mathbf{t}_0$, it follows that its visible part
$\mathcal{P}_{k_1 \ldots \widehat{k_{n-2r}} \ldots k_{n+1}} \cap \mathcal{E} \cap \mathcal{U}_{k_1}
\cap \cdots \cap \mathcal{U}_{k_{r-1}} \cap \mathcal{U}_{k_{r+1}} \cap \cdots \cap \mathcal{U}_{k_{n+1}}$
extends in the direction complementary to that in Proposition~\ref{prop:non-visible_P}, either indefinitely 
or up to a point of $\mathcal{E}$ where it meets $\mathcal{U}_m$ 
with some $m \not\in \{k_1,\ldots,k_{n+1}\}$. We only need to consider the latter case further. 
Then $\{\mathbf{t}_1\} := \mathcal{U}_{k_1} \cap \cdots \cap \mathcal{U}_{k_{r-1}} 
\cap \mathcal{U}_{k_{r+1}} \cap \cdots \cap \mathcal{U}_{k_{n+1}} \cap \mathcal{U}_m \cap \mathcal{E}$ 
is a visible point in $\mathcal{P}_{k_1 \ldots \widehat{k_{n-2r}} \ldots k_{n+1} m} \cap \mathcal{E}$. 
Since the line between $\mathbf{t}_0$ and $\mathbf{t}_1$ is visible, it cannot be part of the 
non-visible half-line determined by Proposition~\ref{prop:non-visible_P} applied to $\mathbf{t}_1$. 
\hfill $\square$
\vskip.2cm

The following proposition shows that the existence of a \emph{visible} critical point requires 
the existence of a \emph{visible} critical point one level higher.

\begin{proposition}
\label{prop:vis_neigb-levels}
Let $n \leq M$ and $t^{(r)}_0 \in \mathbb{R}$, $r=n,\ldots,M$.
If all points $\mathbf{t}_{k_1 \ldots k_{n+1}}(t^{(n+1)}_0,\ldots,t^{(M)}_0)$, where  
$\{ k_1, \ldots, k_{n+1} \} \supset \{ l_1, \ldots, l_n \}$, are non-visible, then the line 
$\{ \mathbf{t}_{l_1 \ldots l_n}(t^{(n)},t^{(n+1)}_0,\ldots,t^{(M)}_0) \, | \, t^{(n)} \in \mathbb{R} \}$, 
is non-visible. 
\end{proposition}
\noindent
\textit{Proof:} Let $\mathcal{E}$ again denote the set $\{ \mathbf{t} \in \mathbb{R}^M \, | \, 
t^{(n+1)} = t^{(n+1)}_0, \ldots, t^{(M)} = t^{(M)}_0 \}$ and let 
$\mathbf{t}_0 \in \mathcal{P}_{l_1 \ldots l_n} \cap \mathcal{E}$ be visible. 
Since $n \leq M$, a critical point exists one level higher, given by 
$\{\mathbf{t}_1\} = \mathcal{P}_{k_1 \ldots k_{n+1}} \cap \mathcal{E}$ with some $k_1,\ldots,k_{n+1}$ 
such that $\{l_1, \ldots, l_n\} \subset \{ k_1, \ldots, k_{n+1} \}$. If this point is visible, 
the proposition holds. If $\mathbf{t}_1$ is non-visible, then 
$\mathbf{t}_1 \in \mathcal{U}_m \cap \mathcal{E}$ with some $m \not\in \{ k_1, \ldots, k_{n+1} \}$. 
Clearly, it cannot coincide with $\mathbf{t}_0$. By continuity, on the line segment between 
$\mathbf{t}_0$ and $\mathbf{t}_1$ a \emph{visible} critical point then exists, which is 
$\mathbf{t}_{l_1 \ldots l_n m}(t^{(n+1)}_0,\ldots,t^{(M)}_0)$. This proves that  
if $\mathbf{t}_{l_1 \ldots l_n}(t^{(n)}_0,t^{(n+1)}_0,\ldots,t^{(M)}_0)$ is visible, 
then there is a visible critical point 
$\mathbf{t}_{k_1 \ldots k_{n+1}}(t^{(n+1)}_0,\ldots,t^{(M)}_0)$ with
$\{ l_1, \ldots, l_n \} \subset \{ k_1, \ldots, k_{n+1} \}$.
Our assertion is the negation of this statement. 
\hfill $\square$
\vskip.2cm

Without restriction of generality we can choose 
\bez
       p_1 < p_2 < \cdots < p_{M+1} \; .
\eez 
There is only a single critical value $t^{(M)}_{1,\ldots,M+1}$. As a meeting point of \emph{all} 
phases, $\mathbf{t}(t^{(M)}_{1,\ldots,M+1})$ is visible. 
Proposition~\ref{prop:t^(n-1)-ordering} shows that, for $1 \leq i < j \leq M+1$, 
\bez
      t^{(M-1)}_{1,\ldots, \hat{j}, \ldots,M+1} \lessgtr
      t^{(M-1)}_{1,\ldots, \hat{i}, \ldots,M+1}
                        \qquad \mbox{for} \qquad  
    t^{(M)} \lessgtr t^{(M)}_{k_1\ldots k_{M+1}}  \; .
\eez
According to Propositions~\ref{prop:non-visible_P} and \ref{prop:visible_P}, only the 
half-lines 
\bez
 &&  \{ \mathbf{t}_{1, \ldots, \widehat{M-2r}, \ldots, M+1}(t^{(M)})  \, | \, 
      t^{(M)} > t^{(M)}_{k_1\ldots k_{M+1}} \} \subset \mathcal{P}_{1, \ldots, \widehat{M-2r}, \ldots, M+1}
        \qquad \quad  r=0,\ldots,\lceil (M+1)/2 \rceil -1 \, , \\
 &&  \{ \mathbf{t}_{1, \ldots, \widehat{M-2r+1}, \ldots, M+1}(t^{(M)})  \, | \, 
      t^{(M)} < t^{(M)}_{k_1\ldots k_{M+1}} \} \subset \mathcal{P}_{1, \ldots, \widehat{M-2r+1}, \ldots, M+1}
        \quad  r=0,\ldots,\lfloor (M+1)/2 \rfloor
\eez
are visible. We note that the two sets of lines are complementary, and each of them  
is visible in exactly one of the two half-spaces (corresponding to $t^{(M)} > t^{(M)}_{k_1\ldots k_{M+1}}$, 
respectively $t^{(M)} < t^{(M)}_{k_1\ldots k_{M+1}}$). Proceeding in this way, we find that 
each critical 2-plane $\mathcal{P}_{k_1 \ldots k_{M-1}}$ is visible in some region of $\mathbb{R}^M$, 
and so forth. 
Since $\mathcal{P}_{k_1 \ldots k_{M+1}}$ is contained in all critical planes, they all 
contain visible points. 

The next result is particularly helpful.

\begin{proposition}
\label{prop:exhaust}
All \emph{non-visible} critical points are obtained by application of Proposition~\ref{prop:non-visible_P} 
only to \emph{visible} critical points, and by  Proposition~\ref{prop:vis_neigb-levels}.\footnote{We  
conjecture that the reference to Proposition~\ref{prop:vis_neigb-levels} can be dropped, provided 
the application of Proposition~\ref{prop:non-visible_P} is extended to \emph{all} critical points. } 
\end{proposition}
\noindent
\textit{Proof:} Let $\mathbf{t}_0 = \mathbf{t}_{l_1 \ldots l_n}(t_0^{(n)}, \ldots, t_0^{(M)})$
be non-visible. If there is no visible critical point of the form
$\mathbf{t}_{k_1 \ldots k_{n+1}}(t_0^{(n+1)}, \ldots, t_0^{(M)})$, with
$\{l_1, \ldots, l_n\} \subset \{k_1, \ldots, k_{n+1}\}$,  then the non-visibility of $\mathbf{t}_0$ 
is a consequence of Proposition~\ref{prop:vis_neigb-levels}. 
If there is a visible critical point of the above form, then there is also a visible critical point 
$\mathbf{t}_1$ such that no other visible critical point exists on 
the line segment joining $\mathbf{t}_0$ and $\mathbf{t}_1$. Since $\mathbf{t}_0$ cannot lie on the visible side of
$\mathbf{t}_1$ as determined by Proposition~\ref{prop:visible_P}, it lies on the non-visible side of 
$\mathbf{t}_1$ as determined by Proposition~\ref{prop:non-visible_P}. Hence the non-visibility of 
$\mathbf{t}_0$ is a consequence of Proposition~~\ref{prop:non-visible_P}.
\hfill $\square$
\vskip.2cm

The chains of rooted binary trees describing line soliton solutions can be constructed from the 
knowledge of the ``visible'' critical values $t^{(n)}_{k_1,\ldots,k_{n+1}}$, 
$n=1,\ldots,M$, and their order (determined top down via Proposition~\ref{prop:t^(n-1)-ordering}). 
For $t^{(n+1)}$ from the interval between two of its critical 
values (formally including $\pm \infty$), the corresponding visible critical values of $t^{(n)}$ 
are obtained from all critical values simply by deleting all those that are non-visible by an application 
of the rules of Propositions~\ref{prop:t^(n-1)-ordering}, \ref{prop:non-visible_P} and 
\ref{prop:vis_neigb-levels} (where the latter may not be necessary).
\vskip.2cm

Of course, one can establish further useful results about the visibility or non-visibility of critical points. 
The following is an example. 

\begin{proposition} 
\label{prop:2step_non-vis}
(1) Let $0 \leq r < s \leq \lceil (n+1)/2 \rceil -1$ and $t^{(n+1)}_0 < t^{(n+1)}_{k_1 \ldots k_{n+2}}(t^{(n+2)}_0,\ldots, t^{(M)}_0)$. Then the whole line 
$\{ \mathbf{t}_{k_1 \ldots \widehat{k_{n+1-2s}} \ldots \widehat{k_{n+1-2r}}\ldots k_{n+2}}
(t^{(n)},t^{(n+1)}_0, \ldots, t^{(M)}_0) \, | \, t^{(n)} \in \mathbb{R} \}$ is non-visible. \\
(2) Let $0 \leq r < s \leq \lfloor (n+1)/2 \rfloor$ and 
$t^{(n+1)}_0 > t^{(n+1)}_{k_1 \ldots k_{n+2}}(t^{(n+2)}_0,\ldots, t^{(M)}_0)$. Then the whole line 
$\{ \mathbf{t}_{k_1 \ldots \widehat{k_{n+2-2s}} \ldots \widehat{k_{n+2-2r}}\ldots k_{n+2}}
(t^{(n)},t^{(n+1)}_0, \ldots, t^{(M)}_0) \, | \, t^{(n)} \in \mathbb{R} \}$ is non-visible.
\end{proposition}
\noindent
\textit{Proof:}
We only prove (1). If $t^{(n+1)}_0 < t^{(n+1)}_{k_1 \ldots k_{n+2}}(t^{(n+2)}_0,\ldots, t^{(M)}_0)$, 
then Proposition~\ref{prop:t^(n-1)-ordering} implies
\bez
    t_{k_1 \ldots \widehat{k_{n+1-2r}} \ldots k_{n+2}}(t^{(n+1)}_0, \ldots, t^{(M)}_0) 
  < t_{k_1 \ldots \widehat{k_{n+1-2s}} \ldots k_{n+2}}(t^{(n+1)}_0, \ldots, t^{(M)}_0) \; .
\eez
$\mathbf{t}_{k_1 \ldots \widehat{k_{n+1-2s}} \ldots \widehat{k_{n+1-2r}}\ldots k_{n+2}}
(t^{(n)},t^{(n+1)}_0, \ldots, t^{(M)}_0)$, where 
$t^{(n)} < t_{k_1 \ldots \widehat{k_{n+1-2s}} \ldots k_{n+2}}(t^{(n+1)}_0, \ldots, t^{(M)}_0)$, 
is non-visible by application of Proposition~\ref{prop:non-visible_P}. But, again as a consequence 
of Proposition~\ref{prop:non-visible_P}, it is also non-visible for 
$t^{(n)} > t_{k_1 \ldots \widehat{k_{n+1-2r}} \ldots k_{n+2}}(t^{(n+1)}_0, \ldots, t^{(M)}_0)$.
\hfill $\square$

\begin{example}
Let $M=5$ and $t^{(5)} < t^{(5)}_{123456}$. Applying Proposition~\ref{prop:2step_non-vis} (part 1)
with $n=4$, 
we find that the events associated with the critical times $t_{1246}, t_{2346}, t_{2456}$ are 
non-visible. Since these are all possible critical times at which $y_{246}$ can coincide with 
other critical $y$-values, and since there is no corresponding node in the initial rooted 
binary tree, it cannot appear during any line soliton evolution (with $t^{(5)} < t^{(5)}_{123456}$). 
The non-visibility of $y_{246}$ also follows by an application of Proposition~\ref{prop:vis_neigb-levels}.
As a consequence, the left tree in Fig.~\ref{fig:y246&135} does not appear in Fig.~\ref{fig:T4_t4deform1}. \\
\noindent
If $t^{(5)} > t^{(5)}_{123456}$, Proposition~\ref{prop:2step_non-vis} (part 2) with $n=4$ 
shows that $t_{1235}, t_{1345}, t_{1356}$ are non-visible. This in turn implies that 
$y_{135}$ can never show up, which excludes the right tree in Fig.~\ref{fig:y246&135}, which 
indeed does not appear in Fig.~\ref{fig:T4_t4deform2}.
\begin{figure}[H] 
\begin{center} 
\resizebox{!}{1.5cm}{
\includegraphics{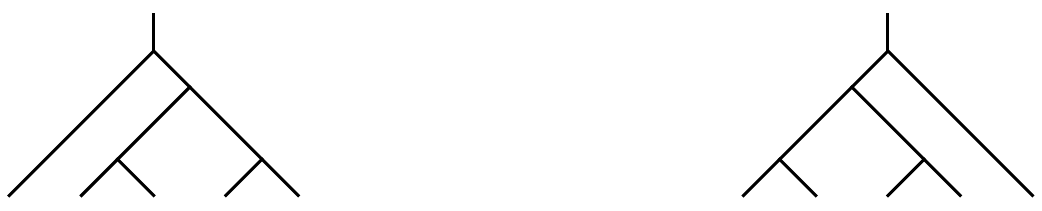}
}
\parbox{15cm}{
\caption{Rooted binary trees possessing a node with $y_{246}$, respectively $y_{135}$. \label{fig:y246&135} }
}
\end{center} 
\end{figure}
\end{example}

\section{A finer classification in terms of trees with levels}
\label{AppB}
\begin{figure}[H] 
\begin{center} 
\resizebox{16.cm}{!}{
\includegraphics{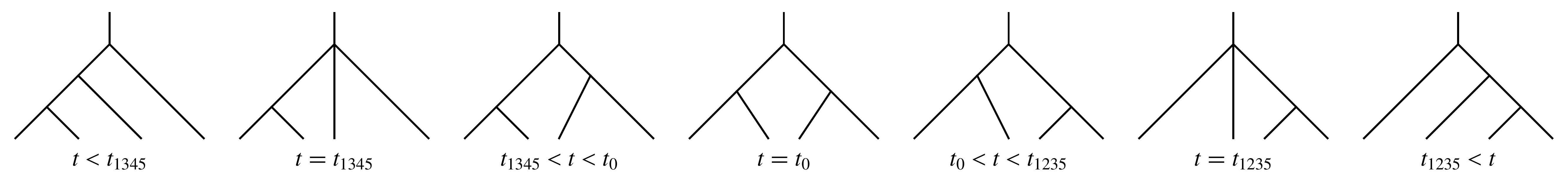}
}
\parbox{15cm}{
\caption{A finer description of the evolution as compared with that given 
by the second chain in Fig.~\ref{fig:T3_chains}. The third and the fifth tree  
are the two trees with levels associated with the tree in the middle. \label{fig:treeswithlevels} }
}
\end{center} 
\end{figure}

\setcounter{equation}{0}
\setcounter{theorem}{0}
A finer description of line soliton evolutions can be achieved by using the refinement of 
(rooted) binary trees to ``trees with levels'' \cite{Loday+Ronco98}. Fig.~\ref{fig:treeswithlevels} 
shows such a refinement of the second chain in Fig.~\ref{fig:T3_chains}, including also the degenerate 
trees at $t=t_{1345}$ and $t=t_{1235}$ which are not binary. Here we took into account that 
a time $t_0$ exists at which the two subtrees appearing between $t_{1345}$ and $t_{1235}$ have the same 
height, i.e. the same $y$-value. The third and the fifth tree  
are the two trees with levels associated with the forth tree in Fig.~\ref{fig:treeswithlevels}. 
Setting $t^{(5)}$ and higher variables to zero, the condition $y_{ijk} = y_{lmn}$ determines 
the ``critical'' time
\bez
       t_{ijk;lmn} = - \frac{1}{p_i + p_j + p_k -p_l -p_m -p_n} \Big( c_{ijk} - c_{lmn} 
               + [h_2(p_i,p_j,p_k) - h_2(p_l,p_m,p_n)] \, t^{(4)} \Big) \, ,
\eez
provided that $p_i + p_j + p_k \neq  p_l + p_m + p_n$. 

\begin{example}
Let $M=4$. 
In the case considered in Fig.~\ref{fig:treeswithlevels}, we have $t_0 = t_{123;345}$ and
\bez
    t_0 - t_{1345} = \frac{(p_4-p_2)(p_5-p_2)}{p_4-p_1 + p_5-p_2} \, (t^{(4)} - t^{(4)}_{12345}) \, , \quad
    t_{1235} - t_0 = \frac{(p_4-p_1)(p_4-p_2)}{p_4-p_1 + p_5-p_2} \, (t^{(4)} - t^{(4)}_{12345}) \; .
\eez
Assuming $p_1 < \cdots < p_5$, these expressions are both positive since the chain is only 
realized if $t^{(4)} > t^{(4)}_{12345}$ (right chain in Fig.~\ref{fig:T3}). Furthermore,
\bez
    x_{345}(t_0) - x_{123}(t_0) = \frac{(p_4-p_1)(p_4-p_2)(p_5-p_1)(p_5-p_2)}{p_4-p_1 + p_5-p_2} \, 
      (t^{(4)} - t^{(4)}_{12345}) > 0 \; .
\eez
\hfill $\square$
\end{example}

\begin{figure}[H] 
\begin{center} 
\resizebox{8.cm}{!}{
\includegraphics{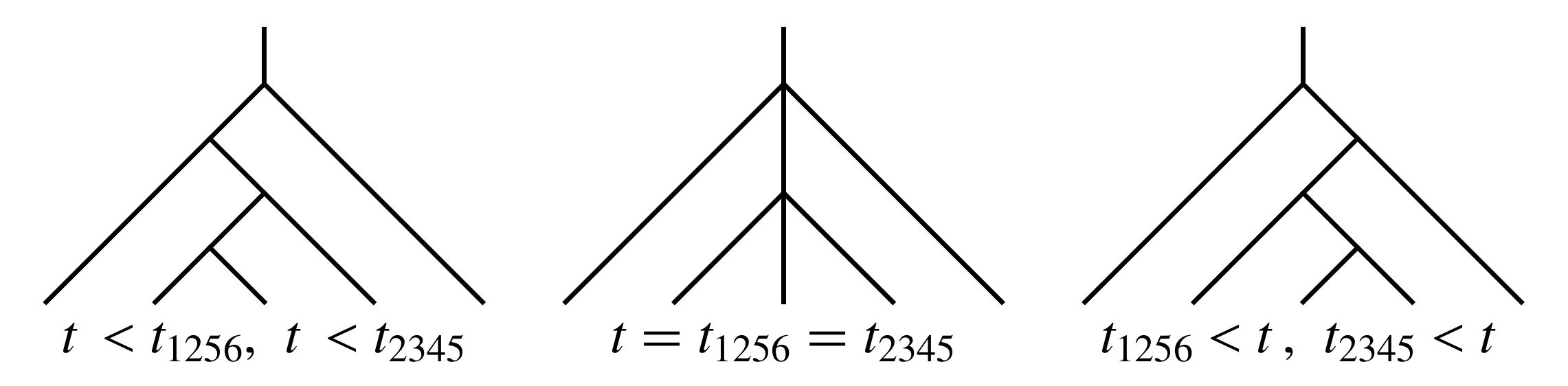}
}
\parbox{15cm}{
\caption{A transition through a coincidence of two critical times, $t_{1256}$ and $t_{2345}$, corresponding 
to the critical value $t^{(4)}_{1256;2345}$. This is a simultaneous rotation with respect to two 
different nodes.  \label{fig:simrot} }
}
\end{center} 
\end{figure}

After introduction of $t^{(5)}$, the analogous condition $t_{ijkl} = t_{mnrs}$ determines 
the following critical value of $t^{(4)}$,
\bez
   t^{(4)}_{ijkl;mnrs} &=& - \frac{1}{p_i + p_j + p_k + p_l - p_m -p_n - p_r - p_s} \Big( c_{ijkl} - c_{mnrs} \\
         &&      + [h_2(p_i,p_j,p_k,p_l) - h_2(p_m,p_n,p_r,p_s)] \, t^{(5)} \Big) \, ,
\eez
provided that $p_i + p_j + p_k + p_l \neq  p_m + p_n + p_r + p_s$.\footnote{It should now be obvious how this extends to a formula for corresponding critical values of $t^{(n)}$, $n>2$.}
See Fig.~\ref{fig:simrot} for an example. 
A further useful formula is
\bez
    t_{ijkl}-t_{mnrs} =
       (p_m+p_n+p_r+p_s-p_i-p_j-p_k-p_l)(t^{(4)}-t^{(4)}_{ijkl;mnrs}) \; .
\eez
Depending on the order of the $p$'s, and whether $t^{(4)} > t^{(4)}_{ijkl;mnrs}$ or $t^{(4)} < t^{(4)}_{ijkl;mnrs}$,
this determines the relative order of $t_{ijkl}$ and $t_{mnrs}$.

\begin{example}
Let $M=5$. 
The additional critical values of $t^{(4)}$ can be used to refine Figs.~\ref{fig:T4_t4deform1} and \ref{fig:T4_t4deform2}. For $t^{(5)} < t^{(5)}_{12345}$, they have to satisfy the inequalities
$t^{(4)}_{1256;2345} < t^{(4)}_{12345} < t^{(4)}_{12356} < t^{(4)}_{1236;3456} < t^{(4)}_{13456}$. 
Indeed, we find
\bez
    t^{(4)}_{1236;3456} - t^{(4)}_{12356}
  &=& - \frac{(p_4-p_1)(p_4-p_2)}{p_4 + p_5 - p_1 - p_2} \, (t^{(5)} - t^{(5)}_{123456}) > 0 \, , \\
    t^{(4)}_{13456} - t^{(4)}_{1236;3456} 
  &=& - \frac{(p_4-p_2)(p_5-p_2)}{p_4 + p_5 - p_1 - p_2} \, (t^{(5)} - t^{(5)}_{123456}) > 0 \, ,
\eez
so that $t^{(4)}_{1236;3456}$ always exists. This is not so for $t^{(4)}_{1256;2345}$. Firstly, 
it is only defined if $p_1+p_6 \neq p_3+p_4$. Secondly, 
\bez
    t^{(4)}_{12345} - t^{(4)}_{1256;2345} 
  = \frac{(p_6-p_3)(p_6-p_4)}{p_1 + p_6 - p_3 - p_4} \, (t^{(5)} - t^{(5)}_{123456}) 
\eez
is positive only if $p_1 + p_6 < p_3 + p_4$ holds. 

For $t^{(5)} > t^{(5)}_{12345}$, the inequalities  
$t^{(4)}_{1256;2345} < t^{(4)}_{23456} < t^{(4)}_{12456} < t^{(4)}_{1234;1456} < t^{(4)}_{12346}$ 
have to be satisfied. We find 
\bez
    t^{(4)}_{1234;1456} - t^{(4)}_{12456}
  &=& \frac{(p_5-p_3)(p_6-p_3)}{p_5 + p_6 - p_2 - p_3} \, (t^{(5)} - t^{(5)}_{123456}) > 0 \, , \\
    t^{(4)}_{12346} - t^{(4)}_{1234;1456} 
  &=& \frac{(p_5-p_2)(p_5-p_3)}{p_5 + p_6 - p_2 - p_3} \, (t^{(5)} - t^{(5)}_{123456}) > 0 \, ,
\eez
so that $t^{(4)}_{1234;1456}$ always exists. But $t^{(4)}_{1256;2345}$ only shows up if 
\bez
    t^{(4)}_{23456} - t^{(4)}_{1256;2345} 
  = \frac{(p_3-p_1)(p_4-p_1)}{p_1 + p_6 - p_3 - p_4} \, (t^{(5)} - t^{(5)}_{123456}) 
\eez
is positive, which requires $p_1 + p_6 > p_3 + p_4$.
The possible orders of the critical $t^{(4)}$-values are summarized in Table~\ref{table:t4rels}. 
\begin{table}[H]
\begin{center}
\begin{tabular}{|l|l|l|} 
\hline
     $t^{(5)} < t^{(5)}_{12345}$ 
   & $p_1 + p_6 < p_3 + p_4$
   & $t^{(4)}_{1256;2345} < t^{(4)}_{12345} < t^{(4)}_{12356} < t^{(4)}_{1236;3456} < t^{(4)}_{13456}$ \\
\hline 
   & $p_1 + p_6 > p_3 + p_4$
   & $t^{(4)}_{12345} < t^{(4)}_{12356} < t^{(4)}_{1236;3456} < t^{(4)}_{13456}$  \\
\hline 
     $t^{(5)} > t^{(5)}_{12345}$ 
   & $p_1 + p_6 > p_3 + p_4$ 
   & $t^{(4)}_{1256;2345} < t^{(4)}_{23456} < t^{(4)}_{12456} < t^{(4)}_{1234;1456} < t^{(4)}_{12346}$  \\
\hline
   & $p_1 + p_6 < p_3 + p_4$
   & $t^{(4)}_{23456} < t^{(4)}_{12456} < t^{(4)}_{1234;1456} < t^{(4)}_{12346}$   \\
\hline
\end{tabular}
\parbox{15cm}{
\caption{$M=5$. The conditions under which the additional critical values $t^{(4)}_{1256;2345}$ 
and $t^{(4)}_{1236;3456}$, or both, are realized, and the corresponding order of critical values.
 \label{table:t4rels}  }
}
\end{center}
\end{table}
The additional critical values of $t^{(4)}$ moreover allow us to express the conditions 
in Table~\ref{table:T4conditions} under which a line soliton solution corresponds to one of the  
maximal chains in $\bbT_4$ in terms of inequalities involving only $t^{(4)}$ and its critical values. 
Here we use some of the above and similar expressions for the differences of critical values of $t^{(4)}$. 
The results are collected in Table~\ref{table:T4_t4conds}. 
\begin{table}[H]
\begin{center}
\begin{tabular}{|c|l|l|} 
\hline
 1 & $t_{1234}, t_{1245},t_{2345},t_{1256},t_{2356},t_{3456}$ 
   & $t^{(4)} < \min\{ t^{(4)}_{12345}, t^{(4)}_{23456}, t^{(4)}_{2345;1256} \}$ \\
\hline 
 2 & $t_{1234}, t_{1245},t_{1256},t_{2345},t_{2356},t_{3456}$ 
   & $t^{(4)}_{2345;1256} < t^{(4)} < \min\{ t^{(4)}_{12345}, t^{(4)}_{23456} \}$ \\
\hline
 3 & $t_{1234}, t_{1245},t_{1256},t_{2456},t_{2346}$ 
   & $t^{(4)}_{23456} < t^{(4)} < \min\{ t^{(4)}_{12345}, t^{(4)}_{12456} \}$  \\
\hline
 4 & $t_{1234}, t_{1456},t_{1246},t_{2346}$ 
   & $t^{(4)}_{12456} < t^{(4)} < \min\{ t^{(4)}_{12346}, t^{(4)}_{1234;1456} \}$   \\
\hline
 5 & $t_{1456}, t_{1234},t_{1246},t_{2346}$ 
   & $t^{(4)}_{1234;1456} < t^{(4)} < t^{(4)}_{12346}$   \\
\hline
 6 & $t_{1456}, t_{1346},t_{1236}$ 
   & $t^{(4)} > \max\{ t^{(4)}_{12346}, t^{(4)}_{13456} \}$    \\
\hline
 7 & $t_{1345}, t_{1356},t_{3456},t_{1236}$ 
   & $t^{(4)}_{12345} < t^{(4)} < \min\{ t^{(4)}_{12356}, t^{(4)}_{23456} \}$  \\
\hline
 8 & $t_{1345}, t_{1356},t_{1236},t_{3456}$ 
   & $t^{(4)}_{12356} < t^{(4)} < \min\{ t^{(4)}_{13456}, t^{(4)}_{1236;3456} \}$  \\
\hline
 9 & $t_{1345}, t_{1235},t_{1256},t_{2356},t_{3456}$ 
   & $t^{(4)}_{1236;3456} < t^{(4)} < t^{(4)}_{13456}$   \\
\hline
\end{tabular}
\parbox{15cm}{
\caption{The sequences of critical times determining the nine maximal chains in the Tamari lattice 
$\bbT_4$, and the conditions under which 
they are realized by line soliton solutions, here expressed in terms of $t^{(4)}$ and its 
extended set of critical values. Note that the latter are functions of $t^{(5)}$ and, depending 
on the value of $t^{(5)}$, they satisfy certain inequalities.  
See also Fig.~\ref{fig:T4_fine_structure}. \label{table:T4_t4conds} }
}
\end{center}
\end{table}
\begin{figure}[H] 
\begin{center} 
\resizebox{9.5cm}{!}{
\includegraphics{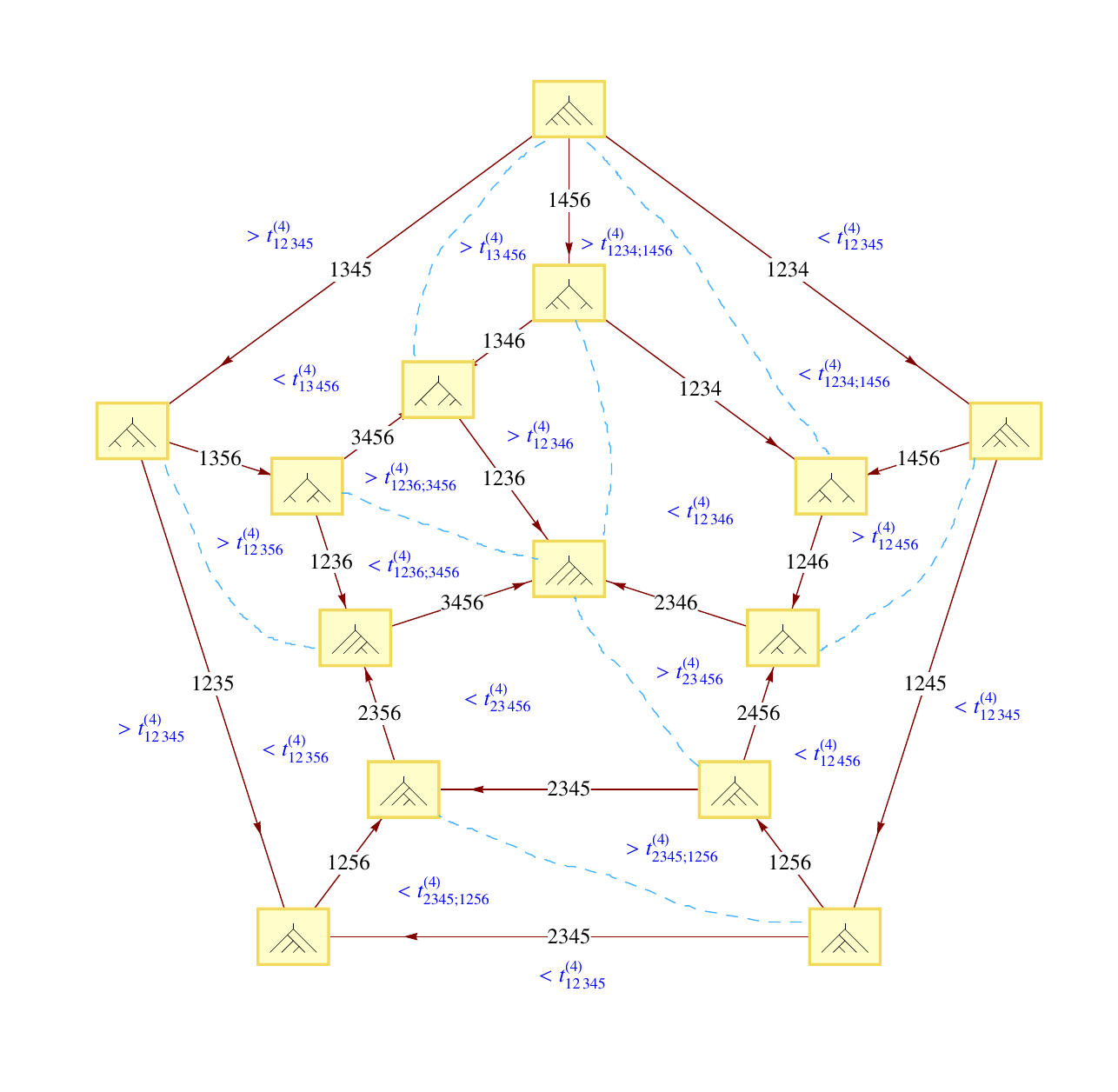}
}
\parbox{15cm}{
\caption{A representation of the Tamari lattice $\bbT_4$ and the conditions on $t^{(4)}$ under which 
the respective chains are realized by line soliton solutions. Here e.g. $>t^{(4)}_{12346}$ stands for 
$t^{(4)}>t^{(4)}_{12346}$. A number $ijkl$ assigned to an edge represents a critical transition 
time $t_{ijkl}$. On a dashed line, $t^{(4)}$ is equal to a critical value, and this corresponds 
to a direct transition, skipping a next neighbor on a maximal chain. For any of the 
additional critical values that take care of trees with levels, this is a transition in a 
tetragon (whereas for an ordinary critical value it takes place in a pentagon). 
   \label{fig:T4_fine_structure} }
}
\end{center} 
\end{figure}
\end{example}

\section{A symbolic representation of trees with levels, and a relation between 
permutohedra and Tamari lattices}
\label{AppC}
\setcounter{equation}{0}
\setcounter{theorem}{0}
This appendix presents some results that should also be of interest beyond the line soliton 
classification problem. We should stress, however, that not all statements are accompanied by 
a rigorous proof. 

\subsection{A poset structure for permutohedra}
\label{subsec:perm}
Let us assign to each node of a rooted binary tree a separate level and number the levels from top to bottom. 
The node on level $i$ will then be represented by the natural number $n_i$ if it lies on 
the $n_i$-th edge, where the edges are consecutively numbered from left to right along the level. 
The highest node (root node) thus always corresponds to $n_1 =1$. 
In this way any rooted binary tree with levels \cite{Loday+Ronco98} (see also Appendix~\ref{AppB}) 
and with $r$ (internal) nodes is uniquely represented by a sequence of 
natural numbers $n_1,n_2,\ldots,n_r$ with $n_i \leq i$, $i=1,\ldots,r$, and any such sequence defines 
a rooted binary tree with levels. 
Hence we have a bijection between the set of rooted binary trees with levels and with $r$ nodes, and 
the set 
\bez
   \mathfrak{S}_r = \{ \mathbf{n} = (n_1,n_2,\ldots,n_r) \, | \, n_i \in \mathbb{N}, \, n_i \leq i,  
                       \, i=1,\ldots,r \} \; .
\eez 
This set has $r!$ elements. 
For example, the chain consisting of the first, third, fifth and last tree in Fig.~\ref{fig:treeswithlevels} 
corresponds to the chain $(1,1,1) \longrightarrow (1,2,1) \longrightarrow (1,1,2) \longrightarrow (1,2,3)$. 
The left tree in Fig.~\ref{fig:simrot} corresponds to $(1,1,2,2)$, the third to $(1,2,2,3)$.

On $\mathfrak{S}_r$ we define an action of the permutation group $\mathcal{S}_r$ as follows. 
Let $\sigma \, : \, \mathbb{N} \times \mathbb{N} \rightarrow \mathbb{N} \times \mathbb{N}$ be given by
\bez
     \sigma(m,n) = \left\{  \begin{array}{l} (n,m+1) \\ (n-1,m) \end{array} \right. \quad
                   \mbox{if} \quad \begin{array}{l} m \geq n \\ m < n \end{array} \; .
\eez
Clearly, $\sigma$ is involutory: $\sigma^2 = \mathrm{id}$. 
For $s=1,\ldots, r-1$, let $\sigma_s \, : \, \mathfrak{S}_r \rightarrow \mathfrak{S}_r$ be the map 
given by application of $\sigma$ to the $s$-th pair, counted \emph{from right to left}, in the 
sequence of natural numbers defining an element of $\mathfrak{S}_r$, i.e.
\bez
   \sigma_s(n_1,\ldots, n_r) 
 = \left\{ \begin{array}{l} (n_1,\ldots, n_{r-s-1}, n_{r-s+1}, n_{r-s}+1, n_{r-s+2}, \ldots, n_r) \\
   (n_1,\ldots, n_{r-s-1}, n_{r-s+1}-1, n_{r-s}, n_{r-s+2}, \ldots, n_r) \end{array} \quad 
   \mbox{if} \quad \begin{array}{l} n_{r-s} \geq n_{r-s+1} \\ n_{r-s} < n_{r-s+1} \end{array} \right. \; .
\eez
Then we have the relations
\bez
    \sigma_s^2 = \mathrm{id} \, , \qquad 
    \sigma_s \, \sigma_{s+1} \, \sigma_s = \sigma_{s+1} \, \sigma_s \, \sigma_{s+1} \, , \qquad
    \sigma_s \, \sigma_{s'} = \sigma_{s'} \, \sigma_s \quad \mbox{if} \quad |s-s'|>1 \, ,
\eez
and we have an action of the symmetric group $\mathcal{S}_r$ on $\mathfrak{S}_r$. 

Let $\sigma^H$ be the restriction of $\sigma$ to 
$H = \{ (n_1,n_2) \in \mathbb{N} \times \mathbb{N} \, | \, n_1 \geq n_2 \}$. 
Defining  
\bez
     \mathbf{n} \prec \mathbf{n}' \qquad \mbox{if} \quad \mathbf{n}' = \sigma^H_s(\mathbf{n}) 
     \quad \mbox{for some } \, s   \, ,
\eez
we obtain in an obvious way a partial order $\preceq$ on $\mathfrak{S}_r$. 
Then $(1,\ldots,1)$ is minimal and $(1,2,\ldots,r)$ is maximal with respect to this partial order. 
This results in a poset underlying the \emph{permutohedron} of order $r$ 
\cite{Bowman72, Ziegler95}.\footnote{See 
also \cite{Loday+Ronco98} for a way to associate a permutation with each tree with levels, and hence 
with any sequence in $\mathfrak{S}_r$ for some $r \in \mathbb{N}$. }  
\vskip.1cm

\begin{figure}[H] 
\begin{center} 
\resizebox{!}{4.cm}{
\includegraphics{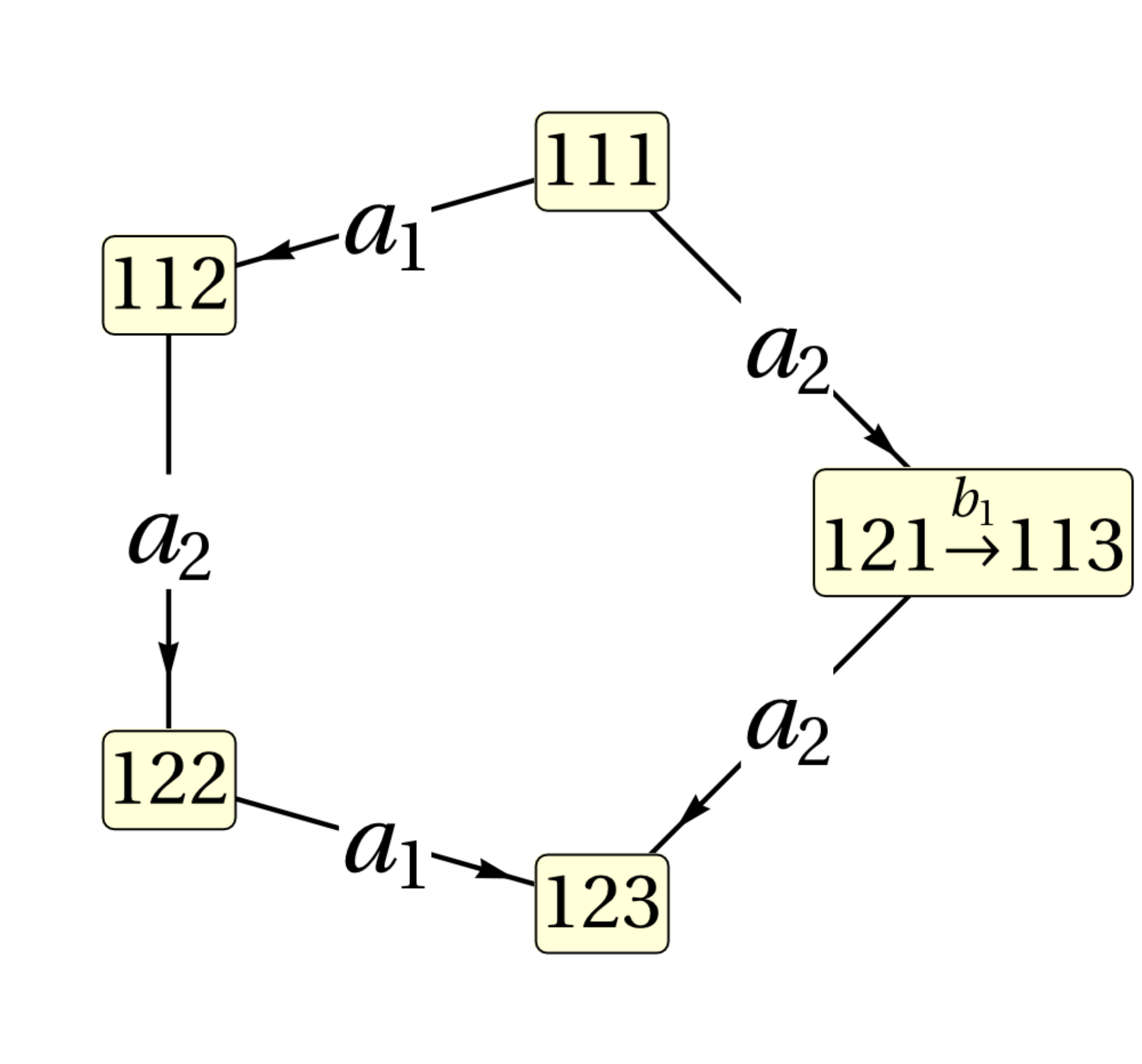}
}
\parbox{15cm}{
\caption{The Tamari lattice $\bbT_3$ as a collapsed permutohedron of order three. The sequences of integers 
at the nodes represent rooted binary trees. Here e.g. $112$ stands for $(1,1,2) \in \mathfrak{S}_3$. \label{fig:P3->T3} }
}
\end{center} 
\end{figure}

For what follows it is convenient to split the operation $\sigma^H$ into two operations
$a$ and $b$, according to a split of $H$ into its diagonal part and the rest. Hence, for 
$m,n \in \mathbb{N}$ we have
\bez
     a(n,n) = (n,n+1) \, , \qquad 
     b(m,n) = (n,m+1) \quad \forall m > n \; .
\eez
As a consequence of their origin, the operations $a_s$ and $b_s$ satisfy the braid relation
\be
     a_s a_{s+1} a_s = a_{s+1} b_s a_{s+1} \; .              \label{braid_rules1}
\ee
This is only defined on a subsequence of the form $n,n,n$ (with the last $n$ at position $s$, 
counted from the right). 
For $r=3$, (\ref{braid_rules1}) applied to the minimal element $(1,1,1)$ generates the whole poset 
underlying the permutohedron of order three, 
see Fig.~\ref{fig:P3->T3}. We observe that it collapses to the Tamari lattice $\bbT_3$ if we 
identify $(1,2,1)$ and $(1,1,3)$, which are related by $b_1$ and which are trees with levels 
having the same underlying rooted binary tree. 
In addition, we have the identities
\be
    b_s b_{s+1} a_s = a_{s+1} b_s b_{s+1} \, , \qquad
    a_s b_{s+1} b_s = b_{s+1} b_s a_{s+1} \, , \qquad
    b_s b_{s+1} b_s = b_{s+1} b_s b_{s+1} \, ,                 \label{braid_rules2}
\ee
(where the first relation is only defined on $m,n,n$ with $m >n$, the second only on $m,m,n$ with $m>n$, 
and the third only on $k,m,n$ with $k>m>n$), 
and also
\be
     a_s a_{s'} = a_{s'} a_s \, , \quad
     a_s b_{s'} = b_{s'} a_s \, , \quad 
     b_s b_{s'} = b_{s'} b_s   \qquad \mbox{for} \quad  |s-s'|>1 \; . \label{braid_rules3}
\ee

\begin{proposition} 
\label{prop:perm_maxchain}
A special maximal chain in the permutohedron poset $(\mathfrak{S}_r, \preceq)$ is obtained by application 
of\footnote{The brackets are only used to display the structure of these expressions more clearly. }
\bez
     a_1 (a_2 a_1) (a_3 a_2 a_1) \cdots (a_{r-2} \cdots a_1) (a_{r-1} \cdots a_1) 
\eez
to the minimal element $11\ldots 1$ (with $r$ times $1$). Its length is $\frac{1}{2}(r-1)r$.
\end{proposition}
\noindent
\textit{Proof:} 
Stepwise application of $a_{r-1} \cdots a_1$ yields 
$(1,\ldots,1) \stackrel{a_1}{\longrightarrow} (1,\ldots,1,2) \stackrel{a_2}{\longrightarrow} (1,\ldots,1,2,2)
\stackrel{a_3}{\longrightarrow} \cdots \stackrel{a_{r-1}}{\longrightarrow} (1,2,\ldots,2)$. 
Application of the next subsequence leads to 
$(1,2,\ldots,2) \stackrel{a_1}{\longrightarrow} (1,2,\ldots,2,3) \stackrel{a_2}{\longrightarrow} \cdots 
\stackrel{a_{r-2}}{\longrightarrow} (1,2,3,\ldots,3)$. Continuing in this way, we finally obtain 
the maximal element $(1,2,\ldots,r)$. 
The total number of $a$'s in the sequence is $\sum_{n=1}^{r-1} n = (r-1) r/2$.  
\hfill $\square$

\begin{remark} 
The application of an $a$ or $b$ to an element $\mathbf{n} \in \mathfrak{S}_r$ raises the \emph{weight} 
$|\mathbf{n}| = n_1 + \cdots + n_r$ by $1$. 
In order to get from $(1,\ldots, 1)$, which has weight $r$, to $(1,2,\ldots, r)$ with weight $r(r+1)/2$, 
we need $r(r+1)/2-r = r(r-1)/2$ operations of the type $a$ or $b$. This shows that all chains in the 
permutohedron have the same length, namely $r(r-1)/2$. 
\end{remark}

\begin{figure}[H] 
\begin{center} 
\resizebox{!}{8.cm}{
\includegraphics{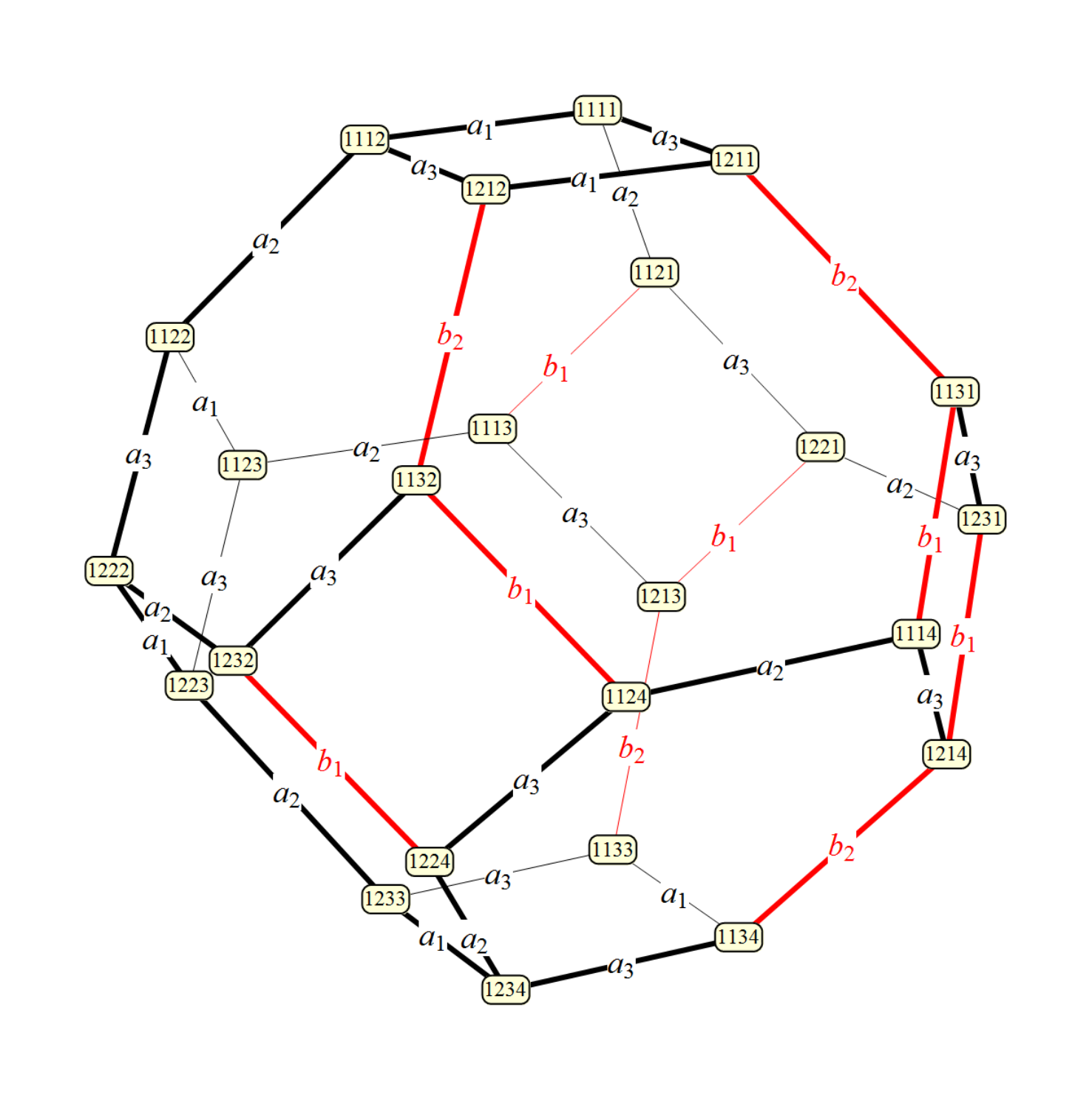}
}
\parbox{15cm}{
\caption{A poset structure for the permutohedron of order four. 
Here e.g. $1211$ stands for $(1,2,1,1) \in \mathfrak{S}_4$.
\label{fig:permutohedron} }
}
\end{center} 
\end{figure}
Fig.~\ref{fig:permutohedron} shows the permutohedron of order four (i.e. $r=4$), supplied with the 
poset structure introduced above. The 16 maximal chains are generated via application of the above 
braid relations to the sequence $a_1 a_2 a_1 a_3 a_2 a_1$ that determines a maximal chain according to 
proposition~\ref{prop:perm_maxchain}:\footnote{If a sequence of $a$'s and $b$'s maps the 
minimal element $(1,\ldots,1)$ to the maximal element $(1,2,\ldots,r)$ of $\mathfrak{S}_r$, this 
remains true for any sequence obtained from it via application of the braid rules. Hence every 
sequence obtained in this way again generates a maximal chain in $\mathfrak{S}_r$. }  \\
$a_1 a_2 a_1 a_3 a_2 a_1$, $a_1 a_2 a_3 a_1 a_2 a_1$, $a_2 b_1 a_2 a_3 a_2 a_1$, $a_1 a_2 a_3 a_2 b_1 a_2$,
$(a_2 b_1 a_3 b_2 a_3 a_1, a_2 a_3 b_1 b_2 a_3 a_1)$, \\
$(a_2 b_1 a_3 b_2 a_1 a_3, a_2 a_3 b_1 b_2 a_1 a_3, 
a_2 a_3 a_2 b_1 b_2 a_3)$, $(a_1 a_3 b_2 a_3 b_1 a_2, a_1 a_3 b_2 b_1 a_3 a_2)$, \\
$(a_3 a_1 b_2 a_3 b_1 a_2, a_3 a_1 b_2 b_1 a_3 a_2, a_3 b_2 b_1 a_2 a_3 a_2)$, 
$(a_3 b_2 a_3 b_1 b_2 a_3, a_3 b_2 b_1 a_3 b_2 a_3)$. \\
Here we grouped those chains together that are related by a braid relation which only involves $b$'s. 
We shall see that also this permutohedron can be collapsed to the corresponding Tamari lattice $\bbT_4$.

\subsection{From permutohedra to Tamari lattices}
\begin{figure}[H] 
\begin{center} 
\resizebox{!}{2.5cm}{
\includegraphics{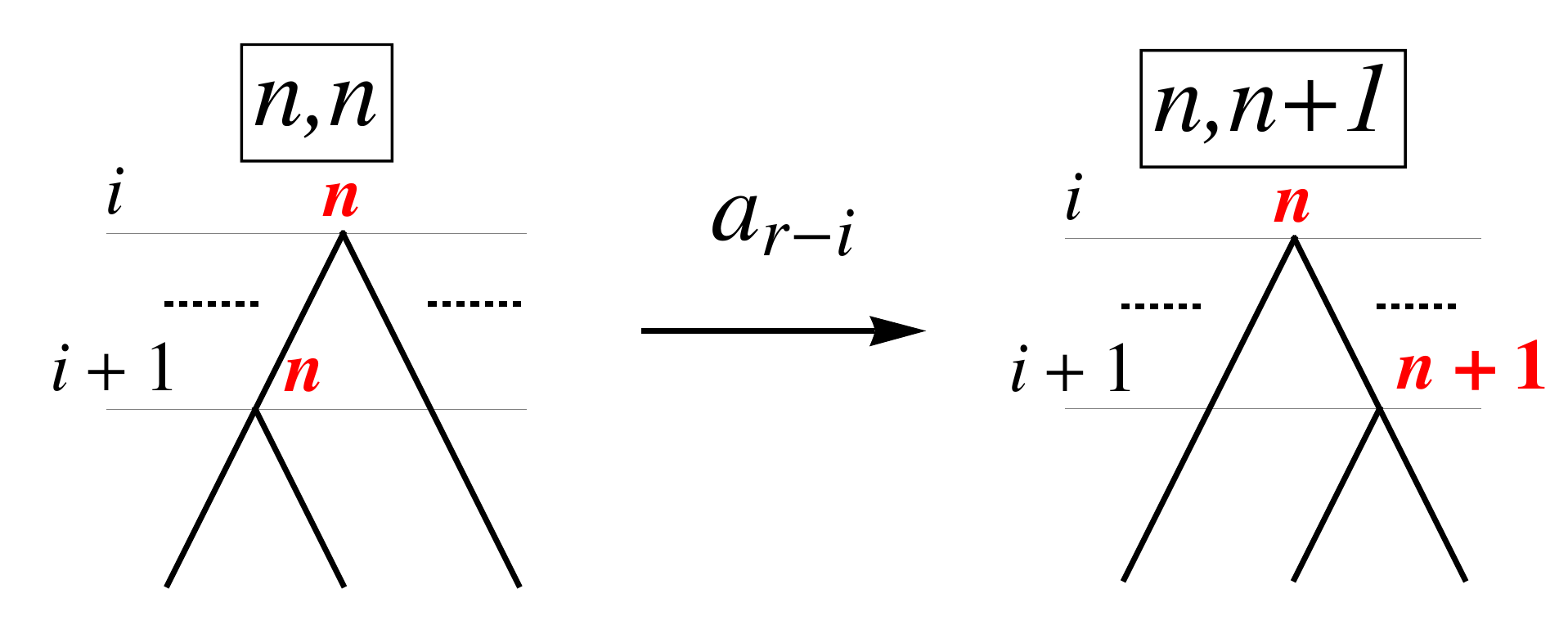}
}
\parbox{15cm}{
\caption{The operation $a_{r-i}$ amounts to a right rotation taking place 
between the two levels $i$ and $i+1$. For the left tree we have $n_i = n_{i+1} = n$, for the right tree
$n_i=n$ and $n_{i+1} = n+1$. \label{fig:a_tree-rotation} }
}
\end{center} 
\end{figure}

\begin{figure}[H] 
\begin{center} 
\resizebox{!}{2.5cm}{
\includegraphics{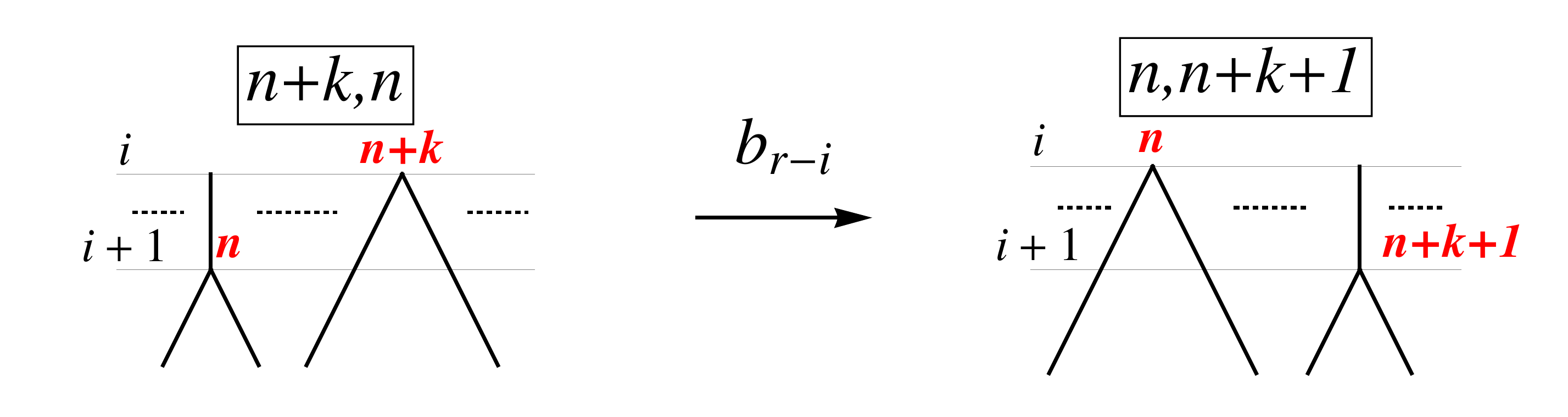}
}
\parbox{15cm}{
\caption{The operation $b_{r-i}$ exchanges the nodes of two consecutive levels. 
For the left tree we have $n_i = n+k$, $k \in \mathbb{N}$, and $n_{i+1} = n$, for the right tree 
$n_i=n$ and $n_{i+1} = n+k+1$.
\label{fig:b_tree-level-exchange} }
}
\end{center} 
\end{figure}

As explained in Fig.~\ref{fig:a_tree-rotation}, the operation $a_s$ corresponds to a right rotation in a 
rooted binary tree, which is the characteristic property of a Tamari lattice.
An application of $b_s$ does \emph{not} change the respective underlying rooted binary tree, but 
only exchanges the associated rooted binary trees \emph{with levels}, see 
Fig.~\ref{fig:b_tree-level-exchange}.

Identifying those rooted binary trees with levels that correspond to the same rooted binary 
tree (without levels), we can use as representative the sequence for which we also have 
$n_i \leq n_{i+1}$ (see also \cite{Stan99v2}). This defines 
a bijection between the set of rooted binary trees with $r$ nodes and 
\bez
    \mathfrak{Y}_r = \{ (n_1,n_2,\ldots,n_r) \, | \, n_i \in \mathbb{N}, \, 
    n_i \leq i \; \mbox{and} \; n_i \leq n_{i+1} \; \forall i \} \; .
\eez
The number of elements of this set is the Catalan number $\mathfrak{c}_r = \frac{1}{r+1} {2r \choose r }$ 
(see exercise 19 in \cite{Stan99v2}). The above partial order on $\mathfrak{S}_r$ induces a 
partial order on $\mathfrak{Y}_r$, and in this way a permutohedron collapses to the corresponding 
Tamari lattice (or associahedron, see also \cite{Tonks97}). 

\begin{remark}
In section~\ref{subsec:first_step} we described the nodes of a rooted binary tree, describing a line 
soliton solution at some event, as coincidences of three phases. Ordering the nodes from top to bottom 
and from left to right, this assigns a sequence $(i_1,j_1,k_1), \ldots, (i_r,j_r,k_r)$ of ordered 
triples of natural numbers, $i_m < j_m < k_m$, to the tree. 
Then the sequence $i_1,i_2,\ldots,i_r$ of the first indices is precisely the sequence of 
natural numbers in $\mathfrak{Y}_r$ that characterizes the tree in the way described above. 
This correspondence does \emph{not} extend to trees with levels. 
\end{remark}

By definition, the operation $a_s$ preserves $\mathfrak{S}_r$ (hence operates on trees with levels), 
but it does not preserve $\mathfrak{Y}_r$. 
We can correct this by application of operations $b_s$ (which are \emph{not} defined 
on $\mathfrak{Y}_r$). 
Indeed, one can show that for any sequence $\mathbf{n} \in \mathfrak{S}_r \setminus \mathfrak{Y}_r$, 
there is a finite combination of $b$'s that transforms it into a sequence in $\mathfrak{Y}_r$. 

In describing Tamari lattices, hence disregarding the refinement to trees with levels, we 
have to regard two sequences of $a$'s and $b$'s as equivalent if they only differ by an application 
of any of the rules (\ref{braid_rules2}), and those in (\ref{braid_rules3}) involving $b$'s.
The restriction of the permutohedron poset to $\mathfrak{Y}_r$ selects those sequences in which 
any application of some $a_s$ that leads out of $\mathfrak{Y}_r$ is immediately corrected by 
$b$'s. Hence these are sequences where all $b$'s are commuted as far as possible to the right, 
using the braid rules that involve $b$, with the exception of (\ref{braid_rules1}). For the 
permutohedron of order four, the 16 maximal chains given in section~\ref{subsec:perm} reduce 
to 9 maximal chains, which (applied to $(1,1,1,1)$) generate the maximal chains of the Tamari 
lattice $\bbT_4$ (cf. Table~\ref{table:T4conditions}). 

Stepwise application of the special sequence of $a$'s in Proposition~\ref{prop:perm_maxchain} to 
the minimal element $(1,\ldots,1)$ actually generates a sequence of elements in $\mathfrak{Y}_r$
(see the proof of the proposition). 
Since the application of $a$ encodes the characteristic property of a Tamari lattice, this 
determines a maximal chain in a Tamari lattice. Its length is $(r-1) r/2$, and this is known 
to be the greatest length of a chain in $\bbT_r$ \cite{Mark92}. 

\begin{proposition} 
\label{prop:shortest_maxTamari-chain}
A shortest maximal chain in the Tamari lattice $\bbT_r$ is obtained by application of 
\bez
     a_{r-1} (b_{r-2}) a_{r-1} (b_{r-3} b_{r-2}) a_{r-1} (b_{r-4} b_{r-3} b_{r-2}) a_{r-1} \cdots
     a_{r-1} (b_1 \cdots b_{r-2}) a_{r-1} 
\eez
to the minimal element $(1,\ldots,1)$ of $(\mathfrak{Y}_r,\preceq)$. 
\end{proposition}
\noindent
\textit{Proof:} 
Application of $b_1 \cdots b_{r-2} a_{r-1}$ yields
$(1,\ldots,1) \stackrel{a_{r-1}}{\longrightarrow} (1,2,1,\ldots,1) 
\stackrel{b_{r-2}}{\longrightarrow} (1,1,3,1,\ldots,1) \stackrel{b_{r-3}}{\longrightarrow} (1,1,1,4,1,\ldots,1) 
\stackrel{b_{r-4}}{\longrightarrow} \cdots \stackrel{b_1}{\longrightarrow} (1,\ldots,1,r)$. 
The next subsequence $b_2 \cdots b_{r-2} a_{r-1}$ maps $(1,\ldots,1,r)$ to $(1,\ldots,1,r-1,r)$. 
Continuing in this way, we finally obtain $(1,2,\ldots,r-1,r)$, the final node of $\bbT_r$. 
Hence the chain is maximal. The total number of $a$'s is $r-1$, 
which is known to be the shortest length of a maximal Tamari chain \cite{Mark92}. 
\hfill $\square$
\vskip.2cm

Two sequences of $a$'s and $b$'s are said to belong to the same \emph{class} if 
they differ only by an application of $a_s a_{s'} = a_{s'} a_s$ for $|s-s'|>1$. 
In particular, for $n>3$, this rule creates further longest 
maximal chains from those in Propositions~\ref{prop:perm_maxchain} and \ref{prop:shortest_maxTamari-chain}. 
The ``pentagon rule'' (\ref{braid_rules1}) changes a sequence (and hence a Tamari chain) 
in a more drastic way (since it changes the number of $a$'s). 

$\bbT_3$ consists of two chains, each of which is a class:
$a_1 a_2 a_1$ and $a_2 b_1 a_2$. For $\bbT_4$ there are six classes: 
(1) $a_1 a_2 a_3 a_1 a_2 a_1$ and $a_1 a_2 a_1 a_3 a_2 a_1$, 
(2) $a_1 a_2 a_3 a_2 b_1 a_2$, (3) $a_2 b_1 a_2 a_3 a_2 a_1$, 
(4) $a_1 a_3 b_2 a_3 b_1 a_2$ and $a_3 a_1 b_2 a_3 b_1 a_2$, 
(5) $a_2 a_3 a_2 b_1 b_2 a_3$, (6) $a_3 b_2 a_3 b_1 b_2 a_3$. 
For $\bbT_5$ there are 25 classes and 94 chains, see Table~\ref{table:T5chains} and Fig.~\ref{fig:superT4}. 
\begin{table}[H]
\begin{center}
\begin{tabular}{|r|l|} 
\hline
 1 & $1234123121$, $1231423121$, $1213423121$, $1231243121$, $1213243121$, $1234121321$, $1231421321$, \\
   &   $1213421321$, $1231241321$, $1213241321$, $1231214321$, $1213214321$  \\
\hline 
 2 & $12341232\mathbf{1}2$, $12314232\mathbf{1}2$, $12134232\mathbf{1}2$, $12312432\mathbf{1}2$, 
     $12132432\mathbf{1}2$   \\
\hline
 3 & $2\mathbf{1}23423121$, $2\mathbf{1}23243121$, $2\mathbf{1}23421321$, $2\mathbf{1}23241321$, 
     $2\mathbf{1}23214321$ \\
\hline
 4 & $12342\mathbf{1}2321$, $1232\mathbf{1}24321$, $12324\mathbf{1}2321$    \\
\hline
 5 & $2\mathbf{1}232432\mathbf{1}2$, $2\mathbf{1}234232\mathbf{1}2$, \\
\hline
 6 & $123413\mathbf{2}3\mathbf{1}2$, $123431\mathbf{2}3\mathbf{1}2$, $123143\mathbf{2}3\mathbf{1}2$,
     $121343\mathbf{2}3\mathbf{1}2$ \\
\hline
 7 & $123423\mathbf{12}31$, $123243\mathbf{12}31$, $1234231\mathbf{12}3$, $1232431\mathbf{12}3$ \\
\hline
 8 & $13\mathbf{2}34\mathbf{1}2321$, $13\mathbf{2}3\mathbf{1}24321$, $31\mathbf{2}34\mathbf{1}2321$, 
     $31\mathbf{2}3\mathbf{1}24321$ \\
\hline
 9 & $23\mathbf{12}343121$, $23\mathbf{12}341321$, $23\mathbf{12}314321$, $23\mathbf{12}134321$  \\
\hline
10 & $1214\mathbf{3}4\mathbf{2}3\mathbf{1}2$, $1241\mathbf{3}4\mathbf{2}3\mathbf{1}2$,  
     $1421\mathbf{3}4\mathbf{2}3\mathbf{1}2$, $4121\mathbf{3}4\mathbf{2}3\mathbf{1}2$,
     $124\mathbf{3}41\mathbf{2}3\mathbf{1}2$, $142\mathbf{3}41\mathbf{2}3\mathbf{1}2$,
     $412\mathbf{3}41\mathbf{2}3\mathbf{1}2$     \\
\hline
11 & $12343\mathbf{2}3\mathbf{12}3$ \\
\hline
12 & $13\mathbf{2}343\mathbf{12}31$, $13\mathbf{2}3432\mathbf{12}3$, $31\mathbf{2}343\mathbf{12}31$, 
     $31\mathbf{2}3432\mathbf{12}3$ \\
\hline
13 & $2\mathbf{1}2343\mathbf{2}3\mathbf{1}2$ \\
\hline
14 & $23\mathbf{12}3432\mathbf{1}2$ \\
\hline
15 & $3\mathbf{2}3\mathbf{12}34321$ \\
\hline
16 & $234\mathbf{123}4121$, $2342\mathbf{123}421$, $23423\mathbf{123}41$, $234232\mathbf{123}4$, 
     $2324\mathbf{123}421$, $23243\mathbf{123}41$, $232432\mathbf{123}4$  \\
\hline
17 & $3\mathbf{2}3432\mathbf{123}4$, $3\mathbf{2}343\mathbf{123}41$, $3\mathbf{2}34\mathbf{123}421$   \\
\hline
18 & $341\mathbf{23}4\mathbf{12}31$, $341\mathbf{23}42\mathbf{12}3$, $3413\mathbf{23}4\mathbf{12}3$,
     $314\mathbf{23}4\mathbf{12}31$, $314\mathbf{23}42\mathbf{12}3$, $3143\mathbf{23}4\mathbf{12}3$   \\
\hline
19 & $234\mathbf{123}42\mathbf{1}2$, $2343\mathbf{123}4\mathbf{1}2$, $2343\mathbf{2}3\mathbf{123}4$   \\
\hline
20 &  $2\mathbf{1}24\mathbf{3}4\mathbf{2}3\mathbf{1}2$, $2\mathbf{1}42\mathbf{3}4\mathbf{2}3\mathbf{1}2$, 
      $42\mathbf{1}2\mathbf{3}4\mathbf{2}3\mathbf{1}2$  \\
\hline
21 & $124\mathbf{3}4\mathbf{2}3\mathbf{12}3$, $142\mathbf{3}4\mathbf{2}3\mathbf{12}3$, 
     $412\mathbf{3}4\mathbf{2}3\mathbf{12}3$ \\
\hline
22 & $14\mathbf{3}4\mathbf{23}4\mathbf{12}3$, $41\mathbf{3}4\mathbf{23}4\mathbf{12}3$, 
     $4\mathbf{3}41\mathbf{23}4\mathbf{12}3$   \\
\hline
23 & $42\mathbf{3}4\mathbf{123}4\mathbf{1}2$, $42\mathbf{3}4\mathbf{2}3\mathbf{123}4$, 
     $24\mathbf{3}4\mathbf{123}4\mathbf{1}2$, $24\mathbf{3}4\mathbf{2}3\mathbf{123}4$ \\
\hline
24 & $343\mathbf{23}4\mathbf{123}4$, $34\mathbf{23}42\mathbf{123}4$, $34\mathbf{23}4\mathbf{123}41$   \\
\hline
25 &  $4\mathbf{3}4\mathbf{23}4\mathbf{123}4$ \\
\hline
\end{tabular}
\parbox{15cm}{
\caption{A representation of the 25 classes of maximal chains of the Tamari lattice $\bbT_5$ in 
terms of the braid operations. 
Here a number $s$ in boldface stands for $b_s$, otherwise for $a_s$. 
\label{table:T5chains} }
}
\end{center}
\end{table}

\section{Tropical approximation}
\label{AppD}
\setcounter{equation}{0}
\setcounter{theorem}{0}
After the rescaling $t^{(n)} \mapsto t^{(n)}/\hbar$ and $c_i \mapsto c_i/\hbar$, 
with a constant $\hbar$, the class of solutions studied in sections~\ref{sec:simplest_class}, 
\ref{sec:evolution} and Appendix~\ref{AppA} is given by
\bez
    \tau = \sum_{i=1}^{M+1} e^{\theta_i/\hbar} \, , \qquad
    \theta_i = \sum_{n=1}^M p_i^{n} \, t^{(n)} + c_i \; . 
\eez
Then we have
\bez
    \lim_{\hbar \to 0} \hbar \, \log \tau 
  = \lim_{\hbar \to 0} \hbar \log\Big(\sum_{i=1}^{M+1} e^{\theta_i/\hbar}\Big) 
  = \max\{\theta_1,\ldots,\theta_{M+1}\}  \, , 
\eez
applying a formula familiar in the context of tropical mathematics\footnote{This 
formula underlies what is called ``Maslov dequantization'' \cite{Litv10}. 
A related method is ``ultra-discretization'' \cite{TTMS96}. }, 
and regarding $\theta_i$ as $\hbar$-independent. The result 
confirms our basic approximation formula in section~\ref{sec:simplest_class}. 
So far we were only interested in the (evolution of the) form of 
line solitons as contours in the $xy$-plane. But it is also of interest 
to find a good approximation for the amplitude $u$ of the KP solution e.g. 
at the meeting points of line soliton branches, hence at the coincidence points of 
phases in the tropical approximation. 
 From
\bez
     \phi = \hbar (\log \tau)_x
  = \frac{1}{\tau} \sum_{i=1}^{M+1} p_i \, e^{\theta_i/\hbar} 
  = \frac{p_k + \sum_{i=1, i \neq k}^{M+1} p_i \, e^{-(\theta_k-\theta_i)/\hbar}}{1 
         + \sum_{i=1,i\neq k}^{M+1} e^{-(\theta_k-\theta_i)/\hbar}} 
     \qquad \quad k=1,\ldots,M+1 \, ,
\eez
we obtain $\lim_{\hbar \to 0} \phi = p_k$ in the $\theta_k$-region, away from 
coincidences of phases. At a visible coincidence $\theta_{k_1}= \cdots = \theta_{k_m}$, 
which is \emph{generic} in the sense that it is not a coincidence of more 
than $m$ phases, we find $\phi=\frac{1}{m} \sum_{i=1}^m p_{k_i}$. Furthermore, 
\bez
    u &=& 2 \hbar^2 (\log \tau)_{xx} = 2 \hbar \, \phi_x
       = \frac{2}{\tau} \sum_{i=1}^{M+1} p_i^2 \, e^{\theta_i/\hbar} 
         - \frac{2}{\tau^2} \Big(\sum_{i=1}^{M+1} p_i \, e^{\theta_i/\hbar} \Big)^2 \\
      &=& \frac{2}{\tau^2} \sum_{1 \leq i < j \leq M+1} (p_j - p_i)^2 \,
          e^{(\theta_i+\theta_j)/\hbar} 
      \, = \, 2 \, \frac{\sum_{i<j} (p_j-p_i)^2 \, e^{-(\theta_k + \theta_l - \theta_i
         - \theta_j)/\hbar}}{\sum_i  e^{-(\theta_k - \theta_i)/\hbar} \, 
          \sum_j  e^{-(\theta_l - \theta_j)/\hbar} } \, ,
\eez
which implies $\lim_{\hbar \to 0} u = \frac{1}{2} (p_k-p_l)^2$ at a visible generic coincidence 
$\theta_k=\theta_l$. 

For an asymptotic soliton branch given by $\theta_m = \theta_{m+1}$ for large negative 
values of $y$, for $\hbar=1$ the above formula implies $u \sim \frac{1}{2} (p_{m+1}-p_m)^2$ 
as $y \to -\infty$, which thus coincides with the tropical value. 
A corresponding relation also holds for the remaining asymptotic soliton branch, given 
by $\theta_1 = \theta_{M+1}$, as $y \to +\infty$. 

More generally, we find
\bez
    \lim_{\hbar \to 0} u = \frac{2}{m^2} \sum_{1 \leq i < j \leq m} 
      (p_{k_j}-p_{k_i})^2 \qquad \mbox{at a visible generic coincidence}
    \quad \theta_{k_1} = \cdots = \theta_{k_m} \; .
\eez
At a highest coincidence, i.e. $\theta_1 = \cdots = \theta_{M+1}$, the tropical value is 
precisely the exact value (i.e. the corresponding value of $u$ for $\hbar=1$). 
This is not so at a (generic) visible lower coincidence. But it is clear from the 
above formula for $u$ that the corrections involve (only) exponentials of negative phase 
differences. Hence the tropical values yield a perfect approximation unless those 
phase differences become extremely small (which means that we are close to a higher 
order coincidence).

\end{appendix}

\end{document}